\documentclass[12pt,a4paper]{article}

\usepackage{amsthm,amsmath,amsfonts,amssymb}
\usepackage{a4,times,latexsym}
\usepackage{eurosym} 

\usepackage{caption} 
\usepackage{subcaption} 

\usepackage{graphicx} 
\usepackage[countmax]{subfloat} 

\usepackage{nohyperref}
\usepackage{xcolor}
\hypersetup{
    colorlinks,
    linkcolor={red!50!black},
    citecolor={blue!50!black},
    urlcolor={blue!80!black}
}

\usepackage{lscape} 
\usepackage{listings} 
\usepackage{setspace} 
\usepackage{url}      
\usepackage{scrpage2} 
\usepackage{makeidx}  
\usepackage{rotating,here} 

\usepackage{lineno} 

\usepackage{tabularx} 
\usepackage{booktabs} 
\newcolumntype{Y}{>{\raggedleft\arraybackslash}X} 

\usepackage{color}
\usepackage{psfrag}
\usepackage{lmodern}
\usepackage[USenglish]{babel}
\usepackage[T1]{fontenc}
\usepackage[latin1]{inputenc}
\usepackage[round,comma]{natbib}
\usepackage{geometry}
\theoremstyle{plain}

\theoremstyle{definition}

\usepackage{tikz,color}
\usepackage{pgflibraryshapes}
\usetikzlibrary{trees,arrows, decorations.markings, positioning}
\definecolor{grau}{rgb}{0.9,0.9,0.9}
\tikzstyle{vecArrow} = [line width=2mm, draw=black, -triangle 60,
                        postaction = {draw, line width=5mm, shorten >=6mm, -},
                        preaction = {decorate}]
\tikzstyle{innerWhite} = [line width=1.9mm, draw=grau, -triangle 60,
                        postaction = {draw, line width=4.5mm, shorten >=5.5mm, -}]

\begin{document}

\renewcommand{\thefootnote}{\arabic{footnote}}

\begin{center}
\large{\bf A New High-Dimensional Time Series Approach for Wind Speed, Wind Direction and Air Pressure Forecasting\\}
\vspace{10mm} \normalsize \today
\small

Daniel Ambach$^1$ and Wolfgang Schmid$^2$ 

\end{center}

\thispagestyle{empty} \small

\normalsize

\begin{abstract}
\noindent Many wind speed forecasting approaches have been proposed in literature. In this paper a new statistical approach for jointly predicting wind speed, wind direction and air pressure is introduced. The wind direction and the air pressure are important to extend the forecasting accuracy of wind speed forecasts. A good forecast for the wind direction helps to bring the turbine into the predominant wind direction. We combine a multivariate seasonal time varying threshold autoregressive model with interactions (TVARX) with a threshold seasonal autoregressive conditional heteroscedastic (TARCHX) model. The model includes periodicity, conditional heteroscedasticity, interactions of different dependent variables and a complex autoregressive structure with non-linear impacts. In contrast to ordinary likelihood estimation approaches, we apply a high-dimensional shrinkage technique instead of a distributional assumption for the dependent variables. The iteratively re-weighted least absolute shrinkage and selection operator (LASSO) method allows to capture conditional heteroscedasticity and a comparatively fast computing time. The proposed approach yields accurate predictions of wind speed, wind direction and air pressure for a short-term period. Prediction intervals up to twenty-four hours are presented.
\end{abstract}

\begin{small}{\bf Keywords:} Wind Speed; Wind Direction; Air Pressure; Heteroscedasticity; Stochastic Modelling; Multivariate Model; Time Series
\end{small}

\bigskip

\begin{bfseries}
Addresses:
\end{bfseries}

\begin{small}
$^{1}$ \textbf{Corresponding Author:} Daniel Ambach, European University Via\-dri\-na, Chair of Quantitative Methods and Statistics,
Post Box 1786, 15207 Frankfurt (Oder), Germany, Tel. +49 (0)335 5534 2983, Fax +49 (0)335 5534 2233, ambach@europa-uni.de.\vspace{0.1cm}

$^2$ Wolfgang Schmid,
European University Via\-dri\-na, Chair of Quantitative Methods and Statistics,
Post Box 1786, 15207 Frankfurt (Oder), Germany, Tel. +49 (0)335 5534 2427, Fax +49 (0)335 5534 2233, schmid@europa-uni.de.\vspace{0.5cm}
\end{small}

\pagestyle{plain}
\setcounter{page}{1}
\onehalfspacing

\section{Introduction}\label{section:Introduction}

\noindent The production of wind power reached 4\% of the overall electricity generation \citep[see][]{WWEA2015}. The \cite{WWEA2014} highlights that 393 Gigawatt of generation capacity is installed and the deployment of wind power exceeds 100 countries. The \cite{europeanwind} depicts that $79.1\%$ of the new installed power generation capacity in the European Union are related to renewable energy. \cite{WWEA2015} point out that the energy transition will increase the level of renewable energy worldwide. \cite{Soman2010} stress the fact why short- to medium-term wind speed forecasts are so important which is related to the electricity market and the load dispatch planning. \cite{jung2014current} present a possible definition for short-term (30min - 6h) and medium-term (6h - 24h) wind speed prediction approaches. On the one hand, these forecasts are used for on-line and off-line decisions of the power generator and on the other hand, they provide necessary information for unit commitment decisions on the spot and over-the-counter (OTC) markets.

Accurate wind speed predictions are important for the appraisal of a possible site of a wind farm and in genaral, to increase the efficiency of wind mills. \cite{zhang2013multivariate} highlight that the Wind Power Density (WPD) is a beneficial and efficient method to evaluate the wind capability at a potential new wind farm site. This function could be used to disclose the available wind power at a site. The WPD is based on the wind speed, the wind direction and the air density. The air density, the temperature and the gas constant form the air pressure. Therefore, it is essential to perform simultaneous predictions of the wind speed, the wind direction and the air pressure. 

Nowadays, it is possible to distinguish between five categories of wind speed forecasting approaches namely physical models, conventional statistical models, spatio-temporal prediction models, artificial intelligence models and combinations of all classes (hybrid structure models) \citep[see][]{Lei2009}. \cite{yesilbudak2013new} argue that physical prediction models use temperature, pressure, orography, roughness and obstacles to perform wind speed forecasts. \cite{Tascikaraoglu2014} describe the use of Numerical Weather Prediction (NWP) models which could be used as input and output for wind speed and other meteorological predictions. \cite{lange2006physical} take the calculated values of a NWP model as input quantities and apply a model to obtain better forecasts. \cite{wu2007} point out that NWP models do not update their predictions very frequently and require a lot of computing ressources. The NWP approach permits forecasts up to several days. One of the most popular NWP approaches is the Weather Research and Forecast (WRF) model \citep[see][]{carvalho2012sensitivity}. Further important attempts are the High Resolution Model (HRM), the Consortium for Small Scale Modeling (COSMO) and the Mesoscale Model 5 (MM5) \citep[see][]{su2014new}. \cite{Tascikaraoglu2014} argue that NWP is in general beneficial for longer prediction horizons, but they disclose some weaknesses as well. Sometimes they fail to capture small scale phenomena. Moreover, \cite{Tascikaraoglu2014} point out that they are not convenient for short-term forecasting and require too much time. However, \cite{jung2014current} summarise that artificial neural networks, combined approaches and statistical models are accurate for short- up to medium-term forecasts. These approaches only need historical data and the dynamics of the underlying process have to be exploited to perform accurate and fast predictions. In the recent past, miscellaneous statistical and hybrid structure approaches for wind speed prediction have been proposed, see for instance \cite{zhu2012short}, \cite{guo2014new} and \cite{ambach2015periodic}. 

In this article we develop a new multivariate statistical approach for short-term forecasting. It is based on time series analysis. \cite{bivona2011stochastic} argue that such models are very useful for short-term predictions.
We derive a new prediction model which includes different meteorological variables and their dynamics. \cite{fang2016improving} derive a wind power prediction model, that uses several numerical weather variables to extend the forecasting performance. 
Apparently, different meteorological variables can extend the accuracy of the wind speed predictions. Thus, \cite{ambach2015space} consider a multivariate modelling approach, that includes wind speed, wind direction and temperature. \cite{smith2017new} develop a stochastic temperature prediction model. Moreover, temperature is said to be negatively correlated with air pressure and it is reasonable to include the latter one instead of the temperature. \cite{erdem2011arma} and \cite{el2008one} perform joint wind speed and wind direction predictions. Furthermore, \cite{jeon2012using} take a bivariate vector autoregressive moving average (VARMA) generalized autoregressive conditional heteroscedastic (GARCH) approach to perform wind speed and wind direction forecasts. \cite{zhang2013multivariate} use a method, which they call the Multivariate and Multimodal Wind Distribution (MMWD) model to calculate wind speed, wind direction and air density predictions. Moreover, \cite{zhu2014incorporating} include the geostrophic wind speed, the temperature and the wind direction to obtain accurate short-term predictions. The wind speed \citep[see][]{calif2014multiscaling,calif2012modeling}, the wind direction \citep[see][]{masseran2015markov,petkovic2015adaptive} and the air pressure \citep[see][]{zhu2014incorporating,jeong2012multisite} are stochastic processes which exhibit different distributional assumptions.
Obviously, there are many possibilities of forecasting meteorological characteristics. Our proposed prediction approach provides accurate joint forecasts of the wind speed, the wind direction and the air pressure. A comparison study emphasizes the accuracy of our new model. Moreover, the forecasting results for short- up to medium-term predictions are compared to different benchmark models and prediction intervals are presented. 

\cite{ambach2015periodic} point out that the wind speed itself is non-normally distributed, periodic, persistent and conditional heteroscedastic. Moreover, similar features are shown for the wind direction and the air pressure. Our new model incorporates all of the previously mentioned features of the meteorological variables. Hence, our considered approach is a multivariate seasonal threshold vector autoregressive model with external regressors (TVARX). The approach is high dimensional and so, it is possible to capture the persistence, periodicity and non-linear effects. The latter effects are modelled by a threshold vector autoregressive process, where the thresholds are related to different percentiles. \cite{ziel2016lasso} use a TAR model as a tool to capture non-linear effects in the data and to perform probabilistic load forecasts. The periodic structure of the data is described by periodic B-spline functions which are used to model time varying autoregressive coefficients. However, our periodic seasonal TVARX model captures the correlation of all dependent variables and uses them to calculate accurate forecasts of the wind speed, the wind direction and the air pressure. 

The conditional variances of the considered meteorological dependent variables show heteroscedastic behaviour \citep[see][]{ewing2006time,ambach2015space}. Thus, we propose a periodic power threshold autoregressive conditional heteroscedastic process with external regressors (TARCHX) for the variance. The model is estimated by means of the least absolute shrinkage and selection operator method, which is proposed by \cite{tibshirani1996regression}. Moreover, we use the recently considered iteratively re-weighted least absolute shrinkage and selection operator (LASSO) approach to estimate the mean model and to include the conditional variance structure \citep[see][]{ziel2015efficient}. This method is fast and does not require a distributional assumption as \cite{evans2014}, \cite{ziel2015efficient} and \cite{ambach2015space} point out.

The supposed model structure of our approach enables the simulation of different sample paths. We consider bootstrap simulation to derive probabilistic forecasts. Consequently, we calculate probabilistic prediction intervals and thus our model has a clear advantage compared to other approaches \citep[see][]{Tascikaraoglu2014}. \cite{Taylor2009}, \cite{hong2016probabilistic} and \cite{iversen2015short} emphasize the benefit of probabilistic forecasts in the context of wind speed and wind power predictions. Our proposed statistical wind speed approach is compared with other benchmarks. A comparison study is provided to show the short-term accuracy of our approach. The novel approach provides superior forecasting results compared to a univariate autoregressive fractional integrated moving average model, a gradient boosting machine (GBM) and an ANN. 
The medium-term forecasts provide a similar result, but it is reasonable to extent the comparison study.

This article begins with the description of the underlying data set in Section \ref{section:Data}. Furthermore, this section discusses all important features of meteorological variables. Section \ref{section:Model} introduces the multivariate modelling approach. Thereafter, Section \ref{sec:outsample} discusses the goodness of fit of our model. Moreover, predictions of the wind speed, the wind direction and the air pressure are provided in Section \ref{sec:outsample} and we analyse the obtained out-of-sample performance. In Section \ref{sec:con} a conclusion of the previous discussed findings is provided.

\section{Data Set Characteristics}\label{section:Data}

\noindent This article considers ten-minute observations of wind speed, wind direction and air pressure of stations located in Berlin-Tegel ($52^\circ$ $34'N$ $13^\circ$ $19'E $), Berlin-Tempelhof ($52^\circ$ $28'N$ $13^\circ$ $24'E $), Berlin-Sch\"onefeld ($52^\circ 23'N$ $13^\circ 32'E $) and Lindenberg ($52^\circ 13'N$ $14^\circ 07'E $). The observed wind speed is measured in meters per second ($m/s$), the wind direction in degree and the air pressure in hectopascal ($hPa$). The complete data set contains a few missing observations ($<0.5\%$) and we close the small gaps by linear interpolation. The stations Berlin-Tegel and Berlin-Tempelhof are situated in Berlin (Germany) in an urbanized region, while Berlin-Sch\"onefeld and Lindenberg are located in a rural plain area in Brandenburg (Germany), near to Berlin. Brandenburg has wide open spaces with fields and forests, but Berlin is an urbanized city. The data is collected $36m$ above mean sea level at Berlin-Tegel, $48m$ at Berlin-Tempelhof, $46m$ at Berlin-Sch\"onefeld and $98m$ at Lindenberg and is provided by the ``Deutscher Wetterdienst'' (DWD). The observed time frame spans from January 1, 2011 to October 20, 2015. Hence, there are approximately 250,000 observations. The in-sample period is chosen from January 1, 2011 to October 18, 2014. Due to the fact, that most of the estimation results look similar we provide detailed results for Berlin-Tegel and Berlin-Tempelhof. The results for the other stations are omitted here to save space, but all results are available upon request.

Table \ref{table:Descriptives} provides descriptive statistics for Berlin-Tegel and Berlin-Tempelhof. The 25\% quantile of the wind direction lies and the 75\% quantile indicates the west wind. Therefore, 50\% of the wind direction is observed between south-east and west. However, the descriptive statistics of the wind direction are not easy to interpret, because it is a circular variable. The average wind speed is approximately 3.5 m/s. Moreover, the wind speed has a standard deviation $\widehat{\sigma}$ of $1.8-2.0$. We do not observe a big difference of the air pressure for both stations. Hence, 50\% of the air pressure at Berlin-Tegel lie in an interval from $1006$ to $1017$ and for Tempelhof the interval ranges from $1004$ to $1015$.

\begin{table}[h]
\centering
\begin{tabular}{lrrrrrr}
  \hline
 &  \multicolumn{3}{c}{\textbf{Berlin-Tegel}}  & \multicolumn{3}{c}{\textbf{Berlin-Tempelhof}}  \\  
 & Direction & Speed & Pressure  & Direction & Speed & Pressure \\ 
  \hline
MIN  & 0.00 & 0.00 & 961.20 & 10.00 & 0.10 & 960.00 \\ 
$25\%$-qu. & 110.00 & 2.00 & 1006.00 & 110.00 & 2.30 & 1004.50 \\ 
MEDIAN & 210.00 & 3.10 & 1011.50 & 200.00 & 3.40 & 1009.90 \\ 
Mean & 195.33 & 3.42 & 1011.42 & 191.78 & 3.63 & 1009.88 \\ 
$75\%$-qu. & 270.00 & 4.50 & 1016.80 & 270.00 & 4.70 & 1015.20 \\ 
MAX  & 360.00 & 18.10 & 1039.00 & 360.00 & 17.00 & 1037.10 \\ 
$\widehat{\sigma}$ &  & 2.00 & 8.56 &  & 1.84 & 8.52 \\ 
   \hline
\end{tabular}
 \caption{Descriptive statistics for wind direction, wind speed and air pressure.}
   \label{table:Descriptives}
\end{table}

\begin{figure}[h]
\centering
 \begin{subfigure}[b]{0.49\textwidth}
 \includegraphics[width=1\textwidth]{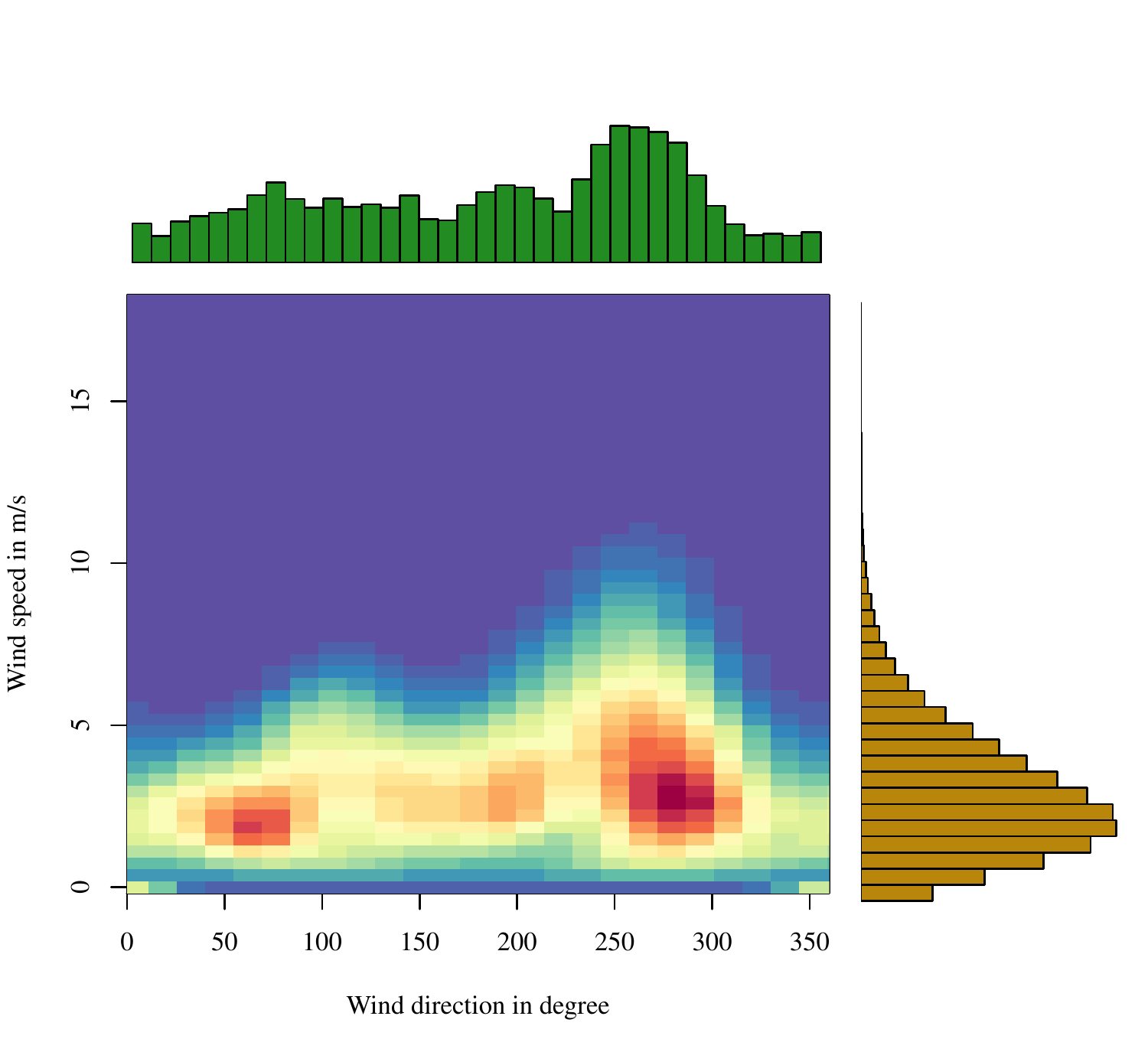}
 \includegraphics[width=1\textwidth]{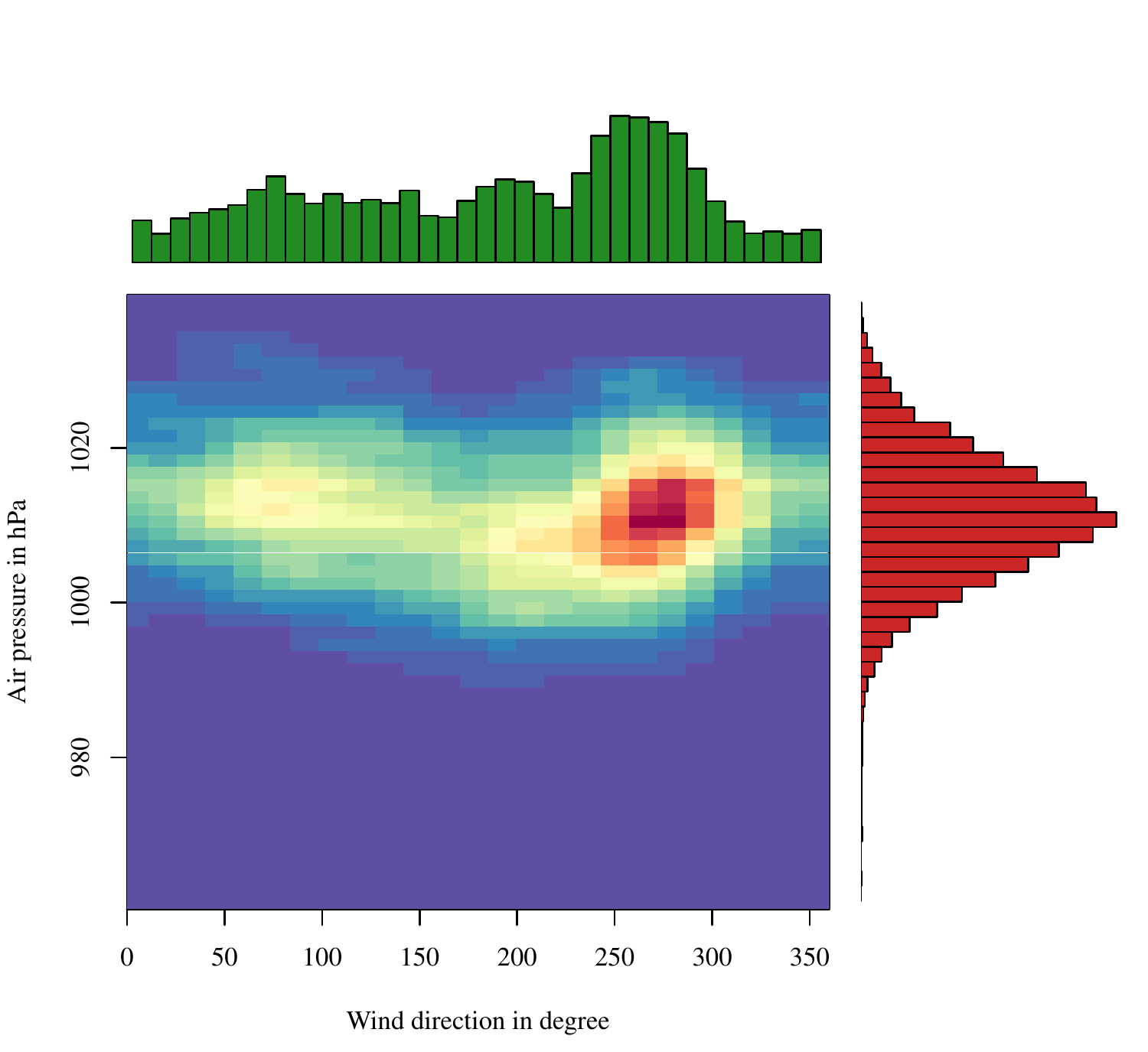}
 \includegraphics[width=1\textwidth]{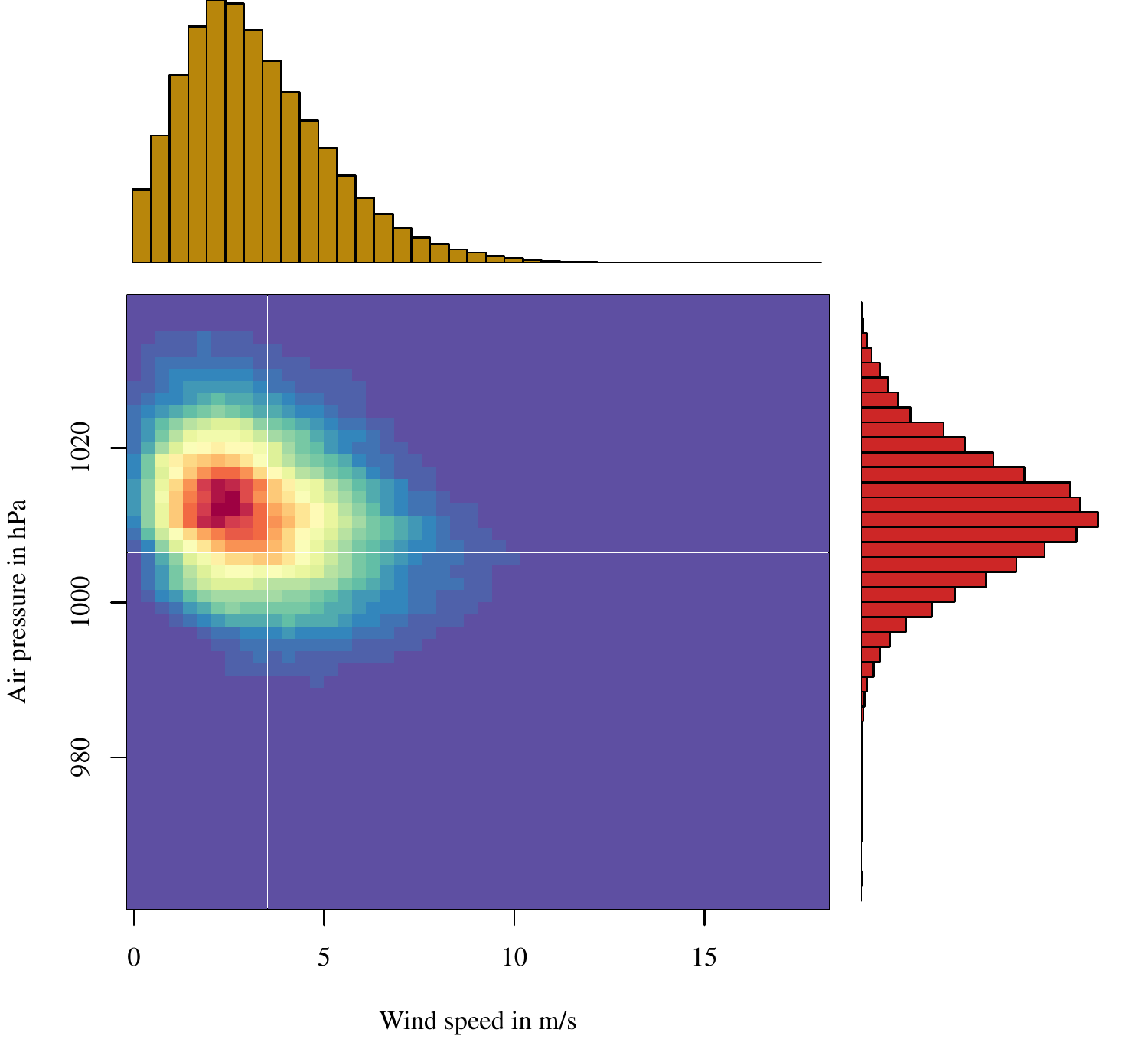}
   \caption{Berlin-Tegel}
\end{subfigure}
\begin{subfigure}[b]{0.49\textwidth}
 \includegraphics[width=1\textwidth]{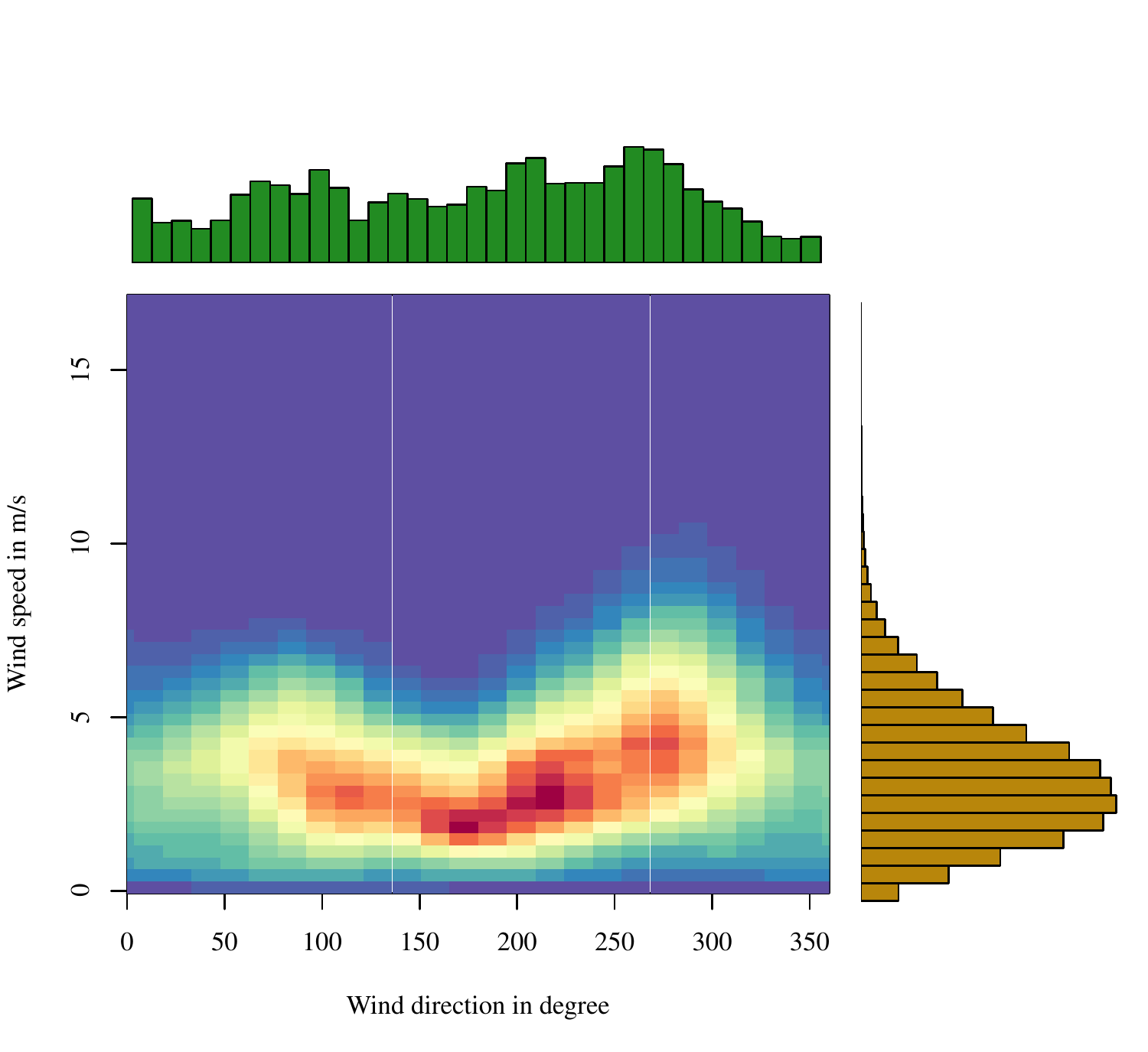}
 \includegraphics[width=1\textwidth]{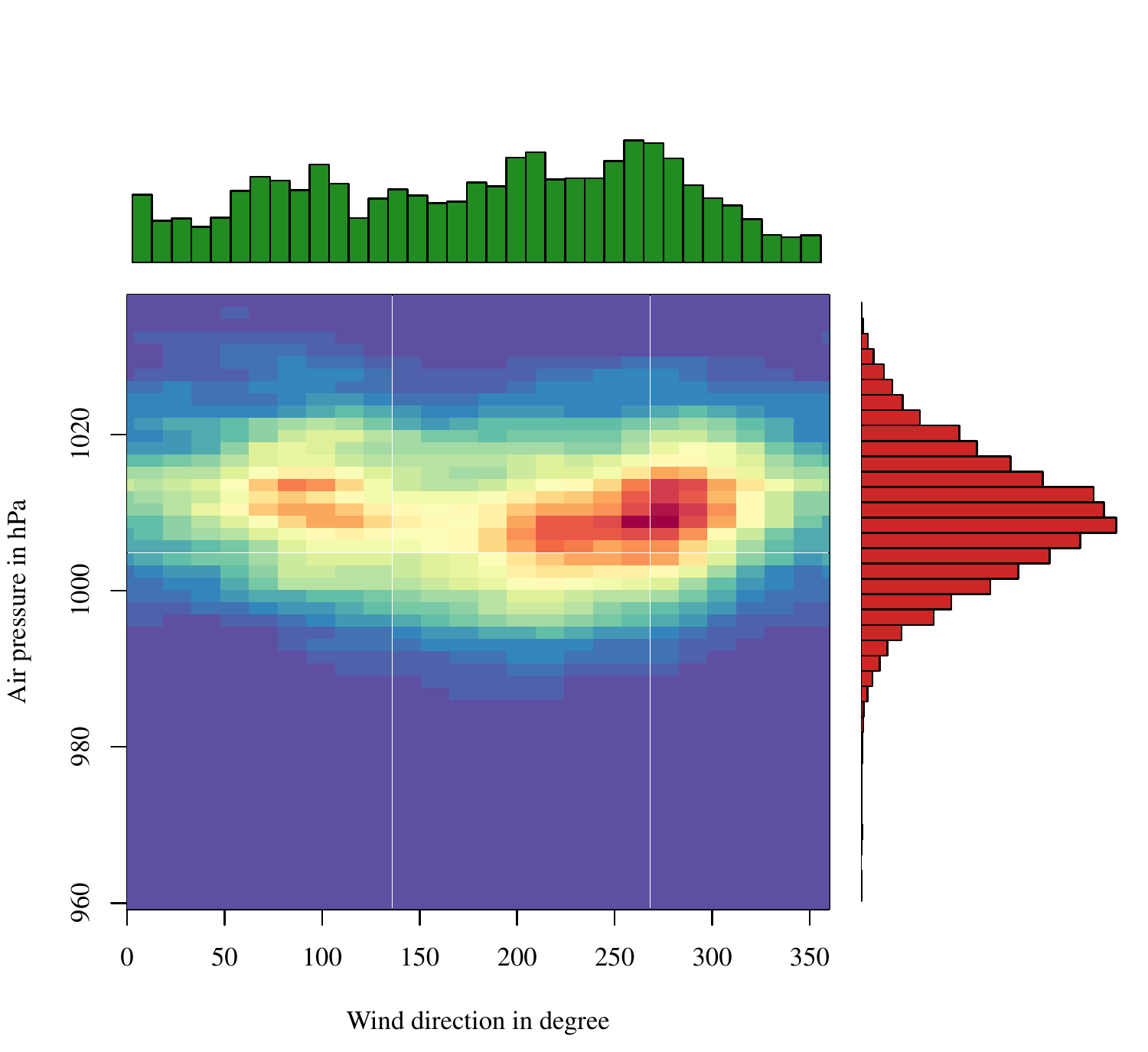}
 \includegraphics[width=1\textwidth]{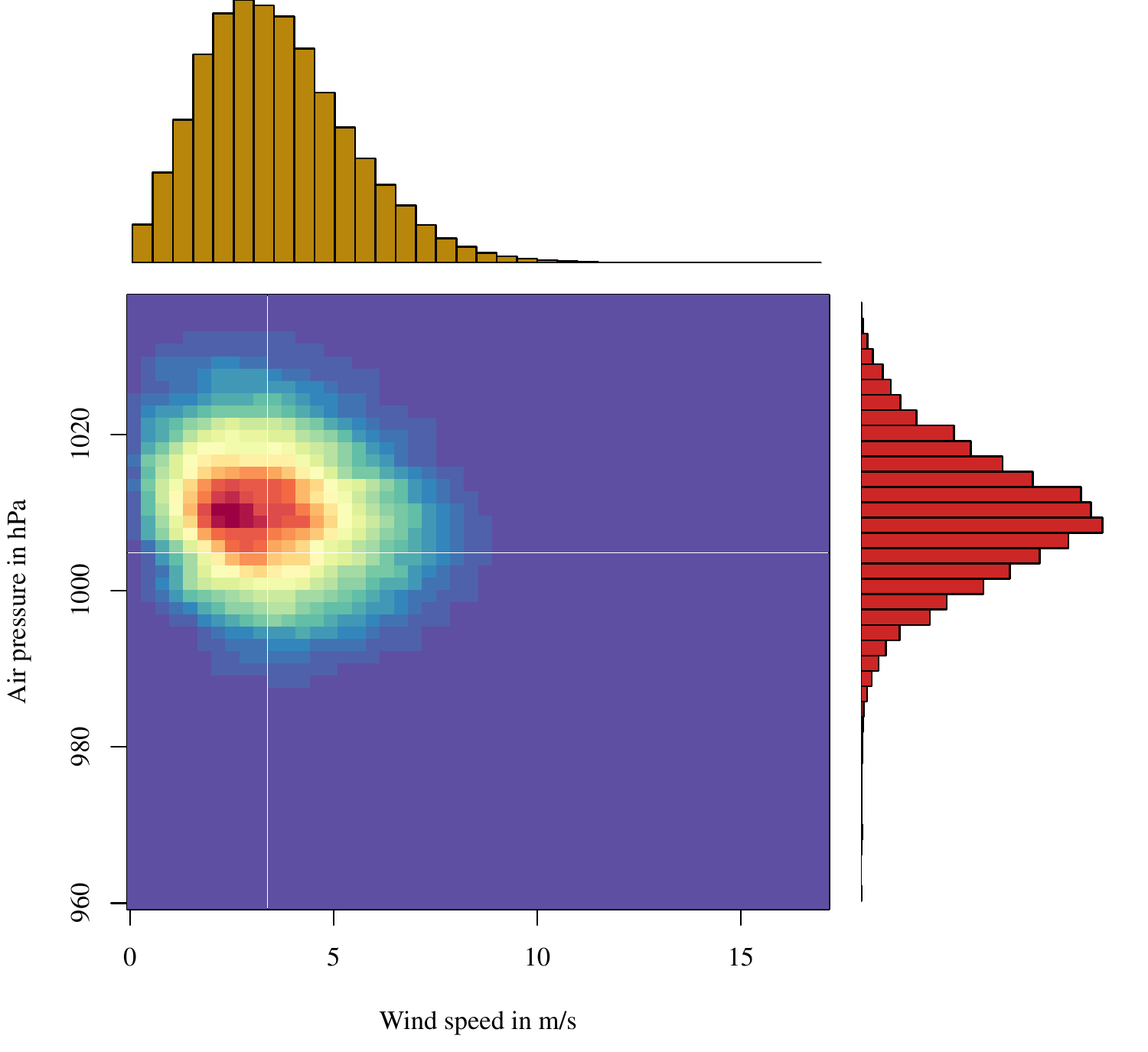}
   \caption{Berlin-Tempelhof}
\end{subfigure}
 \caption{Two dimensional histograms and corresponding histograms for the wind speed and the wind direction (first row), the wind speed and the air pressure (second row) and the wind direction and the air pressure (third row).}
 \label{graph:hist}
\end{figure}

In Figure \ref{graph:hist} two-dimensional histograms for each pair of time series are given. The histogram of the wind direction illustrates non-normally distributed observations, whereas the histogram of the wind speed provides a positive skewed distribution. Figure \ref{graph:hist} shows that the histogram of the air pressure is negatively skewed. The distributional assumptions on the wind speed and the wind direction (first row), the wind speed and the air pressure (second row) and the wind direction and the air pressure (third row) are different. Hence, by means of a correlation test we observe a positive correlation between wind speed and wind direction. The correlation test shows a significant negative correlation between wind speed and air pressure. Finally, the correlation test depicts a small, but significant negative correlation between wind direction and air pressure. Figure \ref{graph:hist} provides incentives to use an estimation method which does not need a distributional assumption, such as the LASSO method. 

\begin{figure}[h]
  \includegraphics[width=.5\textwidth,clip=true]{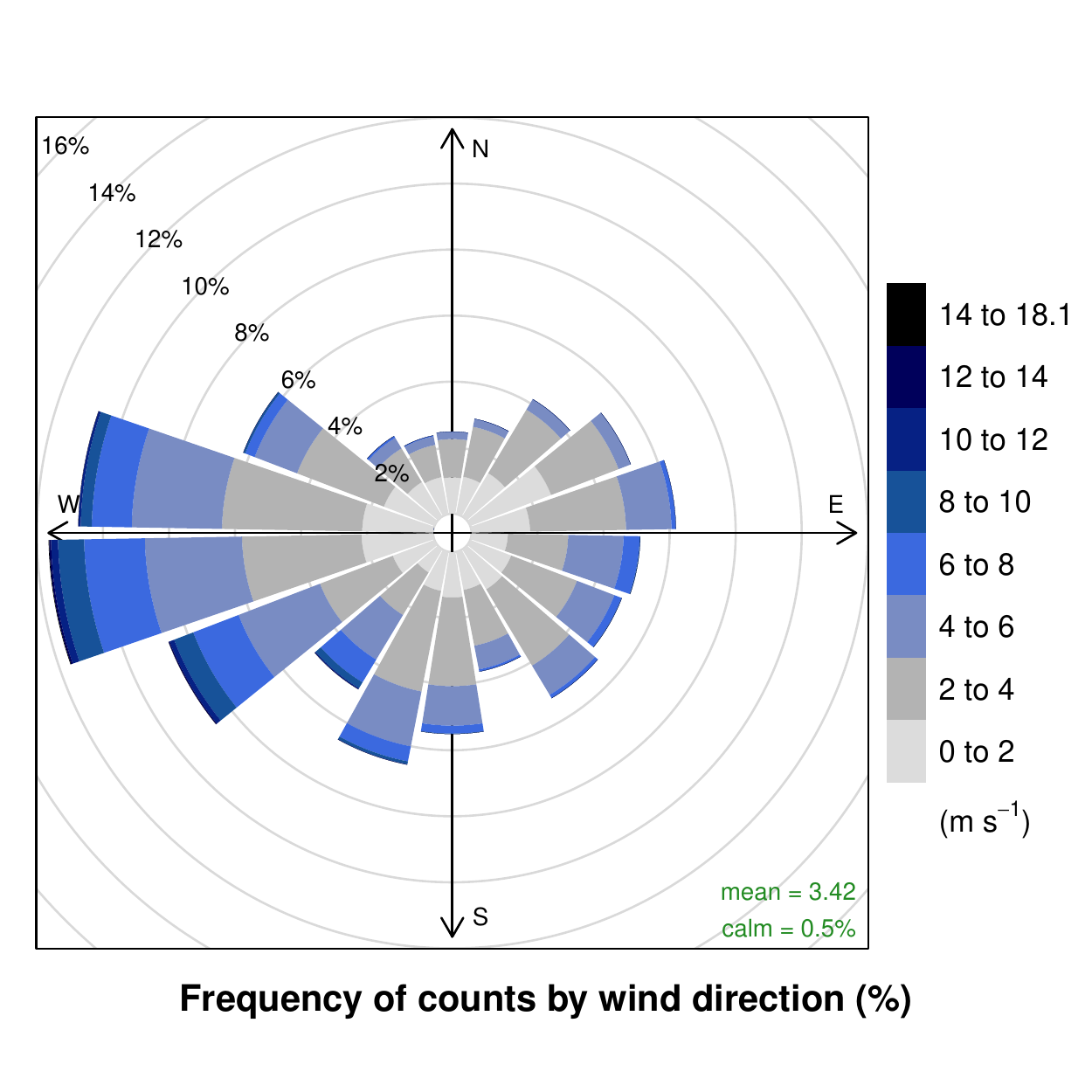}
  \includegraphics[width=.5\textwidth,clip=true]{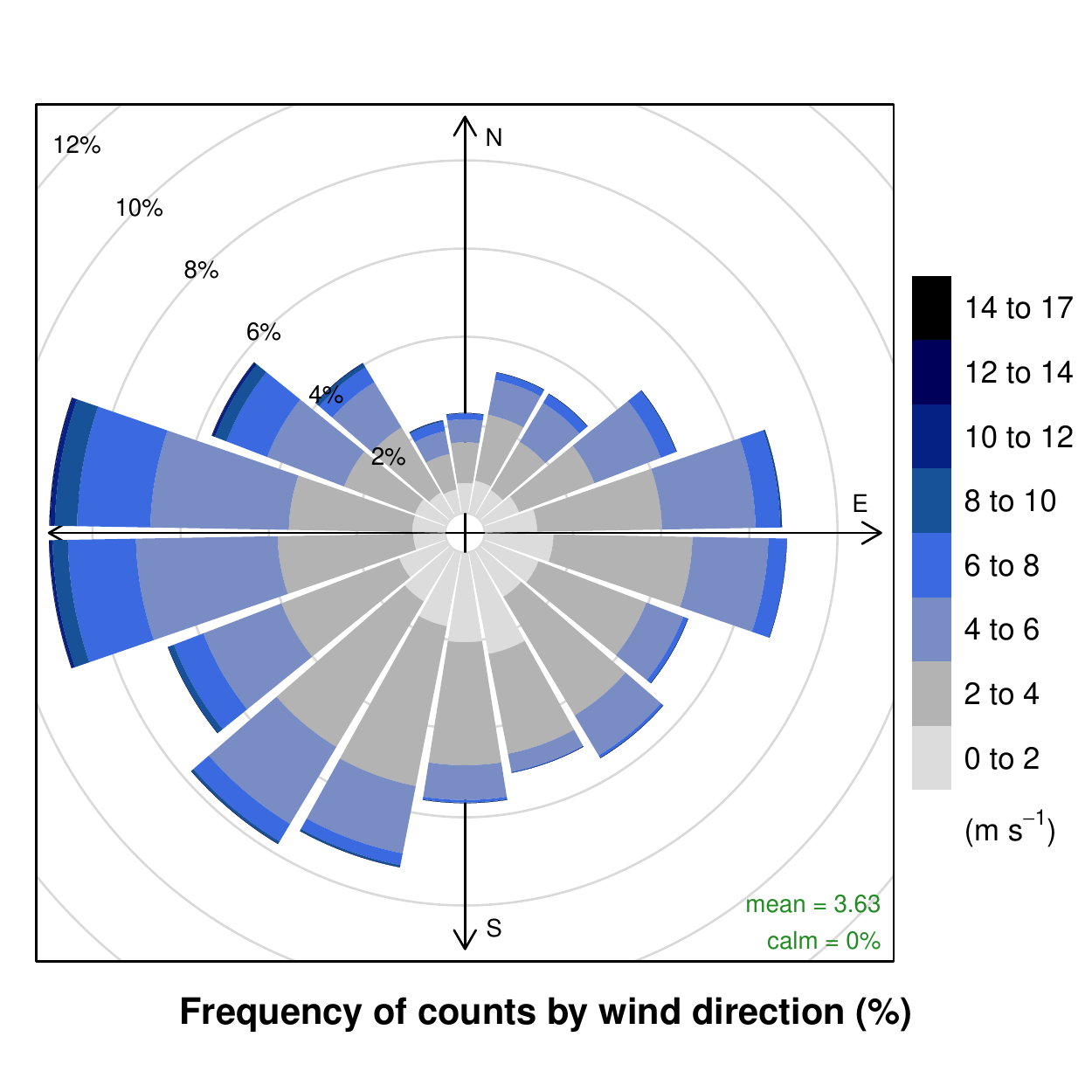}
  \caption{Wind rose, where the wind speed frequencies are plotted by wind direction for station Berlin-Tegel (left) and station Berlin-Tempelhof (right).}\label{graph:rose}
\end{figure}

Hence, Figure \ref{graph:rose} depicts the wind rose for the investigated stations Berlin-Tegel (left) and Berlin-Tempelhof (right). It shows the frequency of counts by wind direction. We observe a high probability for a wind speed which is coming from south-western and western direction. Obviously, the wind speed is directly affected by the wind direction which is related to the Coriolis force within the northern hemisphere. 

\subsection{Detection of Periodic Effects}

The wind speed shows daily and annual periodic effects which are related to the air pressure, local weather conditions and the jet-stream \citep[see][]{Burton2011}. \cite{ambach2015periodic} show the importance of including periodic regressors in the model. In this article the wind direction and the air pressure are included as explanatory variables in the model. Recent literature provides different examples for the existence of periodic effects as, e.g., \cite{carapellucci2013effect} and \cite{silva2016complementarity}. The aforementioned articles focus on daily periods, but there is evidence for annual periodicity as well \citep[see][]{ambach2015short}.

\begin{figure}[h]
  \centering
  \begin{subfigure}[b]{0.49\textwidth}
  \includegraphics[width=1\textwidth]{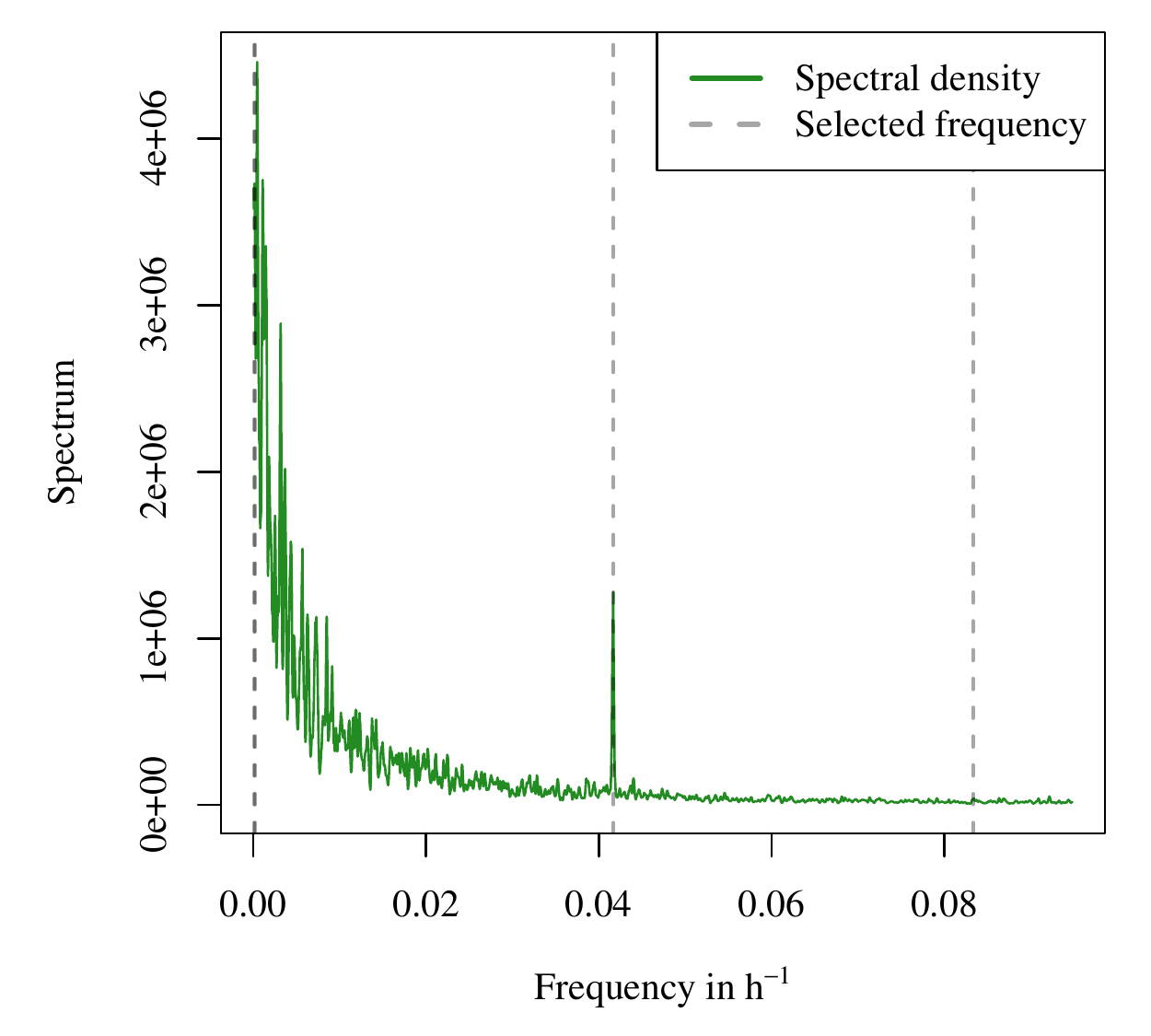} \\
  \includegraphics[width=1\textwidth]{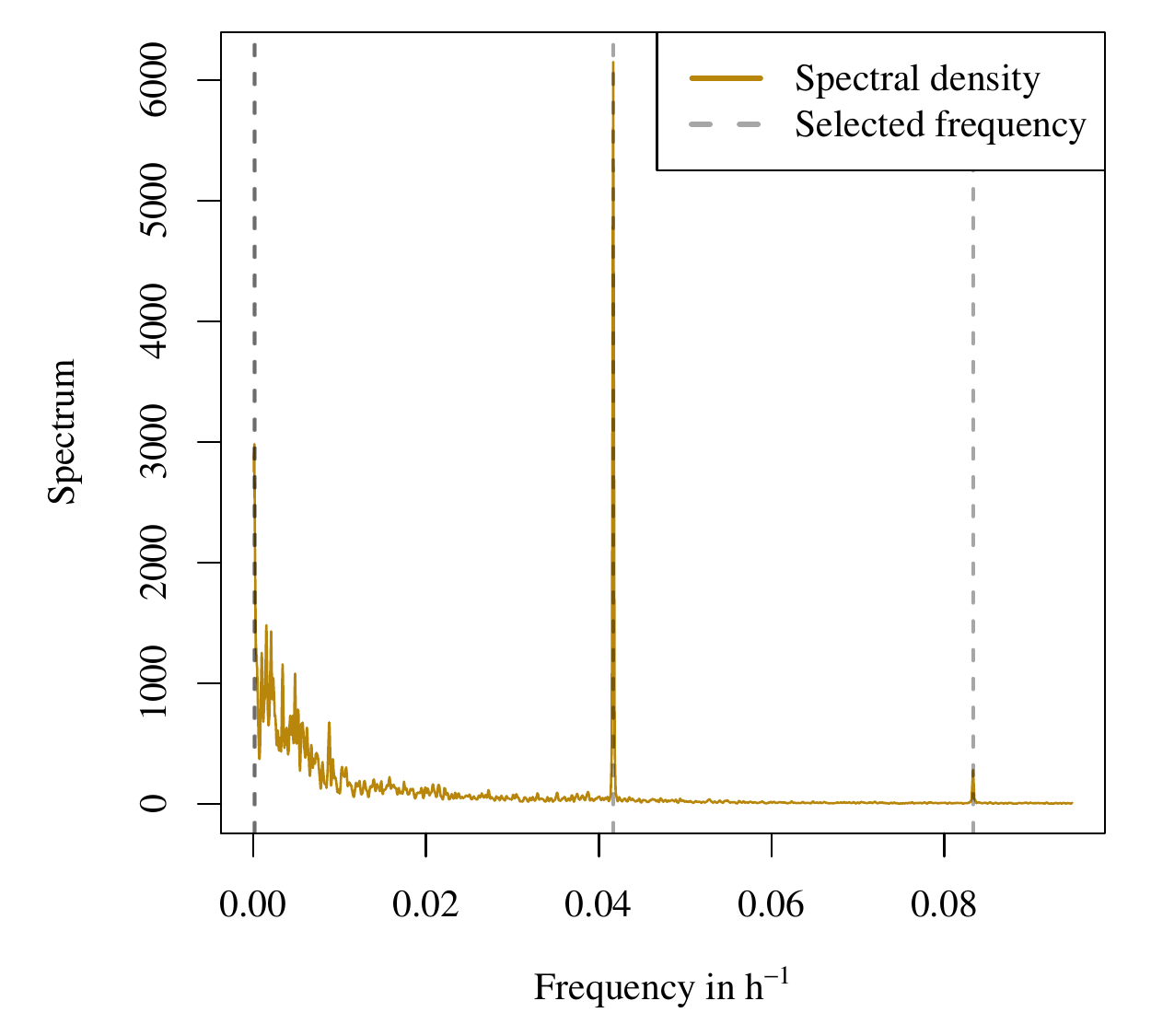} \\
  \includegraphics[width=1\textwidth]{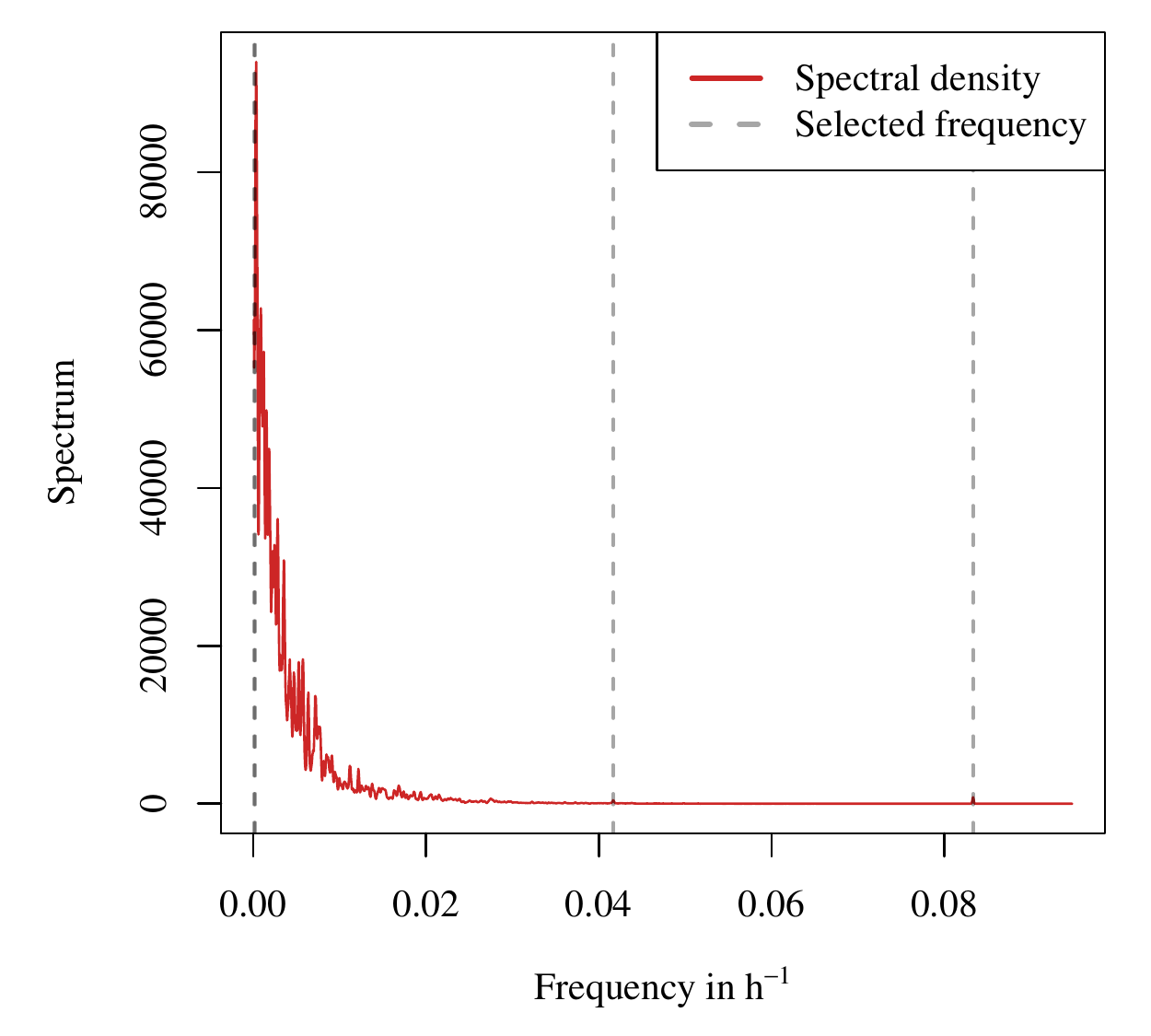}
    \caption{Smoothed Periodogram Berlin-Tegel.}
   \label{graph:Spec1}
  \end{subfigure}
  \begin{subfigure}[b]{0.49\textwidth}
   \includegraphics[width=1\textwidth]{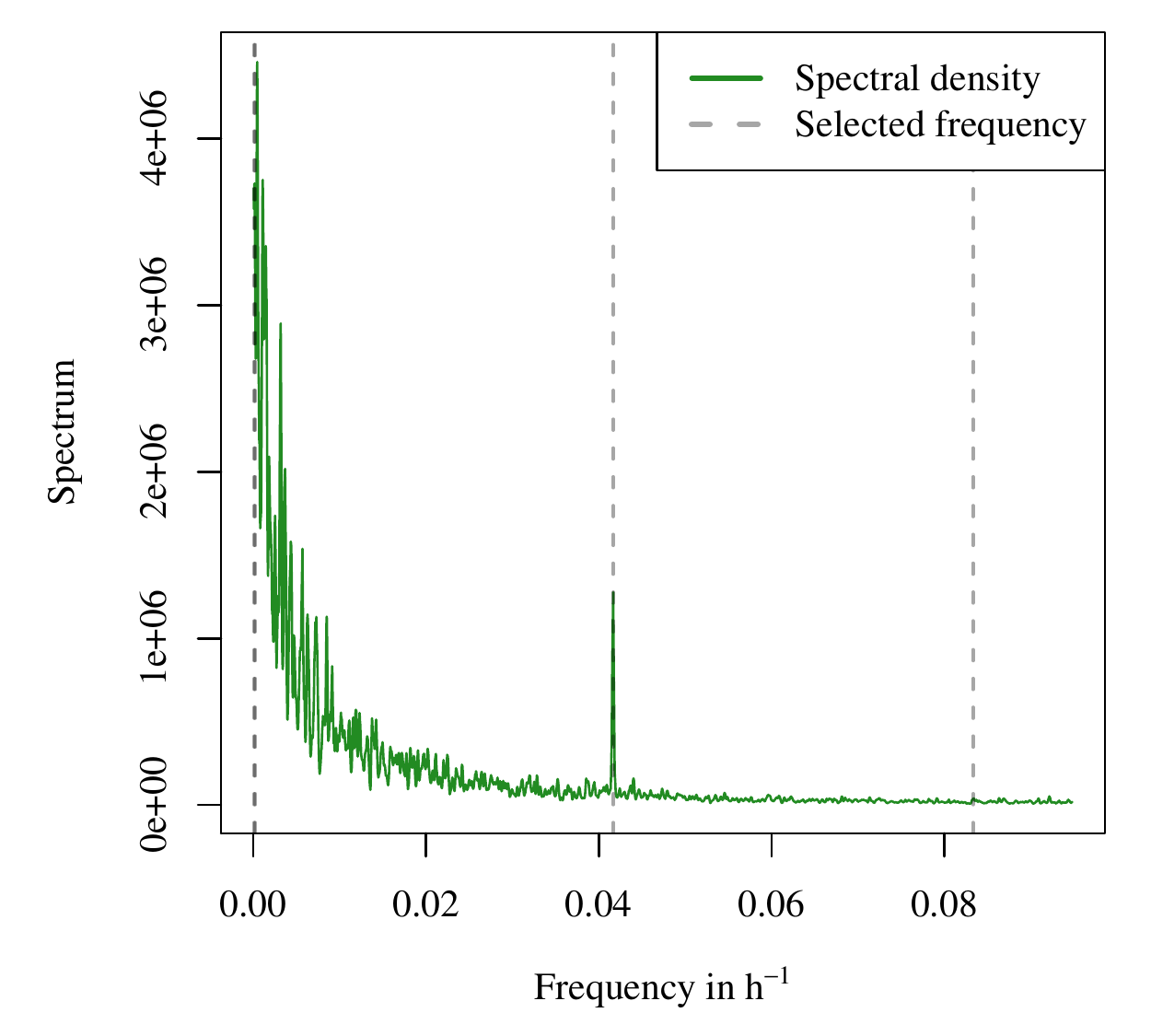} \\
   \includegraphics[width=1\textwidth]{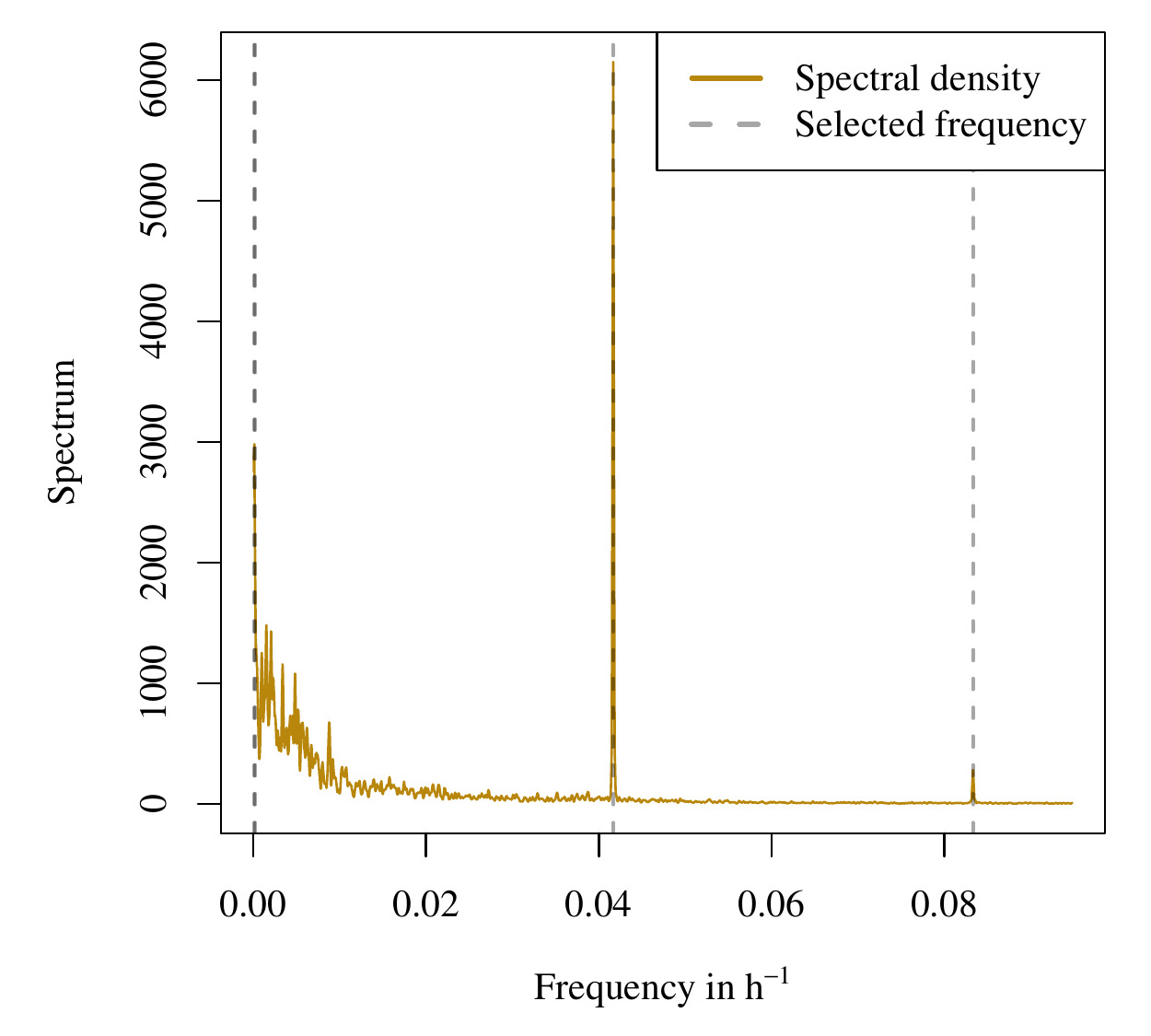} \\
   \includegraphics[width=1\textwidth]{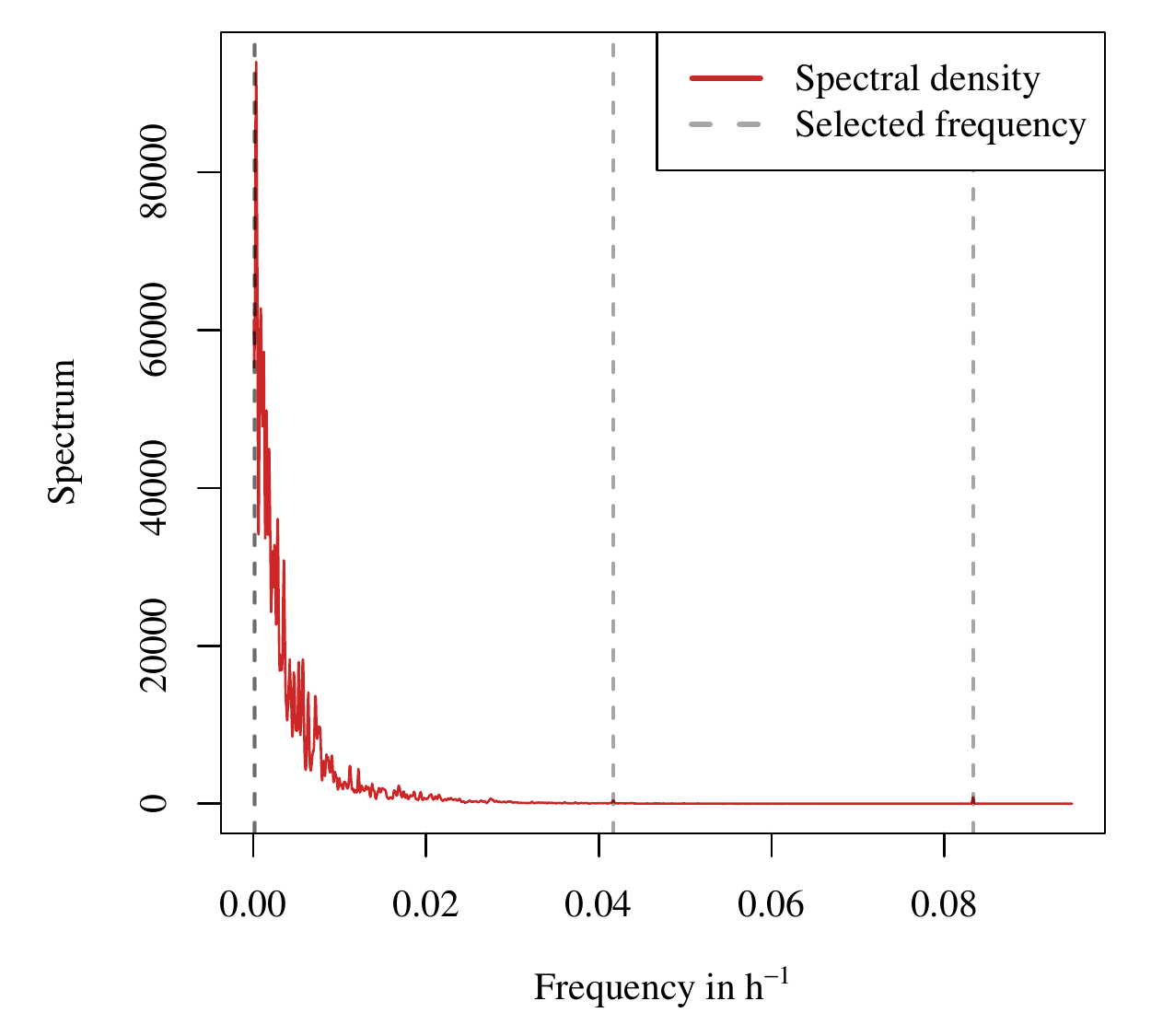}
    \caption{Smoothed Periodogram Berlin-Tempelhof.}
   \label{graph:Spec2}
 \end{subfigure}
  \caption{Estimated spectral density of wind direction (first row), wind speed (second row) and air pressure (third row).}
  \label{graph:Spec}
\end{figure}

The smoothed periodogram provides a tool for the detection of periodic behaviour. The obtained results are shown in Figure \ref{graph:Spec}. There are several high frequencies which might be relevant, but we select only two different periods which are plausible. The choice of the selected frequencies depends on the peak of the estimated spectrum. Next to the point of origin, we detect extremely high periods. The frequencies are scaled in an hourly interval. We select the periods with a peak at an $\omega \in \{0.0416, 0.000019\}$. Transforming these frequencies to a period provides diurnal and annual periods. Hereafter, we transform the frequencies back to the time domain and plug them into our periodic B-spline functions. 

Obviously, our underlying wind speed and wind direction data provide a strong periodic behaviour. The black dotted lines in Figure \ref{graph:Spec} show an annual, half-annual, diurnal and half-daily periods. Obviously, the diurnal frequency for air pressure can be neglected, but we include them into our model and let the LASSO method choose, if the diurnal period is significant. However, periodic B-spline functions help to model all multiples of diurnal and annual periods. The proposed model additionally includes the multiplication of diurnal and annual periodic functions.

\subsection{Analysing the Autocorrelation and Non-Linear Effects}

\begin{figure}[h]
  \centering
  \begin{subfigure}[b]{0.325\textwidth}
  \includegraphics[width=1\textwidth]{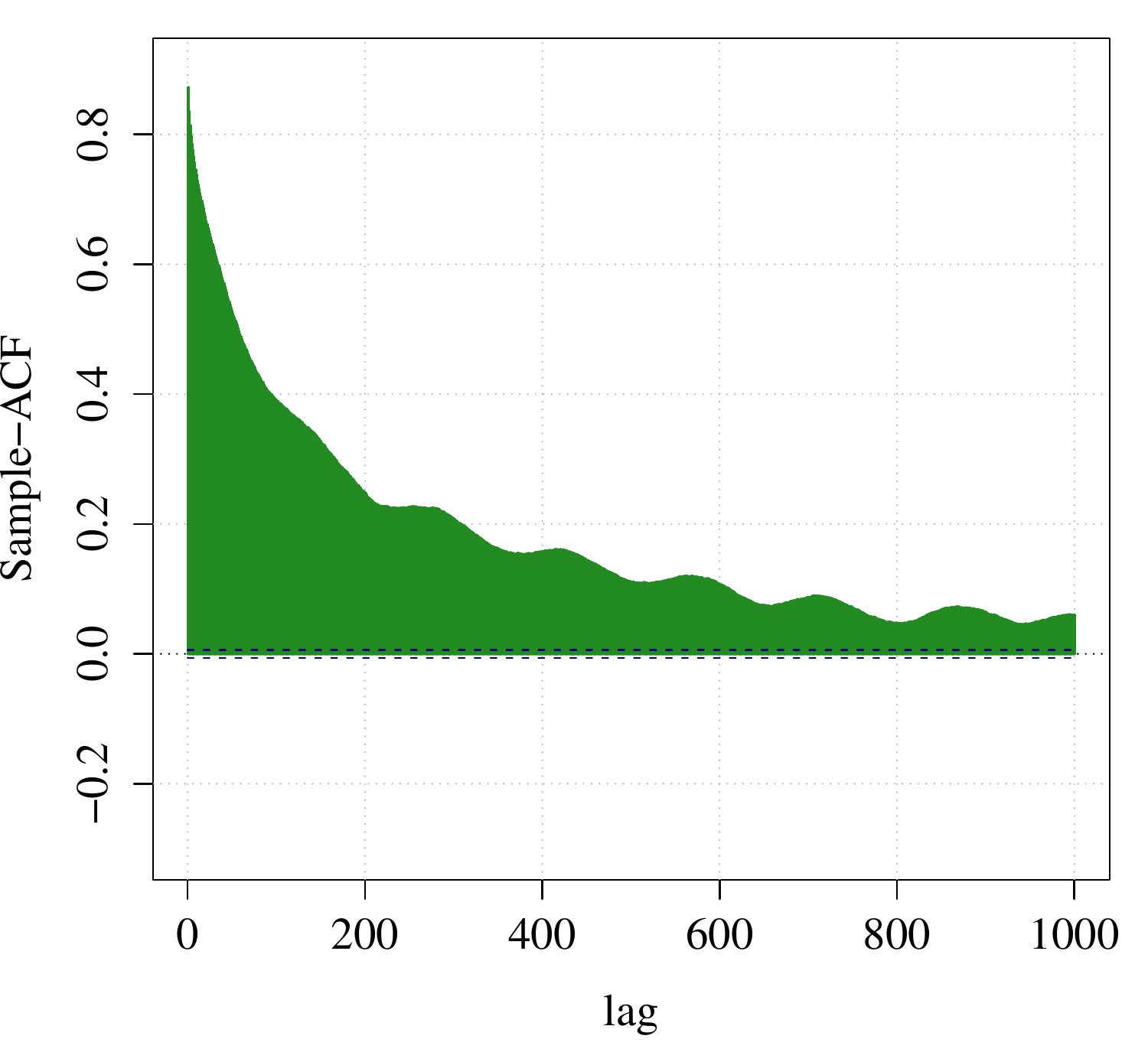} \\
  \includegraphics[width=1\textwidth]{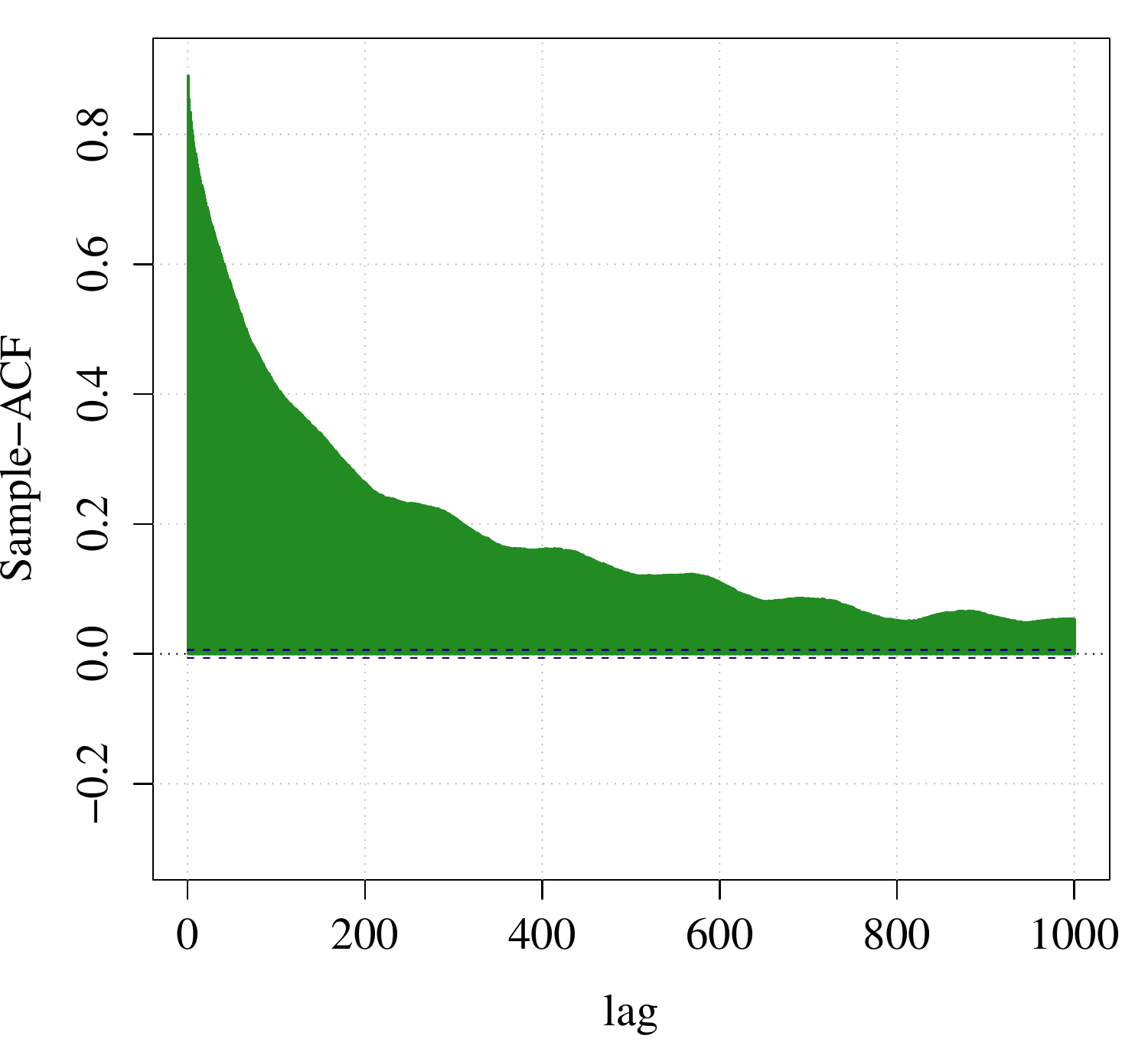}
    \caption{ACF of wind direction.}
  \end{subfigure}
  \begin{subfigure}[b]{0.325\textwidth}
   \includegraphics[width=1\textwidth]{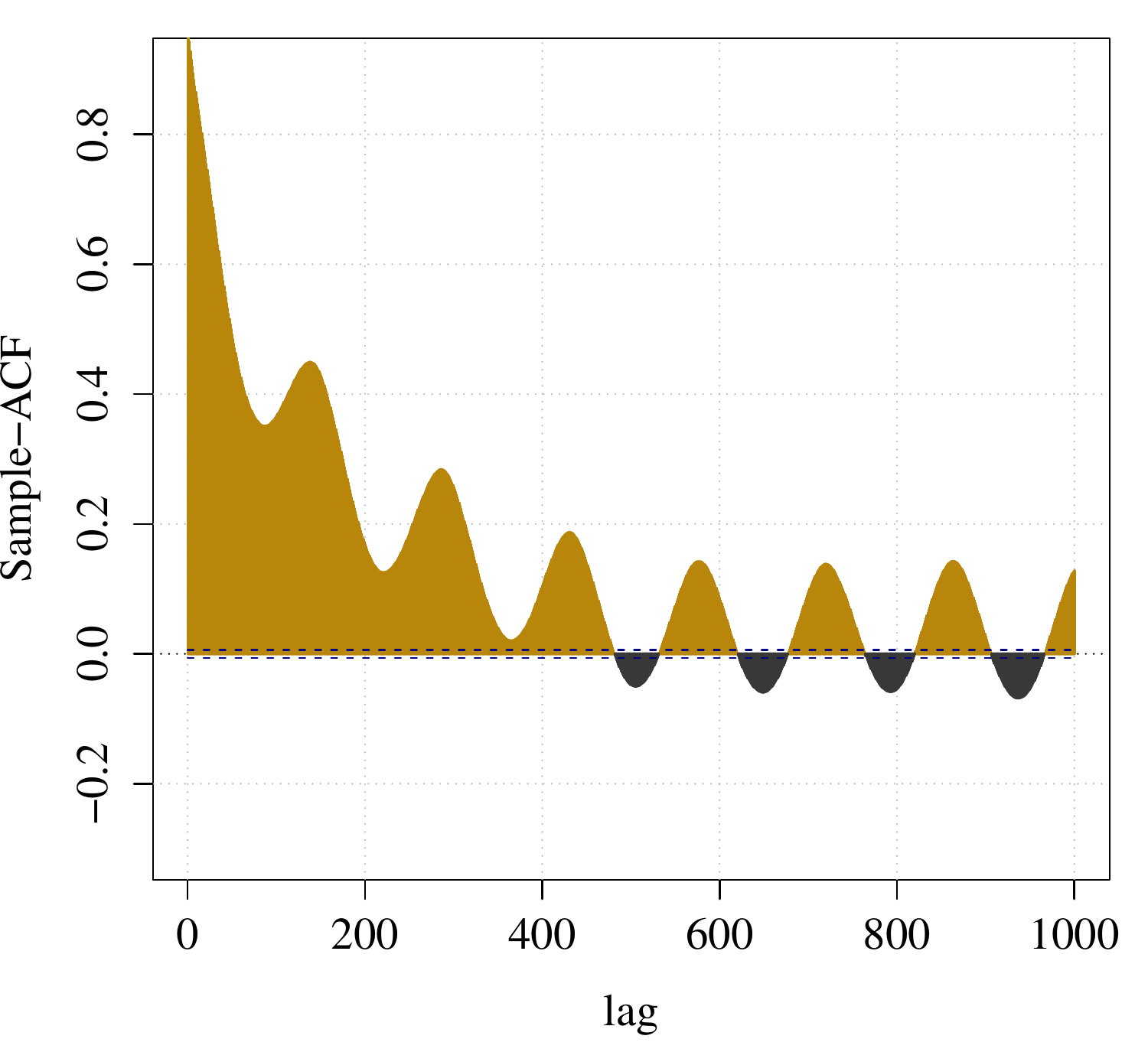} \\
   \includegraphics[width=1\textwidth]{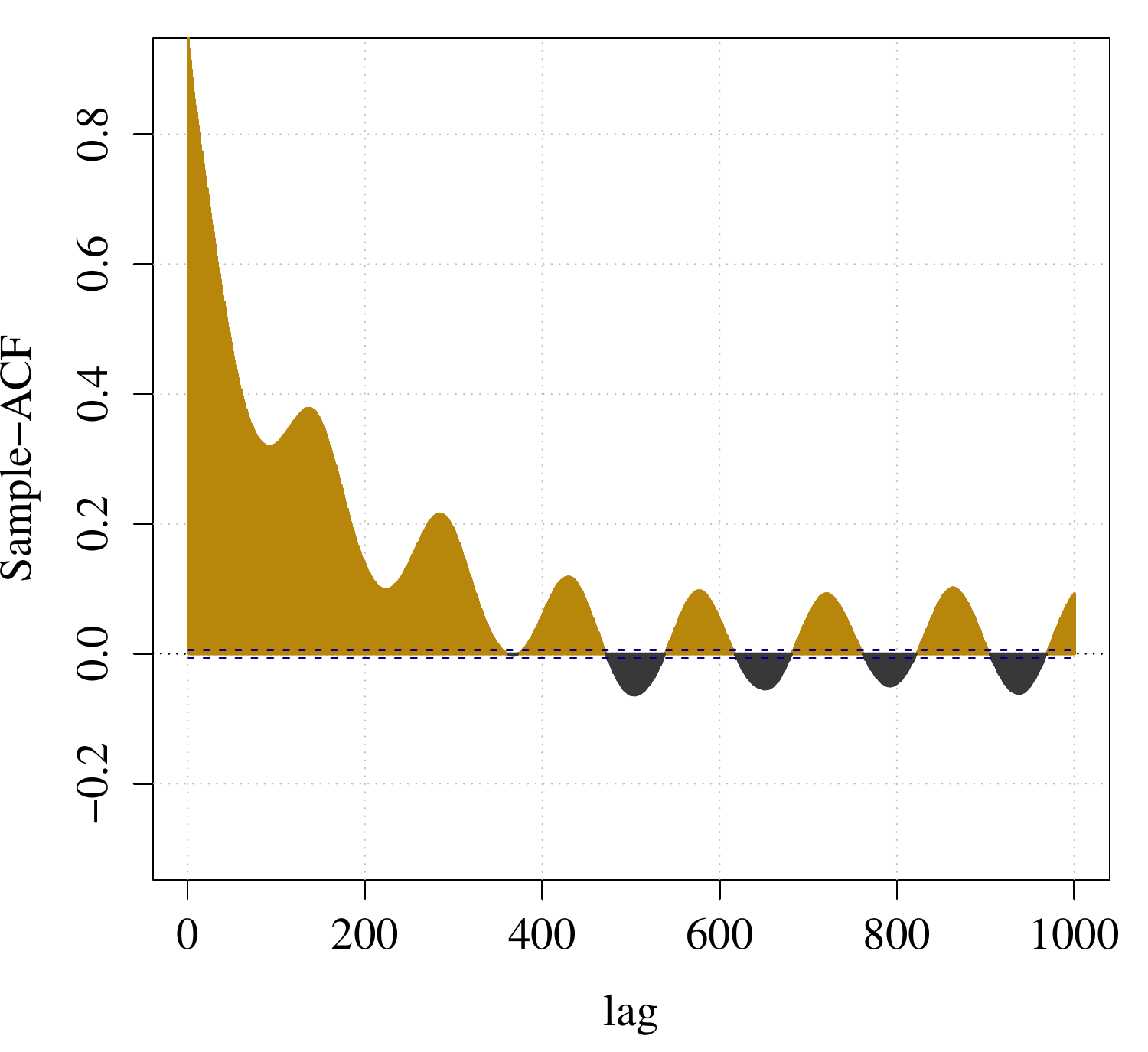}
    \caption{ACF of wind speed.}
 \end{subfigure}
   \begin{subfigure}[b]{0.325\textwidth}
   \includegraphics[width=1\textwidth]{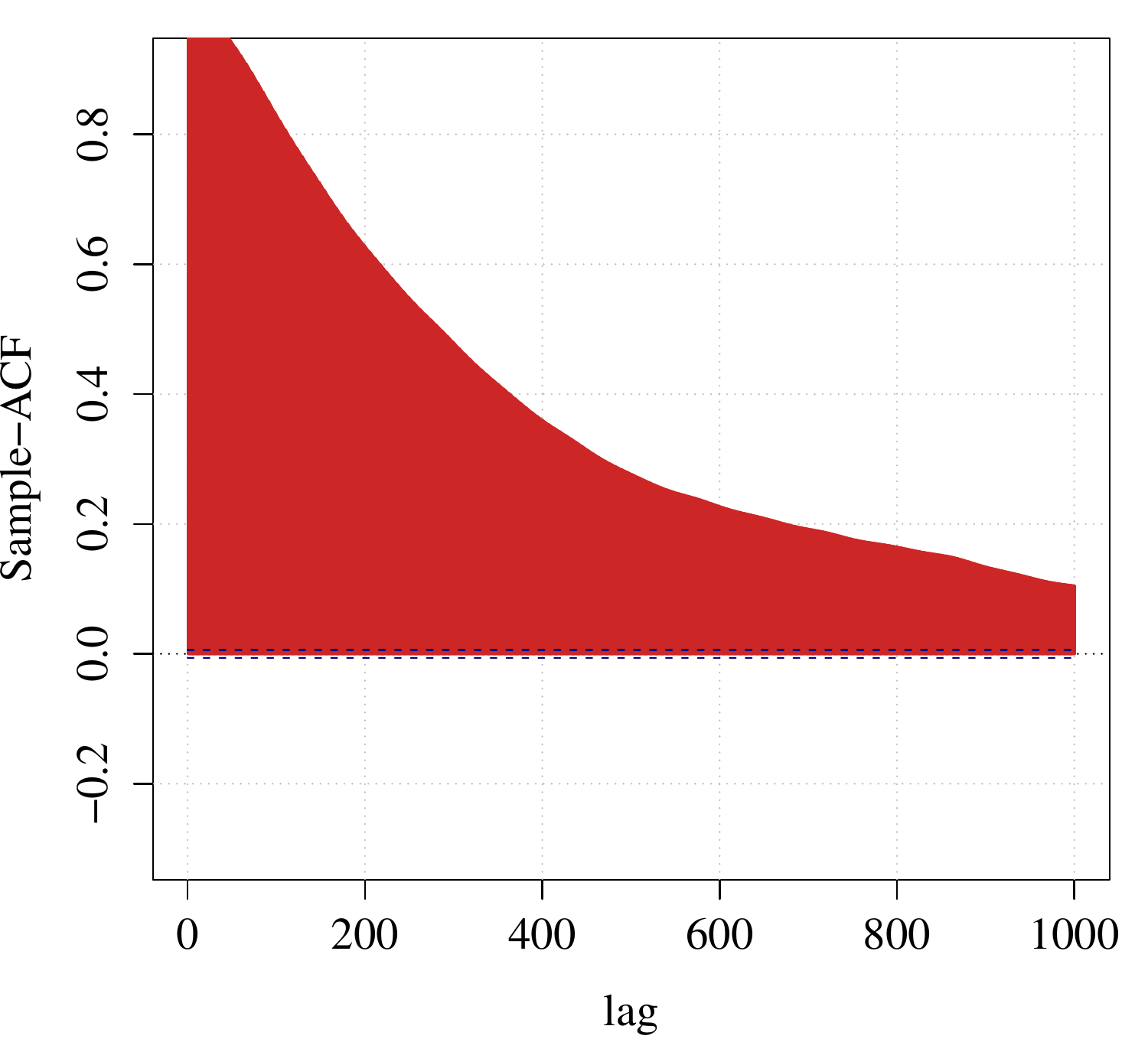} \\
   \includegraphics[width=1\textwidth]{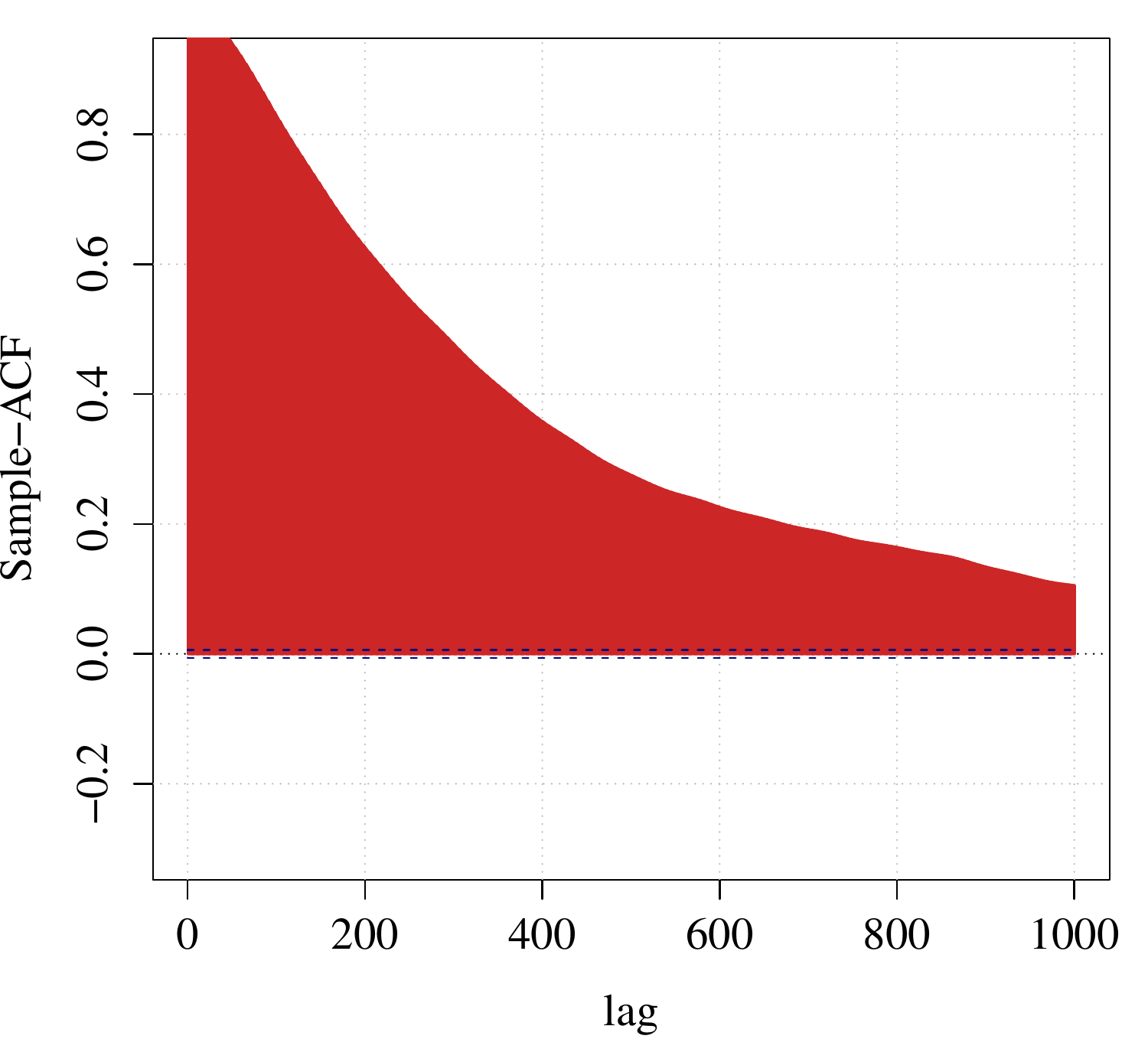}
    \caption{ACF of air pressure.}
 \end{subfigure}
  \caption{ACF of Berlin-Tegel (first row) and Berlin-Tempelhof (second row).}
  \label{graph:acf}
\end{figure}

Besides the periodicity in the data, we have to analyse the autocorrelation of each response variable. The autocorrelation function (ACF) is shown in Figure \ref{graph:acf} and it provides a huge autocorrelation. Figure \ref{graph:acf} supports the hypothesis of high persistence of wind direction, wind speed and air pressure which is directly related to the high-frequency data. 

\begin{figure}[h]
  \centering
  \begin{subfigure}[b]{0.49\textwidth}
  \includegraphics[width=1\textwidth]{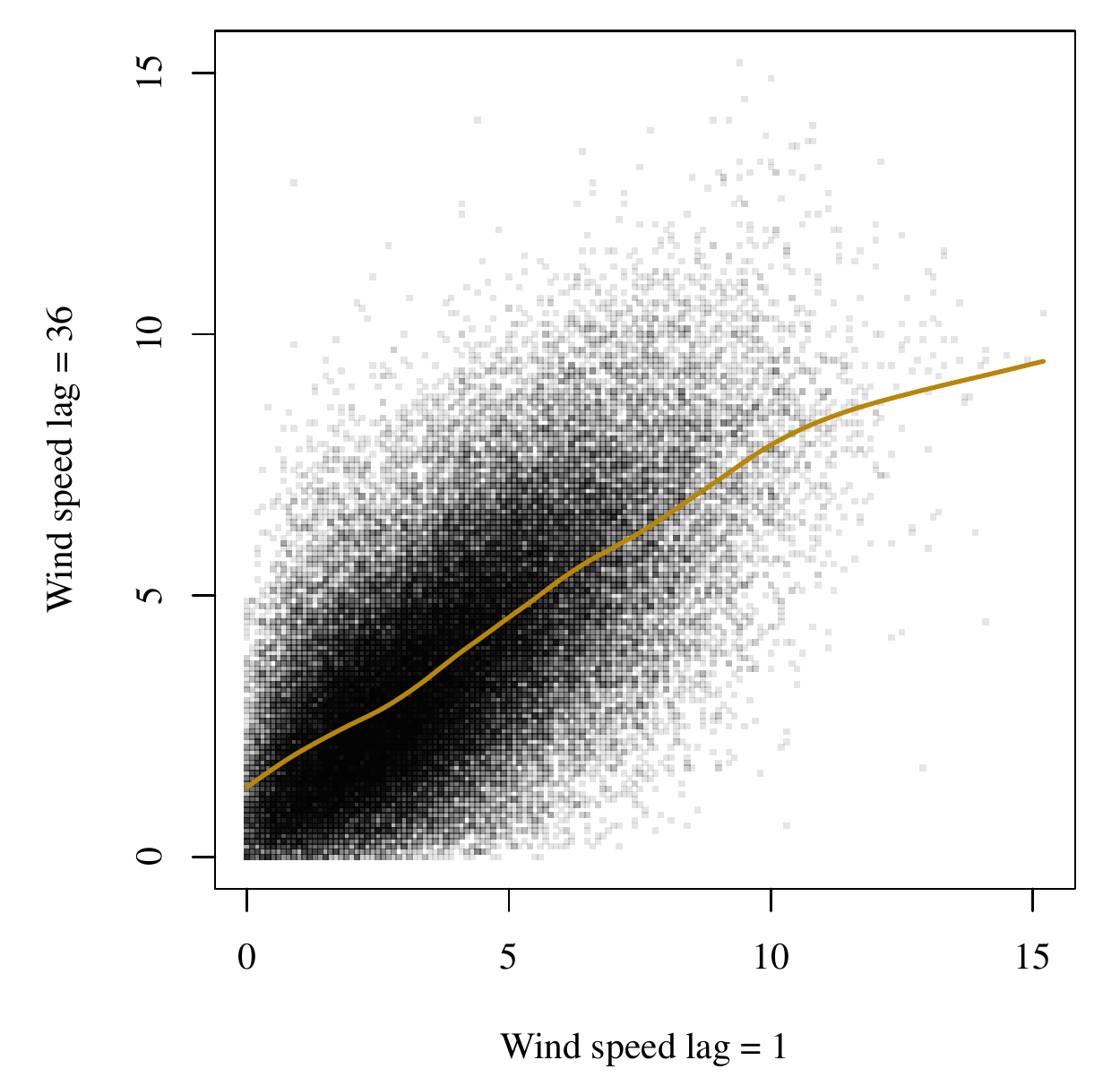} \\
  \includegraphics[width=1\textwidth]{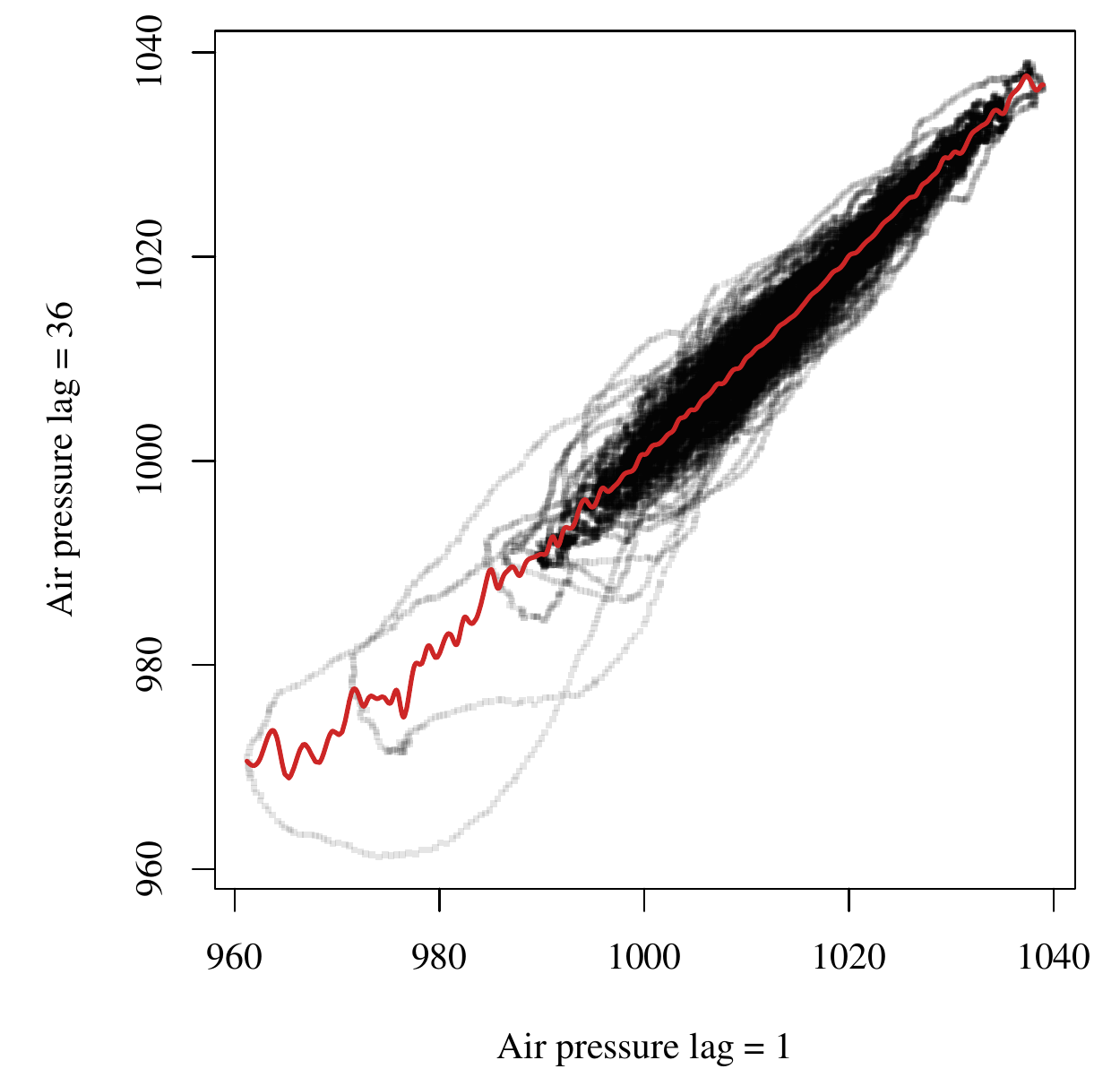}
    \caption{Lag plot for the dependent variable of Berlin-Tegel.}
  \end{subfigure}
  \begin{subfigure}[b]{0.49\textwidth}
   \includegraphics[width=1\textwidth]{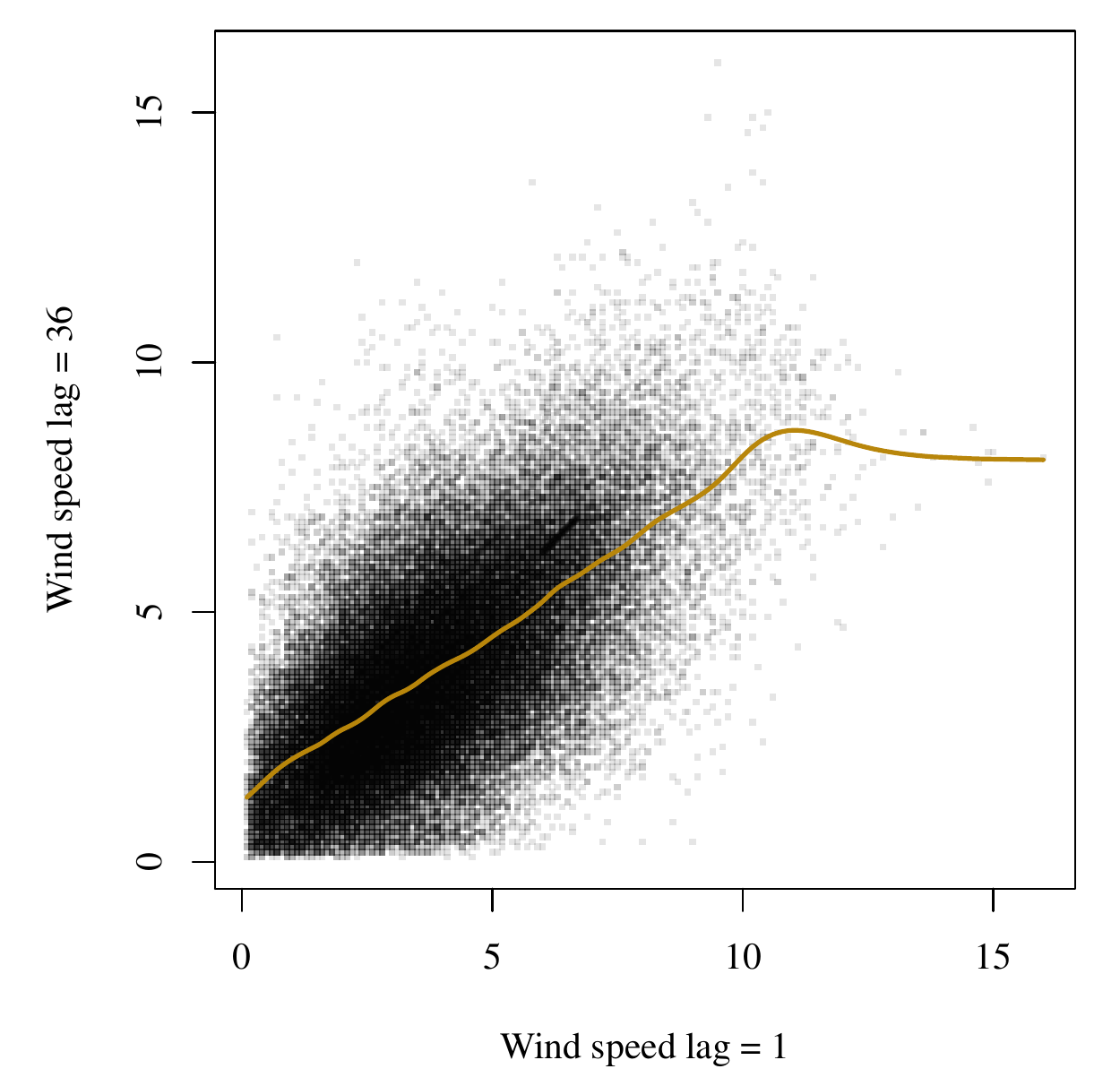} \\
   \includegraphics[width=1\textwidth]{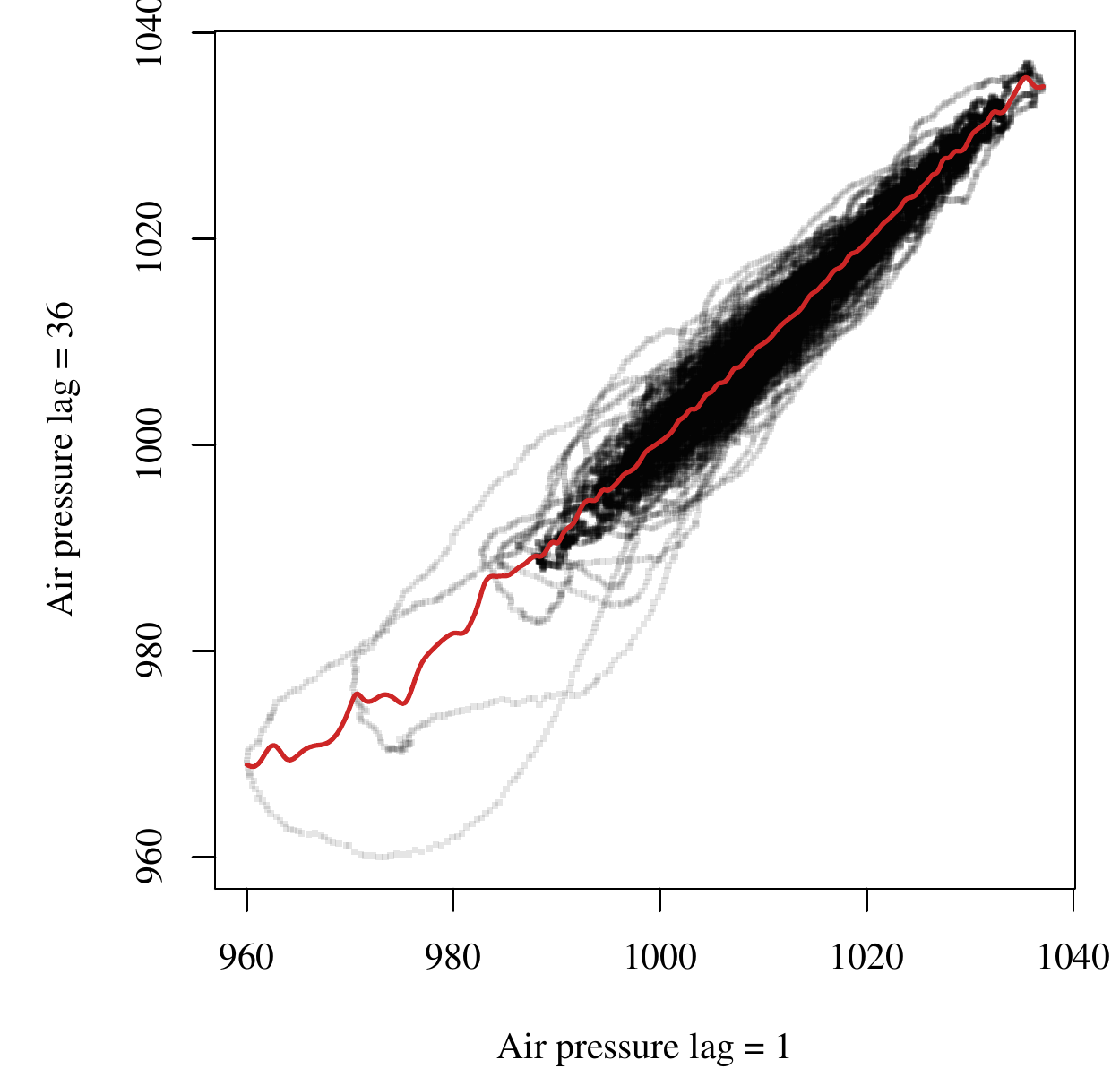}
    \caption{Lag plot for the dependent variable of Berlin-Tempelhof.}
 \end{subfigure}
  \caption{Plot of the lagged wind speed (first row) and the air pressure (second row).}
  \label{graph:thres}
\end{figure}

Furthermore, it might be necessary to incorporate non-linear effects into the new modelling approach if there is a reason for it. \cite{gautama2004delay} describe different methods to check for non-linearities in general, but we have to find an appropriate model for forecasting. Therefore, it is the most convenient way to linearise the non-linear effects by means of threshold autoregression. \cite{chan1990testing} proposes a method to test for threshold autoregressive effects. We use his test to identify the existence of threshold autoregression, but we have to identify the appropriate amount of threshold autoregressive lags. Figure \ref{graph:thres} shows a
scatter plot of the wind speed and the air pressure observation and their corresponding lagged observations. The smoothed spline function within the figure, depicts a region where a linear function would be a good choice. Moreover, we observe that for small and huge observations structural breaks or even non-linear effects are observed. We capture them in our model by including thresholds for every decile. The first lags of wind speed and air pressure do not provide non-linear effects. For wind speed the threshold effect begins with approximately $lag=4$ and for air pressure it begins with $lag=9$. Figure \ref{graph:thres} indicates threshold effects for $lag=36$ (six hours) for small deciles of the air pressure and huge deciles of the wind speed data.

\section{The Multivariate Modelling Equation}\label{section:Model}

\noindent Wind speed is influenced by several explanatory variables such as earth rotation, air pressure and other weather phenomena. Thus, our approach considers the wind direction and the air pressure as dependent variables in a multivariate model. Moreover, both variables are included in one modelling approach. On the one hand, wind direction and air pressure are used to increase the forecasting accuracy of wind speed and on the other hand, these dependent variables are modelled and predicted as well.  \cite{zhang2013multivariate}, \cite{zhu2014incorporating} and \cite{ambach2016vorhersagen} propose different multivariate forecasting approaches to extend the wind speed forecasting precision.  

\begin{figure}[h]
  \centering
  \begin{subfigure}[b]{0.49\textwidth}
  \includegraphics[width=1\textwidth]{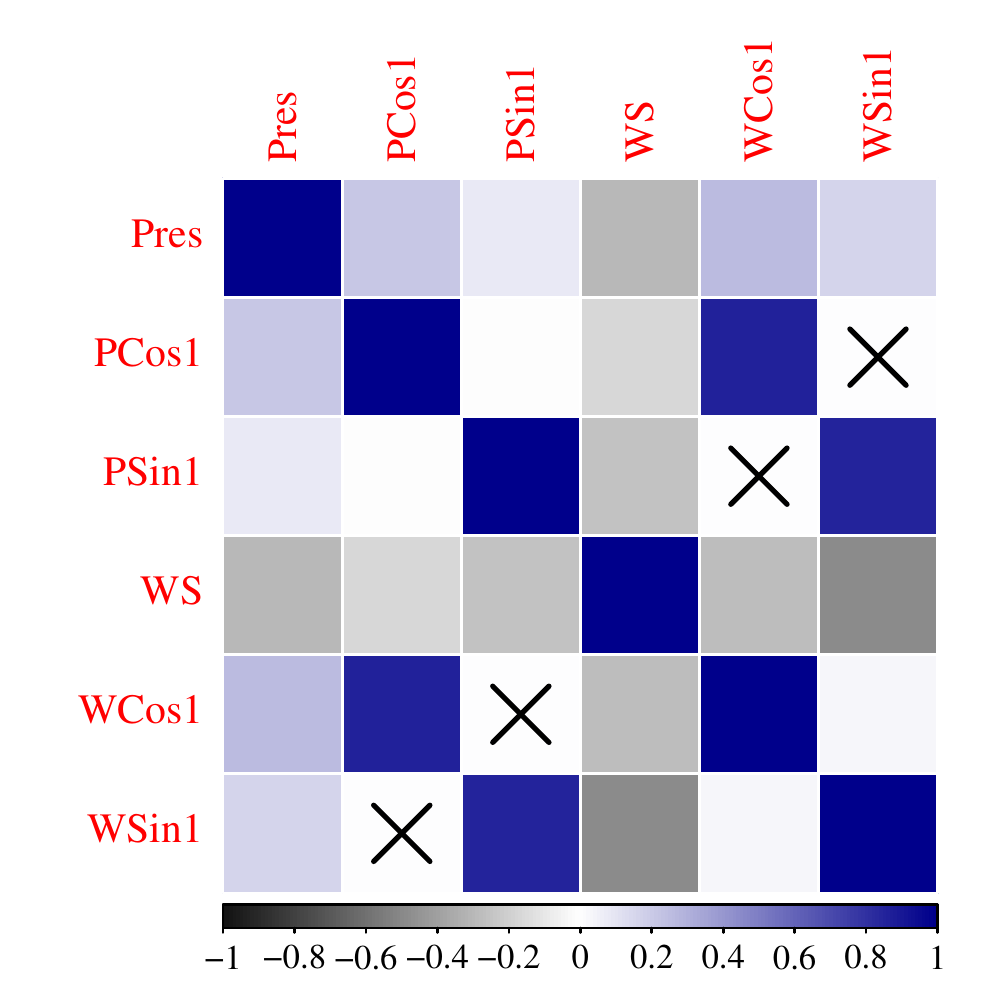} \\
    \caption{Correlation plot Berlin-Tegel.}
  \end{subfigure}
  \begin{subfigure}[b]{0.49\textwidth}
   \includegraphics[width=1\textwidth]{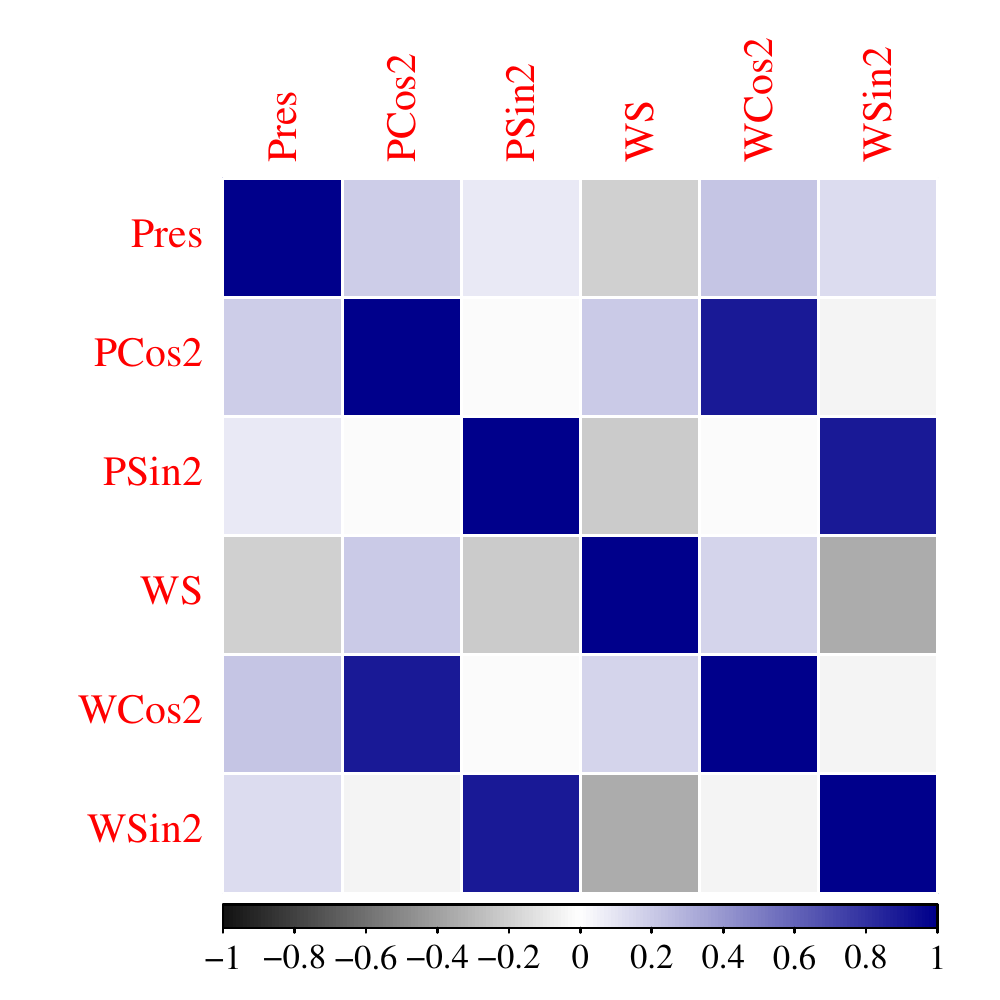} \\
    \caption{Correlation plot Berlin-Tempelhof.}
 \end{subfigure}
  \caption{Plot of pairwise correlation for all dependent variables.}
  \label{graph:cor}
\end{figure}

The wind direction is a circular variable. Therefore, we split it the wind direction into a sine and a cosine part. For that reason we use the following decomposition in Cartesian coordinates: $W_{s,t} = W_t \cdot \sin (D_{t})$ and $W_{c,t} = W_t \cdot \cos (D_{t})$. The wind speed is $W_t$ and $D_t$ is the wind direction. Furthermore, we will also consider the Cartesian coordinates of the air pressure ($P_t$) and the sine and cosine transformed direction, which is given by $P_{s,t} = P_t \cdot \sin (D_{t})$ and $P_{c,t} = P_t \cdot \cos (D_{t})$. Basically, the idea behind this decomposition is to consider the air pressure as a spatial magnitude. Therefore, we use the Cartesian coordinates of the air pressure to derive a pseudo spatial dependent variable. Figure \ref{graph:cor} depicts the correlation between all pairs of dependent variables. The figure illustrates that the correlation between the wind speed and the other dependent variables is significant and negative. The crossed boxes refer to some insignificant correlations.  Moreover, the wind speed and the wind direction are influenced by the air pressure, but not vice versa, because there is no causal relationship. However, wind speed and wind direction are correlated since they are influenced by the Earth's rotation, which is known as the Coriolis force.

The developed model is a seasonal and periodical vector autoregressive model with interactions and thresholds (TVARX). Moreover, the conditional variance is seasonal and so the residual process is modelled by a threshold autoregressive process with interactions (TARCHX). The model is suitable to describe the complete relationship among wind speed, wind direction and air pressure. The multivariate dependent variable $\boldsymbol{Y_t}$ captures the wind speed information $W_t$, the air pressure information $P_{t}$ and the decomposed wind direction (Cartesian coordinate) information $W_{s,t}$, $W_{c,t}$, $P_{s,t}$ and $P_{c,t}$, within a time interval from $t = 1, \ldots, T$. The dependent vector consists of six variables, i.e. $\boldsymbol{Y}_t \in \mathbb{R}^{6}$ and $\mathcal{M} \in \{1,...,6\}$, which is
\begin{equation}\label{eq:1}
\boldsymbol{Y}_t =  \bigg( P_t \; \; \; \; P_{s,t}  \; \; \; \; P_{c,t}  \; \;  \; \;  W_{t}
 \; \;  \; \;  W_{s,t} \; \;\; \; W_{c,t} \bigg)'.
\end{equation}
For each measurement station we propose the same mean and conditional variance model. Thus, the TVARX process is given by 
\begin{eqnarray}\label{eq:TVAR}
                    \boldsymbol{Y_t} &=& \boldsymbol{\upsilon}_t  
                    + \sum_{j_{1}=1}^{{J_{1}}} \boldsymbol{\Theta}_{t,j_{1}} \circ \boldsymbol{A} \boldsymbol{Y}_{t-j_{1}}  
					+ \sum_{j_{2}=1}^{{J_{2}}} \sum_{\boldsymbol{c_{\alpha}} \in \boldsymbol{C}} \boldsymbol{\phi}_{t,j_{2},c_{\alpha}} \circ
					\max \{ \boldsymbol{Y}_{t-j_{2}} ,\boldsymbol{ c_{\alpha}} \}                 
                    + \boldsymbol{\epsilon}_t ,\\
                    \boldsymbol{\epsilon}_t &=&  \boldsymbol{\sigma}_t \boldsymbol{\eta}_t ,
\end{eqnarray}

\begin{eqnarray}\label{eq:TVAR1}
                    \boldsymbol{\upsilon}_t &=& \boldsymbol{\upsilon}_{0}
                    + \sum^{k_1}_{i_1=2} \boldsymbol{\upsilon}_{i_1,1} {B}_{i_1}^{s_1}(t)
                    + \sum^{k_2}_{i_2=2} \boldsymbol{\upsilon}_{1, i_2} {B}_{i_2}^{s_2}(t)
                    + \sum^{k_1}_{i_1=2} \sum^{k_2}_{i_2=2} \boldsymbol{\upsilon}_{i_1, i_2} {B}_{i_1}^{s_1}(t) {B}_{i_2}^{s_2}(t), \\
                    \boldsymbol{\Theta}_{t,j_{1}} & =&  \boldsymbol{\Theta}_{0,j_{1}}
                    + \sum^{k_1}_{i_1=2} \boldsymbol{\Theta}_{i_1,1,j_{1}} {B}_{i_1}^{s_1}(t)
                    + \sum^{k_2}_{i_2=2} \boldsymbol{\Theta}_{1, i_2,j_{1}} {B}_{i_2}^{s_2}(t) \\
                 &&   + \sum^{k_1}_{i_1=2} \sum^{k_2}_{i_2=2} \boldsymbol{\Theta}_{i_1, i_2, j} {B}_{i_1}^{s_1}(t) {B}_{i_2}^{s_2}(t), \\
                    \boldsymbol{\phi}_{t,j_{2},c_{\alpha}}  &=&  \boldsymbol{\phi}_{0,j_{2},c_{\alpha}}
                    + \sum^{k_1}_{i_1=2} \boldsymbol{\phi}_{i_1,1,j_{2},c_{\alpha}} {B}_{i_1}^{s_1}(t)
                    + \sum^{k_2}_{i_2=2} \boldsymbol{\phi}_{1, i_2,j_{2},c_{\alpha}} {B}_{i_2}^{s_2}(t) \\
           &&   + \sum^{k_1}_{i_1=2} \sum^{k_2}_{i_2=2} \boldsymbol{\phi}_{i_1, i_2, j_{2},c_{\alpha}} {B}_{i_1}^{s_1}(t) {B}_{i_2}^{s_2}(t),                
\end{eqnarray}
\noindent where $\{\boldsymbol{\eta}_t\}$ is i.i.d., $\mathrm{E}(\boldsymbol{\eta}_t) = \boldsymbol{0}$ and $\mathrm{Var}(\boldsymbol{\eta}_t) = \boldsymbol{1}$. The matrix $\boldsymbol{A}$ modifies the correlation structure of the autoregressive coefficients. Therefore, we must calculate the Hadamard product. The matrix $\boldsymbol{A}$ is defined by

\begin{equation}
\boldsymbol{A} = \begin{pmatrix}
1 & 1 & 1  & 0 & 0 & 0\\
1 & 1 & 1  & 0 & 0 & 0 \\
1 & 1 & 1  & 0 & 0 & 0 \\
1 & 1 & 1  & 1 & 1 & 1 \\
1 & 1 & 1  & 1 & 1 & 1 \\
1 & 1 & 1  & 1 & 1 & 1 
\end{pmatrix}.
\end{equation}

Moreover, $\boldsymbol{\upsilon}_0$ is an $6 \times 1$ intercept vector and $\boldsymbol{\upsilon}_{i_1, 1}$, $\boldsymbol{\upsilon}_{1, i_2}$ and $\boldsymbol{\upsilon}_{i_1, i_2}$ are $6 \times 1$ periodic coefficient vectors. Furthermore, $\boldsymbol{\Theta}_{0,j_{1}}$, $\boldsymbol{\Theta}_{i_1,1,j_{1}}$, $\boldsymbol{\Theta}_{1,i_2,j_{1}}$ and $\boldsymbol{\Theta}_{i_1,i_2,j_{1}}$ are $6 \times 6$ parameter matrices of autoregressive and periodic autoregressive parameters for lag $j_1 \in \mathbb{N}$. $\boldsymbol{\phi}_{0,j_{2}}$, $\boldsymbol{\phi}_{i_1,1,j_{2}}$, $\boldsymbol{\phi}_{1,i_2,j_{2}}$ and $\boldsymbol{\phi}_{i_1,i_2,j_{2}}$ are $6 \times 6$ parameter matrices of threshold autoregressive and periodic threshold autoregressive parameters for lag $j_2 \in \mathbb{N}$. The maximum vector $\max \{ \boldsymbol{Y}_{t-j_{2}} ,\boldsymbol{ c_{\alpha}} \}$ describes the component-wise maximum between each element of the vector $\boldsymbol{Y}_{t-j_{2}}$ and the vector of the deciles $\boldsymbol{c}$. Finally, $\boldsymbol{C}$ is a vector space of all possible $\alpha$-percentiles. Figure \ref{graph:thres} provides an evidence for some of the following percentiles $\alpha \in \{0.01,0.02,...,0.05,0.1,...,0.95,0.96,...,0.99\}$. An example of the $\alpha$-percentile vector $\boldsymbol{c_{\alpha}}$ is given in equation \eqref{eq:2}. 

\begin{equation}\label{eq:2}
\boldsymbol{c_{\alpha}}' = \bigg( P_{\alpha,t} \,\,\, P_{s,\alpha,t}  \,\,\, P_{c,\alpha,t}  W_{\alpha,t}  \,\,\,  W_{s,\alpha,t}  \,\,\,  W_{c,\alpha,t} \bigg)'.
\end{equation}
Capturing the thresholds in such a way excludes redundant coefficients within the parameter vector $\boldsymbol{\phi}_{t,j_{2},c}$, but the parameter space might be inflated by redundant parameter. Hence, Table \ref{table:Parameter} shows the included threshold autoregressive parameters which are related to a lag of six hours up to one day. 

The periodic functions ${B}_{i_1}^{s_1}(t)$ and ${B}_{i_2}^{s_2}(t)$ are related to the diurnal period $s_1$ and the annual period $s_2$. \cite{Hering2010} and \cite{ambach2015periodic} approximate the periodic functions by Fourier functions. Instead of a Fourier series, we consider periodic B-spline functions which are used to account for a flexible periodic modelling structure. Compared to Fourier series the B-spline functions could be used to describe the time-dependent structure in a better way. The fundamentals of periodic B-spline basis functions are described by \cite{deBoor2001book}. Therefore, it is important and necessary to define a set of equidistant knots. For our model, we consider cubic B-splines, which are suitable, because they are twice continuously differentiable. Besides, we incorporate possible interactions, which are multiplications of diurnal and annual basis functions as described in \cite{ambach2016vorhersagen}. The diurnal and the annual frequency is captured by six equidistant B-spline functions. Finally, \cite{ambach2015periodic} utilize a periodic time series model, where the regression coefficients vary from day to day. 

In the context of wind speed modelling and forecasting it is important to take heteroscedasticity into account as \cite{ambach2015periodic} and \cite{Taylor2009} point out. Therefore, we demand that the conditional variance should follow an ARCH process. According to our data set and the impact of outliers it is better to model the conditional standard deviation $\boldsymbol{\sigma}_t$ instead of the variance. While estimating the power of the ARCH model, we can support the hypothesis of modelling the conditional standard deviation. Hence, $\boldsymbol{\sigma}_t$ and $\boldsymbol{\eta}_t$ are $6 \times 1$ vectors. The conditional standard deviation has to follow a TARCH process \cite[see][]{GJR1993} which is given by
\begin{eqnarray}\label{eq:TARCH}
    \boldsymbol{\sigma}_t &=& \boldsymbol{\beta}_{t} 
    + \sum_{h=1}^{{P}} \boldsymbol{\xi}_{t,h} \circ  \boldsymbol{I}_t^{+}\circ \boldsymbol{\epsilon}_{t-h} 
    + \sum_{l=1}^{{Q}} \boldsymbol{\psi}_{t,l} \circ  \boldsymbol{I}_t^{-} \circ\boldsymbol{\epsilon}_{t-l} ,
 \end{eqnarray} 
\begin{eqnarray}\label{eq:TARCH1}  
\boldsymbol{\beta}_{t} &=& \boldsymbol{\beta}_{0}
               + \sum^{k_1}_{i_1=2} \boldsymbol{\beta}_{i_1,1} {B}_{i_1}^{s_1}(t)
               + \sum^{k_2}_{i_2=2} \boldsymbol{\beta}_{1, i_2} {B}_{i_2}^{s_2}(t)
               + \sum^{k_1}_{i_1=2} \sum^{k_2}_{i_2=2} \boldsymbol{\beta}_{i_1, i_2} {B}_{i_1}^{s_1}(t) {B}_{i_2}^{s_2}(t), \\  
\boldsymbol{\xi}_{t,h} &=&  \boldsymbol{\xi}_{0, h} 
			   + \sum^{k_1}_{i_1=2} \boldsymbol{\xi}_{i_1,1, h} {B}_{i_1}^{s_1}(t) 
			   + \sum^{k_2}_{i_2=2} \boldsymbol{\xi}_{1, i_2, h} {B}_{i_2}^{s_2}(t)
               + \sum^{k_1}_{i_1=2} \sum^{k_2}_{i_2=2} \boldsymbol{\xi}_{i_1, i_2, h} {B}_{i_1}^{s_1}(t) {B}_{i_2}^{s_2}(t), \\  
\boldsymbol{\psi}_{t,l} &=&  \boldsymbol{\psi}_{0, l}  
			 + \sum^{k_1}_{i_1=2} \boldsymbol{\psi}_{i_1, 1, l} {B}_{i_1}^{s_1}(t)
			 + \sum^{k_2}_{i_2=2} \boldsymbol{\psi}_{1, i_2 ,l} {B}_{i_2}^{s_2}(t)
             + \sum^{k_1}_{i_1=2} \sum^{k_2}_{i_2=2} \boldsymbol{\psi}_{i_1, i_2 ,l} {B}_{i_1}^{s_1}(t) {B}_{i_2}^{s_2}(t),   
\end{eqnarray}
\noindent where $\boldsymbol{\beta}_0$ is an $6 \times 1$ intercept vector and $\boldsymbol{\beta}_{i_1, 1}$, $\boldsymbol{\beta}_{1, i_2}$ and $\boldsymbol{\beta}_{i_1, i_2}$ are $6 \times 1$ periodic coefficient vectors within the standard deviation process. Furthermore, $\boldsymbol{\xi}_{0,h}$, $\boldsymbol{\psi}_{0,l}$, $\boldsymbol{\xi}_{i_1,1,h}$, $\boldsymbol{\psi}_{i_1,1,l}$, $\boldsymbol{\xi}_{1,i_2,h}$, $\boldsymbol{\psi}_{1,i_2,l}$ , $\boldsymbol{\xi}_{i_1,i_2,h}$ and $\boldsymbol{\psi}_{i_1,i_2,l}$ are autoregressive and periodic autoregressive parameter vector $6 \times 6$ within the standard deviation. Hence, the $6 \times 1$ vectors of indicator functions $I_t^{+}$ and $I_t^{-}$ are given by
\begin{eqnarray}
\begin{array}[t]{lcr}
 I_{\mathcal{M},t-1}^{+} = \begin{cases} 1 , 
 \qquad
 \boldsymbol{\epsilon}_{\mathcal{M},t-1} > \boldsymbol{0} \\ 0, 
  \qquad
 \boldsymbol{\epsilon}_{\mathcal{M},t-1} \leq \boldsymbol{0} \end{cases} & \text{,}
&  I_{\mathcal{M},t-1}^{-} = \begin{cases} 1, 
\qquad
\epsilon_{\mathcal{M},t-1} \leq 0 \\ 0, 
\qquad
\epsilon_{\mathcal{M},t-1} > 0 \end{cases},
\end{array}
\end{eqnarray}
\noindent where $\mathcal{M}$ depicts one of the six dependent variable and $\mathcal{M} \in \{1,...,6\}$. 
Therefore, we obtain the vector of indicator functions $(\mathcal{M} \times 1)$
$\boldsymbol{I}_t^{+} = \left(I_{1,t}^{+},   \ldots, I_{6,t}^{+},\right)^\prime$ and 
$\boldsymbol{I}_t^{-} = \left(I_{1,t}^{-},   \ldots, I_{6,t}^{-}\right)^\prime$. This indicator function vector is one possibility to model the TARCH process which doubles the parameter space. The conditional standard deviation does only include the most important threshold which is related to negative and positive shocks of the variance \citep[see][]{ambach2015periodic}. The complete autoregressive parametrization is given in Table \ref{table:Parameter}. The dependent variables show a huge autocorrelation structure, which is captured by the included lags in the mean and the conditional variance part. The threshold autoregressive coefficients $\boldsymbol{\phi}_{t,j_2,c}$ increase the parameter space tremendously. For that reason we include only some selected $\alpha$-percentile threshold autoregressive lags. The total amount of parameters for our mean model accounts for roughly 37000 input variables and the variance model includes about 5000 input variables.

\begin{table}[h]
\centering
\begin{tabular}{lll}
  \hline
\textbf{Model} & \textbf{Index} & \textbf{Included lags}  \\ 
  \hline
Mean & ${J_1}$ & $\{1,...,500\}$ and $\{576,720,864,1008\}$ \\
     & ${J_2}$ & $\{1,2,4,9,18,36,72,144\}$ \\
     \hline
Variance & ${P}$ & $\{1,...,40\}$ and $\{140,...,150\}$ \\
     & ${Q}$ & $\{1,...,40\}$ and $\{140,...,150\}$ \\
   \hline
\end{tabular}
 \caption{Included autoregressive and periodic autoregressive lags.}
   \label{table:Parameter}
\end{table}

\subsection{Model Estimation}

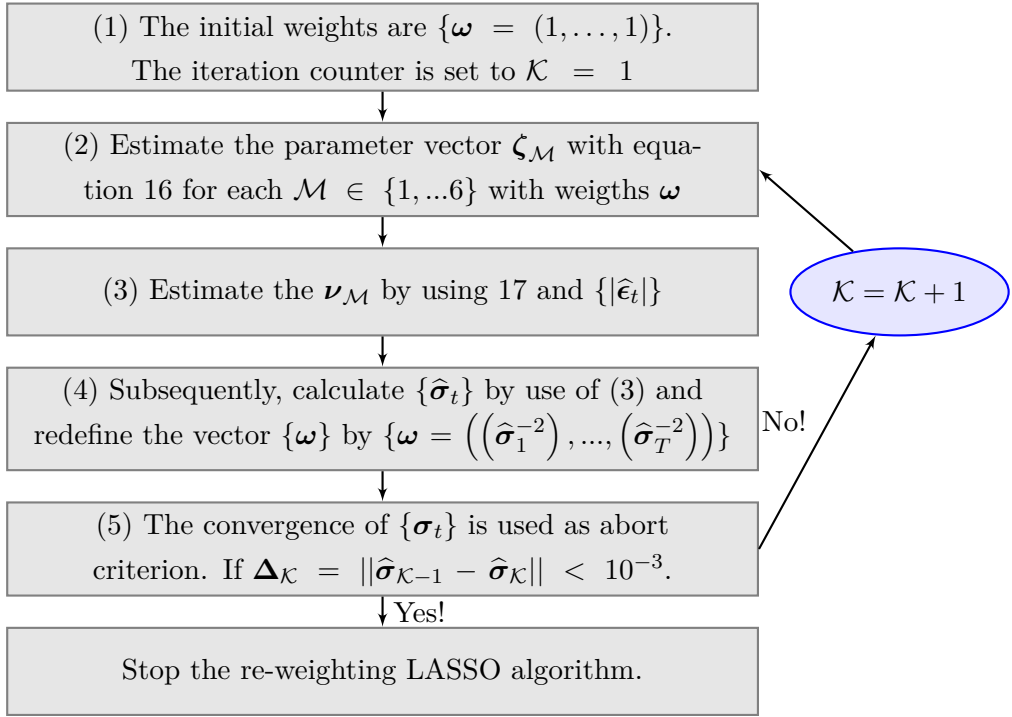
\begin{figure}[H]
\begin{small}
\begin{tikzpicture}
[auto,
decision/.style={diamond, draw=gray, thick, fill=green!10, text width=1em, text badly centered, inner sep=3em},
block/.style ={rectangle, draw=gray, thick, fill=gray!20, text width=25em, text centered, minimum height=3em},
line/.style ={draw, thick, -latex',shorten >=0pt},
cloud/.style ={draw=blue, thick, ellipse, fill=blue!10, minimum height=3em}]
\matrix [column sep=4mm,row sep=4mm]
{
& \node [block] (1) {(1) The initial weights are $\{\boldsymbol{\omega} = (1,\ldots,1)\}$.\\ The iteration counter is set to $\mathcal{K}=1$}; & \coordinate (dummy-write);\\
& \node [block] (2) {(2) Estimate the parameter vector $\boldsymbol{\zeta}_{\mathcal{M}}$ with equation \ref{eq:LASSO} for each $\mathcal{M} \in \{1,...6\}$ with weigths  $\boldsymbol{\omega} $}; & \coordinate (dummy-write-done);\\
& \node [block] (3) {(3) Estimate the $\boldsymbol{\nu}_{\mathcal{M}}$ by using \ref{eq:LASSO1} and $\{|\boldsymbol{\widehat{\epsilon}}_t|\}$}; & 
 \node [cloud, right of=3] (system) {\small{$\mathcal{K} = \mathcal{K} +1$}}; \\
& \node [block] (4) {(4) Subsequently, calculate $\{\boldsymbol{\widehat{\sigma}}_t \}$ by use of (3) and redefine the vector $\{ \boldsymbol{\omega} \}$ by $\{\boldsymbol{\omega} = \left(\left(\boldsymbol{\widehat{\sigma}}_{1}^{-2}\right),...,\left(\boldsymbol{\widehat{\sigma}}_{T}^{-2}\right) \right)\}$ }; & \coordinate (dummy5);\\
& \node [block] (5) {(5) The convergence of $\{ \boldsymbol{{\sigma}}_t \}$ is used as abort criterion. If
$\boldsymbol{\Delta}_{\mathcal{K}} = || \boldsymbol{\widehat{\sigma}}_{\mathcal{K}-1}-\boldsymbol{\widehat{\sigma}}_{\mathcal{K}} || < 10^{-3}$. }; & \coordinate (dummy-review-done);\\
& \node [block] (6) {Stop the re-weighting LASSO algorithm.}; & \\
};
\begin{scope}[every path/.style=line, rounded corners]
\path (1) -- (2);
\path (2) -- node [midway] {} (3);
\path (3) -- (4);
\path (4) -- (5);
\path (5.east) -- node[midway] {No!} (system);
\path (system) -- (2.east);
\path (5) -- node [midway] {Yes!} (6);
\end{scope}
\end{tikzpicture}
\end{small}
  \caption{Estimation scheme.}\label{fig:scheme}

\end{figure}

\noindent The estimation of the complete model is done in two steps. Primarily, the mean is estimated and hereafter the conditional variance process. Therefore, we use an iteratively re-weighted estimation method. \cite{mbamalu1993load} apply such an approach which is based on the ordinary least squares method to predict the wind power. \cite{wagener2012bridge} and \cite{ziel2015iteratively} use an iteratively re-weighted LASSO method in a heteroscedastic setting. We use the LASSO approach to estimate the mean \eqref{eq:TVAR} as well as the standard deviation process \eqref{eq:TARCH}. The iteratively re-weighted LASSO method provides different advantages compared to other estimation methods. First, it is possible to compare parameter estimation and model selection in one algorithm. Secondly, an iteratively re-weighting scheme enables the incorporation of heteroscedastic effects without a specific distributional assumption. Finally, this approach is computationally fast and can be successfully applied for wind speed predictions \citep[see][]{ambach2016vorhersagen}. The mean model is estimated in the first step and afterwards, we re-weight the mean model using $\boldsymbol{\widehat{\sigma}}_t$. Thus, we estimate in the beginning the parameter vector $\boldsymbol{\zeta}_{\mathcal{M}}$ of the TVARX model \eqref{eq:TVAR} 
\begin{equation}\label{eq:LASSO}
\boldsymbol{\hat{\zeta}}_{\mathcal{M}} = \underset{{\tiny{ \boldsymbol{\hat{\zeta}}_{\mathcal{M}}}} \in \mathbb{R}^{V}} {\arg \min} \sum_{t=1}^T({Y_{\mathcal{M},t}} - {\omega_t} \boldsymbol{X}_{\mathcal{M},t,\mu} \boldsymbol{\zeta}_{\mathcal{M},t} )^2 + \lambda_{\mathcal{M},T,\mu} \sum_{v = 1}^{V} |\boldsymbol{\zeta}_{\mathcal{M},v}|,
\end{equation}
where $\omega_t$ is the weight vector $\boldsymbol{\omega} = (\omega_1,...,\omega_T)$, $V \in \mathbb{N}$ is the number of elements of the parameter vector $\boldsymbol{\zeta}_{\mathcal{M}}$ and $\boldsymbol{X}_{\mathcal{M},t,\mu}$ is the matrix of all regressors which are included in \eqref{eq:TVAR}. Figure \ref{fig:scheme} shows the estimation scheme of the re-weighted LASSO method, where the initial weight vector is $\boldsymbol{\omega}_t = (1,...,1)$. The tuning parameter for the mean model is $\lambda_{\mathcal{M},T,\mu} \geq 0$. Subsequently, we have to describe the second LASSO procedure to estimate $\boldsymbol{\widehat{\sigma}}_t$. The parameter vector $\boldsymbol{\nu}_{\mathcal{M}}$ for TARCHX model \eqref{eq:TARCH}  is given by
\begin{equation}\label{eq:LASSO1}
\boldsymbol{\hat{\nu}}_{\mathcal{M}} = \underset{ \boldsymbol{\nu}_{\mathcal{M}} \geq 0}{\underset{{\boldsymbol{\nu}_{\mathcal{M}}} \in \mathbb{R}^{W}}{\arg \min}} \sum_{t=1}^T(\hat{\epsilon}_{\mathcal{M},t} - \boldsymbol{X}_{\mathcal{M},t,\sigma} \boldsymbol{\nu}_{\mathcal{M},t} )^2 + \lambda_{\mathcal{M},T,\sigma} \sum_{w = 1}^{W} |\boldsymbol{\nu}_{\mathcal{M},w}|,
\end{equation}
where $W \in \mathbb{N}$ is the number of elements of the variance parameter vector $\boldsymbol{\nu}_{\mathcal{M}}$, $\lambda_{\mathcal{M},T,\mu} \geq 0$ is the tuning parameter for the variance model and $\boldsymbol{X}_{\mathcal{M},t,\sigma}$ is the matrix of all parameters included in the TARCHX process \eqref{eq:TARCH}. The estimated parameters $\boldsymbol{\nu}_{\mathcal{M}}$ have to be non-negative in order to guarantee that the recurrence equations and therefore the volatility $\boldsymbol{\widehat{\sigma}}_t$ are well defined. The complete estimation scheme for the iteratively re-weighted LASSO method is provided in Figure \ref{fig:scheme} \citep[see][]{efron2004}. The re-weighting procedure is repeated until we archive a kind of convergence. To solve the LASSO optimization, we consider the coordinate descent method which solves the problem for a predetermined parameter grid \citep[see][]{friedman2007pathwise}. The tuning parameters are selected by the Akaike information criterion (AIC). This is related to the fact, that we want to capture as much as possible structure of the underlying process.

An advantage of the iteratively re-weighted LASSO method is that in contrast to a maximum likelihood estimator, we do not need a distributional assumption for the seasonal periodic TVARX-TARCHX model. The covariance structure among all dependent variables are covered by the threshold vector autoregressive mean model. Hence, the threshold autoregressive structure among the conditional standard deviation process covers the dependence structure of the variance. 

\subsection{In-Sample Modelling Results}

\begin{figure}[h]
  \centering
  \begin{subfigure}[b]{0.49\textwidth}
  \includegraphics[width=1\textwidth]{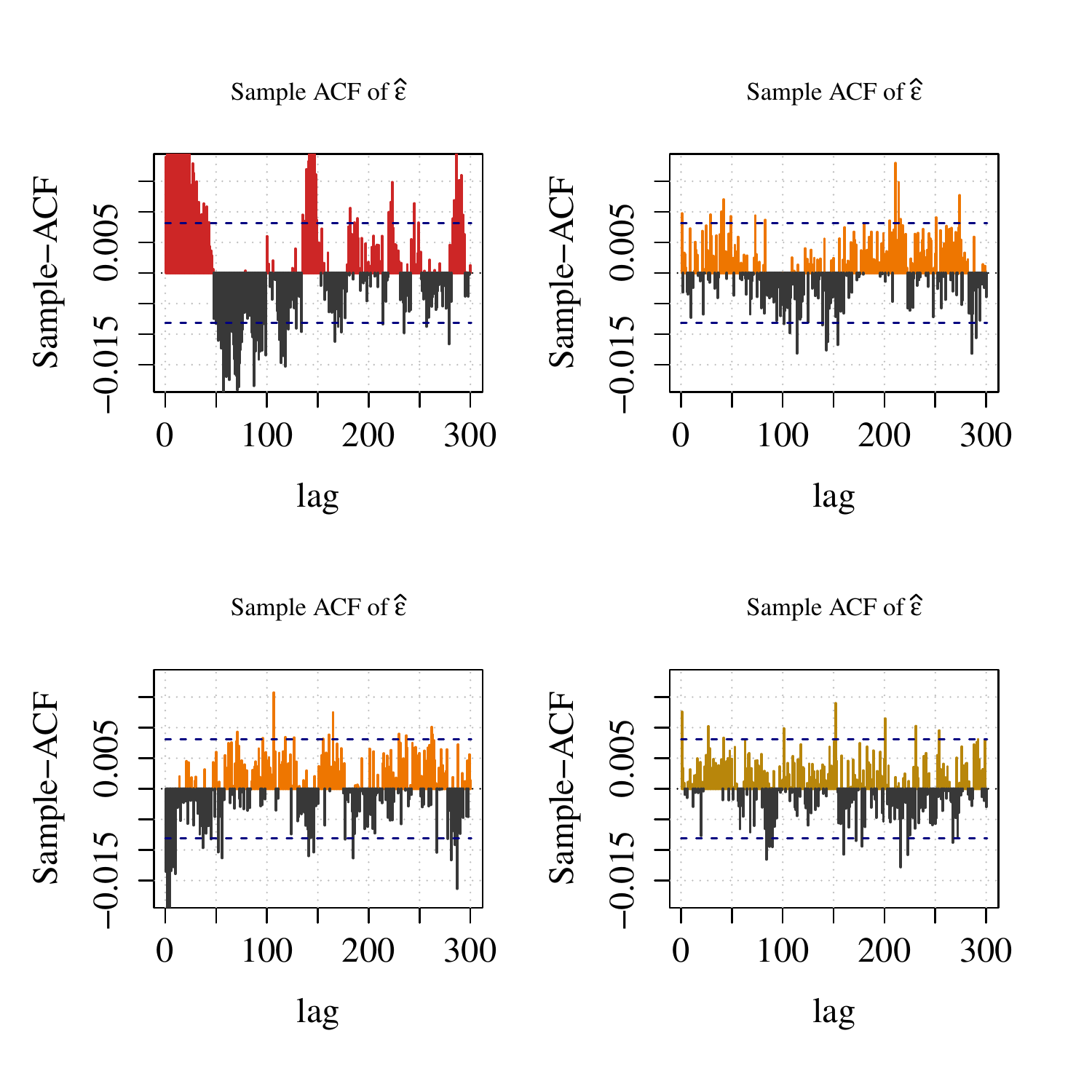} \\
    \caption{ACF and CCF plot of Berlin-Tegel.}
  \end{subfigure}
  \begin{subfigure}[b]{0.49\textwidth}
   \includegraphics[width=1\textwidth]{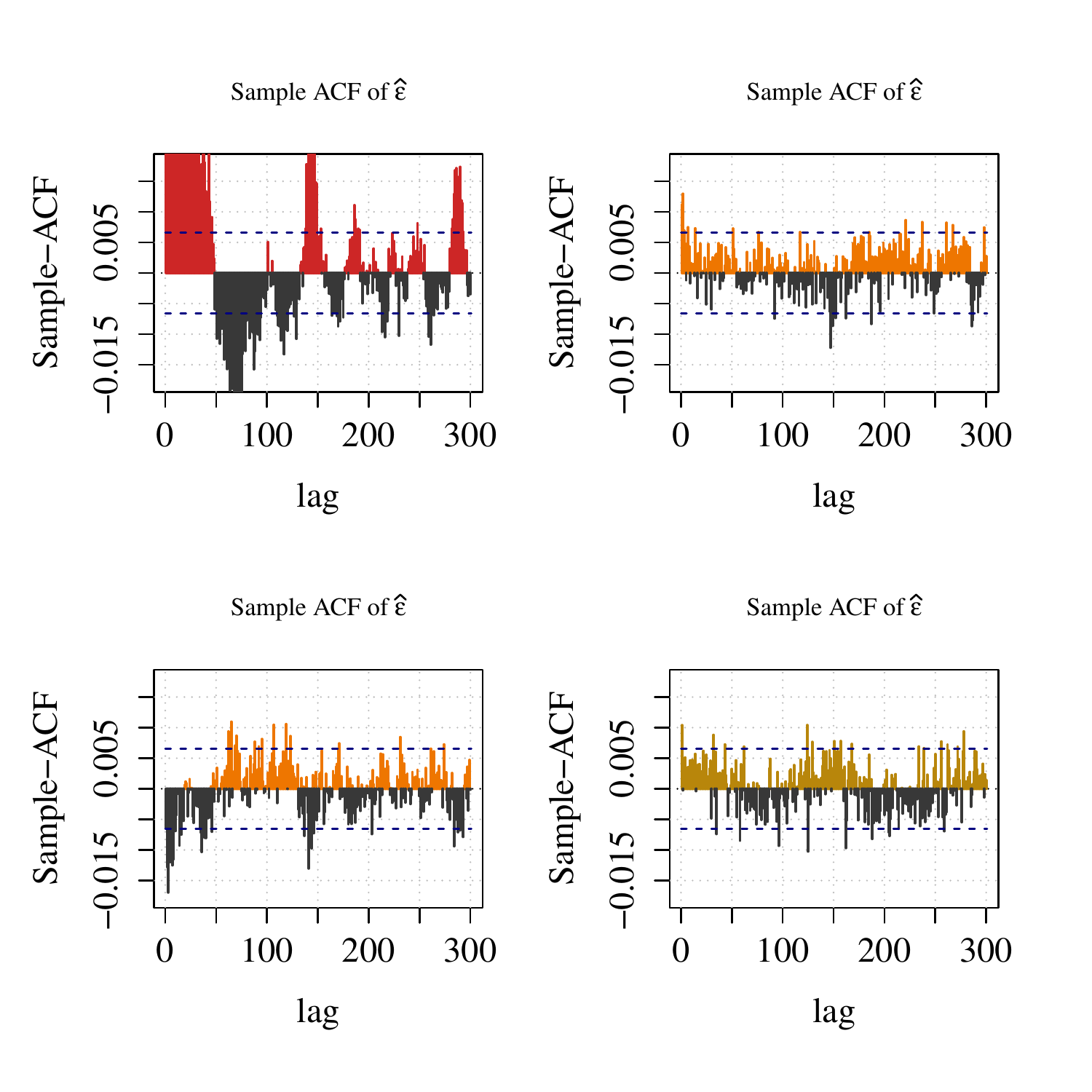} \\
    \caption{ACF and CCF plot of Berlin-Tempelhof.}
 \end{subfigure}
  \caption{Autocorrelation function (ACF) of $\{\boldsymbol{\widehat{\eta}_t}\}$ for the air pressure (top right) and wind speed (bottom left) and the CCF of air pressure and wind speed.}
  \label{graph:acf1}
\end{figure}

\noindent The in-sample modelling results provide satisfactory results. The estimated parameters are selected by means of the LASSO algorithm. Therefore, different periodic, autoregressive or periodic autoregressive coefficient are estimated, but the selected variables and the coefficients may change over time. Besides the model selection, we have to evaluate if the model really removes the autocorrelation and the heteroscedasticity. Figure \ref{graph:acf1} depicts the ACF and CCF plot of the stations Berlin-Tegel and Berlin-Tempelhof for the standardised residuals $\{\boldsymbol{\widehat{\eta}}_t\}$ of the air pressure (red) and wind speed (yellow). For each station four CCF plots are shown in Figure \ref{graph:acf1}. This figure provides some remaining autocorrelation in the bottom right (red) ACF figure. The air pressure process contains different non-linear effects which is not captured by our model so far. Thus, we observe a periodic autoregressive structure within the standardised residuals in the end.

\begin{figure}[h]
  \centering
  \begin{subfigure}[b]{0.49\textwidth}
  \includegraphics[width=1\textwidth]{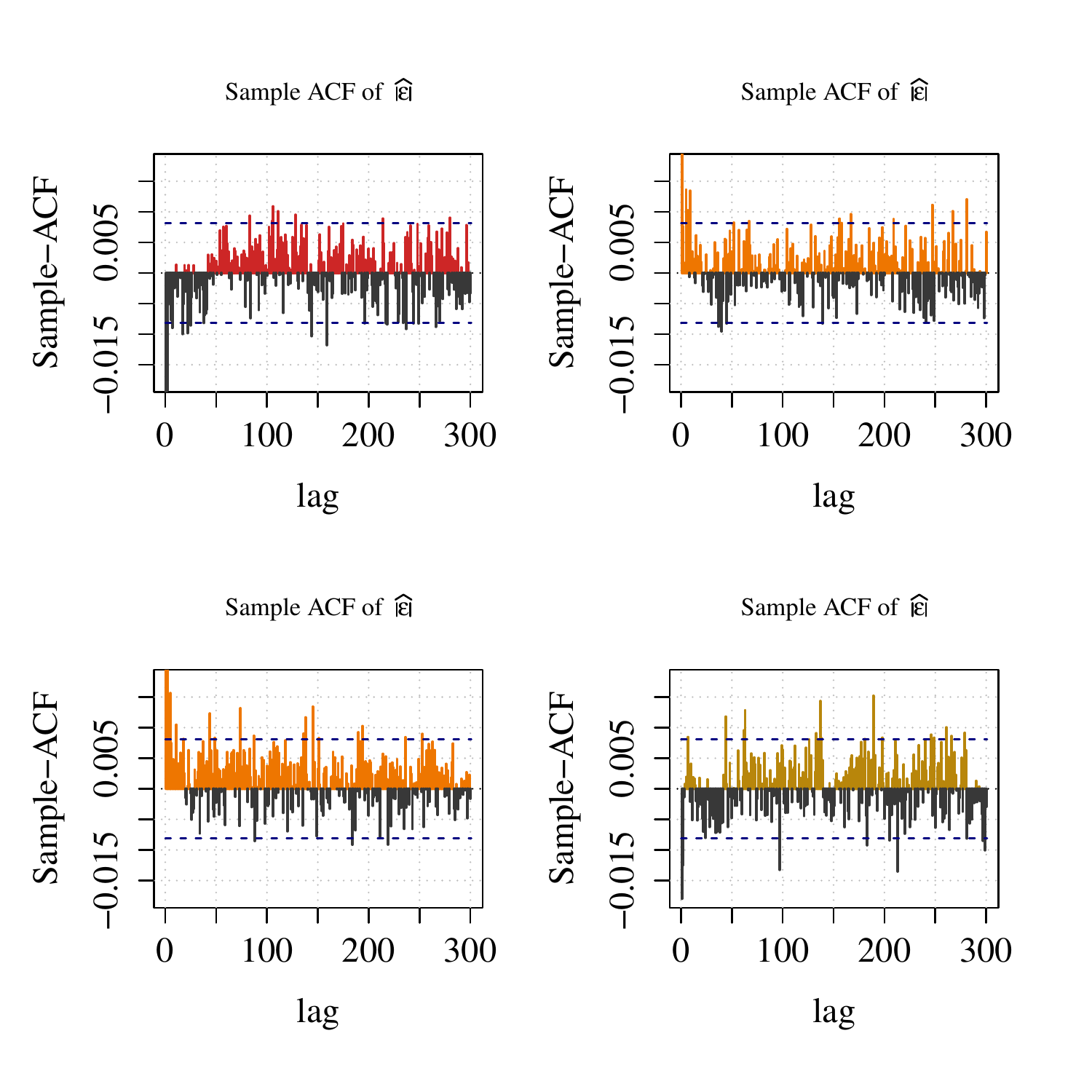} \\
    \caption{ACF and CCF plot of Berlin-Tegel.}
  \end{subfigure}
  \begin{subfigure}[b]{0.49\textwidth}
   \includegraphics[width=1\textwidth]{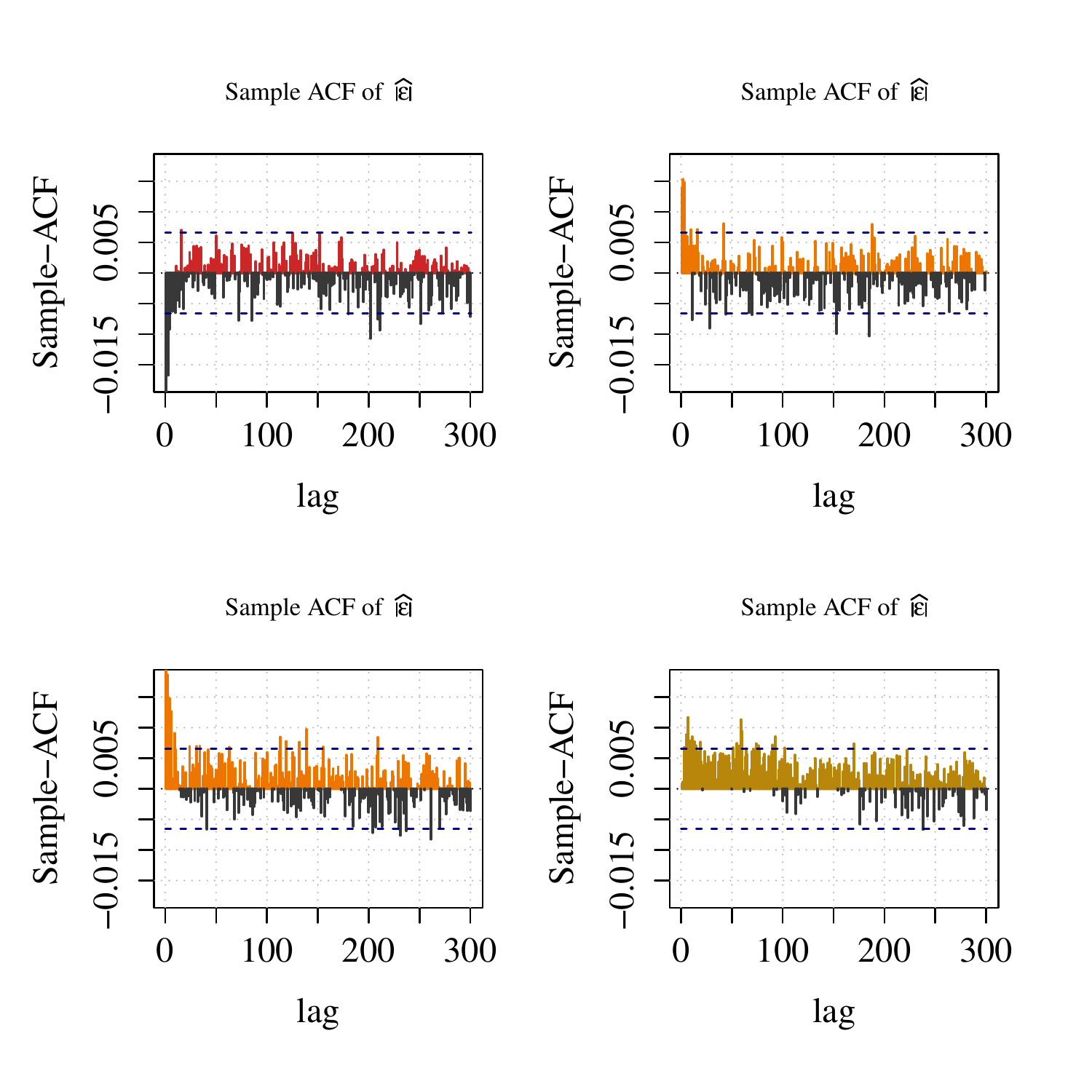} \\
    \caption{ACF and CCF plot of Berlin-Tempelhof.}
 \end{subfigure}
  \caption{Autocorrelation function (ACF) of $\{\boldsymbol{|\widehat{\epsilon}_t|}\}$ for the air pressure (top right) and wind speed (bottom left) and the CCF of air pressure and wind speed.}
  \label{graph:absacf1}
\end{figure}

Figure \ref{graph:absacf1} shows the ACF and CCF of station Berlin-Tegel and Berlin-Tempelhof for the absolute standardised residuals $\{|\boldsymbol{\widehat{\eta}}_t|\}$ of the air pressure and wind speed. The figures \ref{graph:acf1} and \ref{graph:absacf1} provide a minor presence of remaining autocorrelation, as there are few single spikes outside the confidence bands which do not provide a structure. One exception is related to the ACF of standardised residuals $\{\boldsymbol{\widehat{\eta}}_t\}$ for the air pressure. The figures \ref{graph:acf1} and \ref{graph:absacf1} give a hint that autocorrelation has to be considered. The Ljung-Box test for $\{\boldsymbol{\widehat{\eta_t}}\}$ and $\{|\boldsymbol{\widehat{\eta_t}}|\}$ is calculated and we are not able to reject the null hypothesis of independence if we consider a level of significance of $5\%$. Finally, the analysis of autocorrelation and heteroscedasticity suggests a good model fit.

\section{Forecasting Results and Evaluation of the Model}\label{sec:outsample}

In the previous section a multivariate modelling approach is introduced which is now used to predict the wind speed, the wind direction and the air pressure. Subsequently, we shall discuss the out-of-sample performance of the new approach and several benchmark models. The selected benchmarks are recently proposed short-term prediction approaches. Additionally, we compute medium-term forecasts to show the consistency of our prediction results. However, we do not consider long-term forecasts and do not focus on medium-term predictions in this paper. Such forecasts mainly make use of a NWP model or combinations of statistical and NWP models as proposed by \cite{zhao2016improved}.

The out-of-sample forecasting accuracy is measured by means of the root mean square error (RMSE) and the mean absolute error (MAE). Moreover, the probability integral transform (PIT) histogram is calculated to evaluate the sharpness and calibration of the prediction approach \citep[see][]{Gneiting2007}. The PIT histogram is a tool to evaluate the statistical consistency between our calculated probabilistic forecasts and the corresponding observations. Therefore, all actual observations are compared with the predictive distribution of the observations. An ideal forecast has to return a flat PIT histogram. However, if there are bins of the histogram which are very high or low, we observe a misconfiguration of the forecasting model.
The out-of-sample prediction horizon spans from October 19, 2014 to October 20, 2015. For evaluating the accuracy of our forecasts, we select different time-points in the out-of-sample period $\tau^{(i)}$, namely $\tau^{(i)} \in \{\text{ October }19, 2014,00:00$ a.m. ,..., October $20,2015,24:00 \text{ p.m.}\}$ which are chosen by random and $i \in \{1,...,N\}$. In the beginning, we calculate an in-sample model order which is used afterwards to update the underlying forecasting approach. The proposed model as well as all benchmarks are re-estimated with the same part of the information set available at the period $\tau$, namely $\boldsymbol{Y}_{\tau-4\cdot365.25\cdot144+1},...,\boldsymbol{Y}_{\tau}$.

Hence, an appropriate in-sample model structure is chosen and short-term predictions are performed. Besides, we compute medium-term predictions and intervals up to twenty-four hours $\boldsymbol{\widehat{Y}}_{\tau+o|\tau}, o \in \{1,...,144\}$. For each model 1.000 ($N = 1000$) out-of-sample predictions are computed. Finally, we are able to calculate the accuracy forecasting measures for the predicted time point $\tau+o$ which are

\begin{eqnarray}\label{eq:accu}
RMSE_o &=& \sqrt{\frac{1}{N} \sum_{i = 1}^N \left(\boldsymbol{Y}_{\tau^{(i)}+o} - \boldsymbol{\widehat{Y}}_{\tau^{(i)}+o|\tau^{(i)}} \right)^2 } , \\
MAE_o  &=& \frac{1}{N} \sum_{i = 1}^N  \left| \boldsymbol{Y}_{\tau^{(i)}+o} - \boldsymbol{\widehat{Y}}_{\tau^{(i)}+o|\tau^{(i)}}  \right|   ,\\
MAEoA_{o}  &=& \frac{1}{N} \sum_{i = 1}^N  \left| \widehat{\varepsilon}_{\tau^{(i)}+o|\tau^{(i)})}   \right|   .\label{eq:accu3}
\end{eqnarray}

\noindent Naturally, $ \boldsymbol{\widehat{Y}}_{\tau^{(i)}+o}$ is the $o$-step forecast of ${\widehat{Y}}_{\tau^{(i)}+o}$ based on $ \boldsymbol{{Y}}_{\tau^{(i)}-210384},..., \boldsymbol{{Y}}_{\tau^{(i)}}$, $\boldsymbol{Y}_{\tau^{(i)}+o}$ is the true observed vector of dependent variables and $N$ is the number of the out-of sample forecasts. \cite{el2008one} propose a method to calculate the MAE of the wind direction. They argue that the MAE of the wind direction has to be calculated in a different way. Their argumentation is illustrated by a simple example. If we predict a wind direction of $5^{\circ}$ and the observed wind direction is $345^{\circ}$, we will obtain an absolute error of $340^{\circ}$, but indeed the error is $20^{\circ}$. This problem is solved by calculating two errors and evaluating the minimum of both \citep[see][]{el2008one}. Hence, we predict the Cartesian components of the wind direction. Therefore, we transform these components by means of the arc tangent into the wind direction again. Subsequently, we calculate the prediction error of the wind direction as follows 

\begin{eqnarray}
\widehat{er1}_{\tau^{(i)}+o|\tau^{(i)}} &=& \left|\widehat{D}_{\tau^{(i)}+o|\tau^{(i)}}
- D_{\tau^{(i)}+o} \right| ,\\
\widehat{er2}_{\tau^{(i)}+o|\tau^{(i)}} &=& 360^{\circ} - \widehat{er1}_{\tau^{(i)}+o} ,\\
\widehat{\varepsilon}_{\tau^{(i)}+o|\tau^{(i)}} &=& \min\left[\widehat{er1}_{\tau^{(i)}+o|\tau^{(i)}},\widehat{er2}_{\tau^{(i)}+o|\tau^{(i)}} \right].
\end{eqnarray}

\noindent Our multivariate forecasting model provides two predictions of the Cartesian Y-coordinate which are $\widehat{W}_{s,\tau^{(i)}+o|\tau^{(i)}}$ and $ \widehat{P}_{s,\tau^{(i)}+o|\tau^{(i)}}$. The corresponding X-coordinates are given by $\widehat{W}_{c,\tau^{(i)}+o|\tau^{(i)}}$ and $ \widehat{P}_{c,\tau^{(i)}+o|\tau^{(i)}}$. These coordinates are used to calculate $\widehat{D}_{\tau^{(i)}+o|\tau^{(i)}}$. Thus, we obtain two different predictions for the wind direction, which are related to the wind speed $\widehat{D}_{\tau^{(i)}+o|\tau^{(i)}}^{W}$ and the air pressure $\widehat{D}_{\tau^{(i)}+o|\tau^{(i)}}^{P}$. These two forecasts are given by

\begin{equation}
\widehat{D}_{\tau^{(i)}+o|\tau^{(i)}}^{W} = \arctan \left(\frac{\widehat{W}_{s,\tau^{(i)}+o|\tau^{(i)}}}{\widehat{W}_{c,\tau^{(i)}+o|\tau^{(i)}}} \right), \qquad 
\widehat{D}_{\tau^{(i)}+o|\tau^{(i)}}^{P} = \arctan \left(\frac{\widehat{P}_{s,\tau^{(i)}+o|\tau^{(i)}} }{ \widehat{P}_{c,\tau^{(i)}+o|\tau^{(i)}} } \right).
\end{equation}
Hence, it is also possible to derive a second forecast for the air pressure and the wind speed using the following equation
\begin{equation}
 \widehat{P}_{\tau^{(i)}+o|\tau^{(i)}}^\ast = \sqrt{\widehat{P}_{s,\tau^{(i)}+o|\tau^{(i)}}^2 +  \widehat{P}_{c,\tau^{(i)}+o|\tau^{(i)}}^2 } , \qquad
 \widehat{W}_{\tau^{(i)}+o|\tau^{(i)}}^\ast = \sqrt{\widehat{W}_{s,\tau^{(i)}+o|\tau^{(i)}}^2 +  \widehat{W}_{\tau^{(i)}+o|\tau^{(i)}}^2 } .
\end{equation}
The comparison study reports both forecasting results for each response variable. Though, we need a test to determine if the forecasts are significantly different or not. Therefore, we calculate the Diebold-Mariano test \citep[see][]{Diebold2002} to identify a difference between the derived forecasts of the air pressure, the wind speed and the wind direction. We use the accuracy measures given in equations \eqref{eq:accu}-\eqref{eq:accu3} and calculate the differences between the two forecasts of the air pressure, the wind speed and the wind direction. 

Eventually, we calculate the accuracy of each explained variable. Subsequently, the forecasting accuracy of our new multivariate model is compared with miscellaneous benchmark models. Besides the persistent or na\"ive prediction method given by, ($\boldsymbol{Y}_{\tau^{(i)}} = \widehat{\boldsymbol{Y}}_{\tau^{(i)}+o}$), we investigate other benchmark approaches. Two of them are the autoregressive model AR(p) and the vector autoregressive model VAR(p). The latter model uses the vector of dependent variables given in equation \eqref{eq:1} and the Akaike information criterion (AIC) to determine the maximum lag order of the AR and VAR process. The autoregressive models (AR, VAR) are estimated by solving the Yule-Walker equations. The described benchmark models are not sophisticated, but they provide accurate short-term forecasting results. 

Moreover, we consider three additional benchmark methods for our comparison study. The first type of benchmark method is a non-linear artificial neural network (ANN). \cite{doucoure2016time} consider an ANN based model with wavelets to predict the wind speed. We include a single feed-forward neural network which is comparble to \cite{li2010comparing}.  
The second benchmark model is the gradient boosting machine (GBM) which is based on machine learning techniques. The GBM is used for the Global Energy Forecasting Competition GEFCom2012 and 2014 \citep[see][]{hong2014global, landry2016probabilistic}. The GBM is based on a consistent learning framework to fit independent models for each wind zone and it predicts the wind speed, the wind power and the wind direction. For our comparison study, we take the same calibration of the GBM as in \cite{landry2016probabilistic}. The third model is the periodic regression model with autoregressive fractionally integrated moving average errors in the mean part and asymmetric power generalized autoregressive conditional
heteroscedasticity within the conditional variance (ARFIMA-APARCH) with normally distributed residuals as proposed in \cite{ambach2015periodic}. This model incorporates daily varying regression coefficients and is estimated by a quasi-maximum likelihood method. The aforementioned approaches are used to perform predictions for every dependent variable, i.e. the air pressure, the wind direction and the wind speed. 

\begin{figure}[h]
  \includegraphics[width=1\textwidth,clip=true]{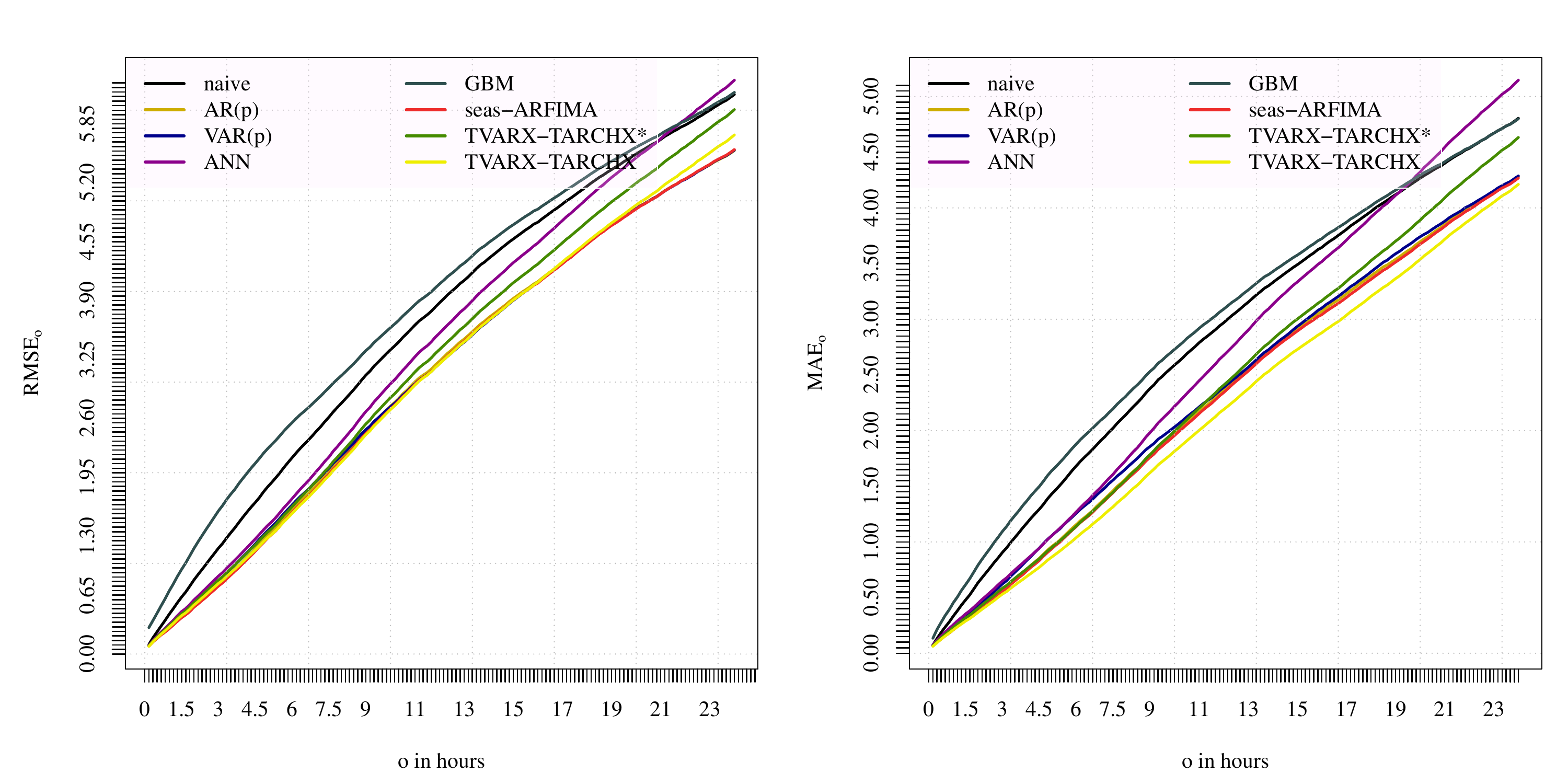}
  \caption{RMSE (first column) and MAE (second column) for the air pressure at station Berlin-Tegel and a forecasting horizon up to one day.}\label{figure:meas1}
  \includegraphics[width=1\textwidth,clip=true]{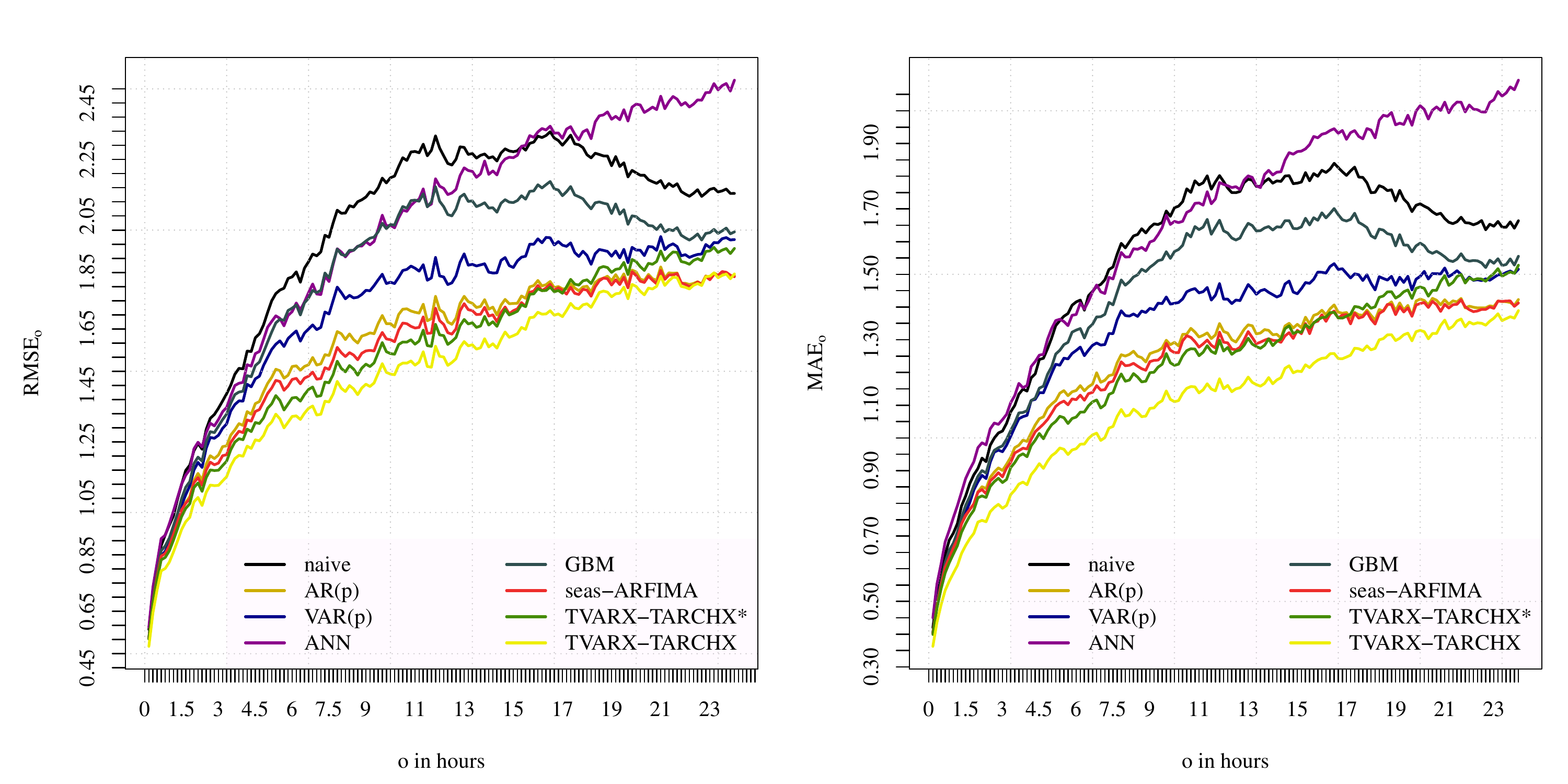}
  \caption{RMSE (first column) and MAE (second column) for the wind speed at station Berlin-Tegel and a forecasting horizon up to one day}\label{figure:meas2}
\end{figure}

Figures \ref{figure:meas1} and \ref{figure:meas2} present the out-of-sample RMSE and MAE results for our wind speed and air pressure response variable for the station Berlin-Tegel. The air pressure predictions $\widehat{P}_{\tau^{(i)}+o}$ of our new model and the benchmark models, i.e., the AR, the VAR, and the seasonal ARFIMA-APARCH approaches, are almost similar up to twenty hours. These forecasting results do not provide a clear result. Moreover, the Diebold-Mariano test does not reject the null hypothesis that the RMSE of all models are different. Using the MAE criterion the best prediction results are obtained for the new periodic seasonal TVARX-TARCHX model. The second forecasting method of the air pressure $\widehat{P}_{\tau^{(i)}+o}^\ast$ which is derived by our new model, provides slightly worse results than $\widehat{P}_{\tau^{(i)}+o}$. 

Furthermore, we take a deeper look at the wind speed forecasting results. Obviously, our new forecasting model outperforms all benchmark approaches. For both performance criteria the best wind speed predictions are obtained by the new forecasting approach based on seasonal TVARX-TARCHX methods

\begin{figure}[h]
  \includegraphics[width=1\textwidth,clip=true]{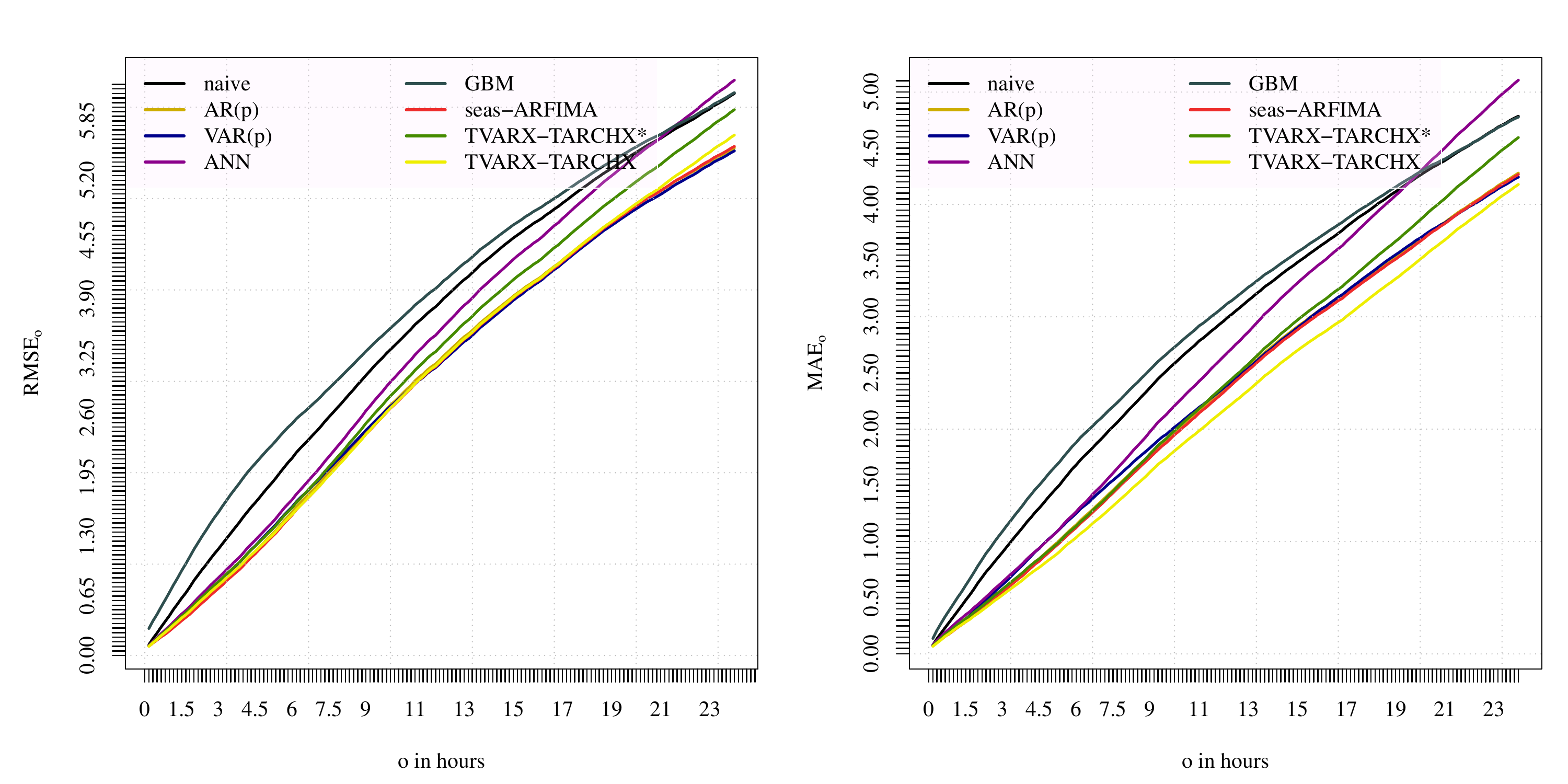}
  \caption{RMSE (first column) and MAE (second column) for the air pressure at station Berlin-Tempelhof and a forecasting horizon up to one day.}\label{figure:meas3}
  \includegraphics[width=1\textwidth,clip=true]{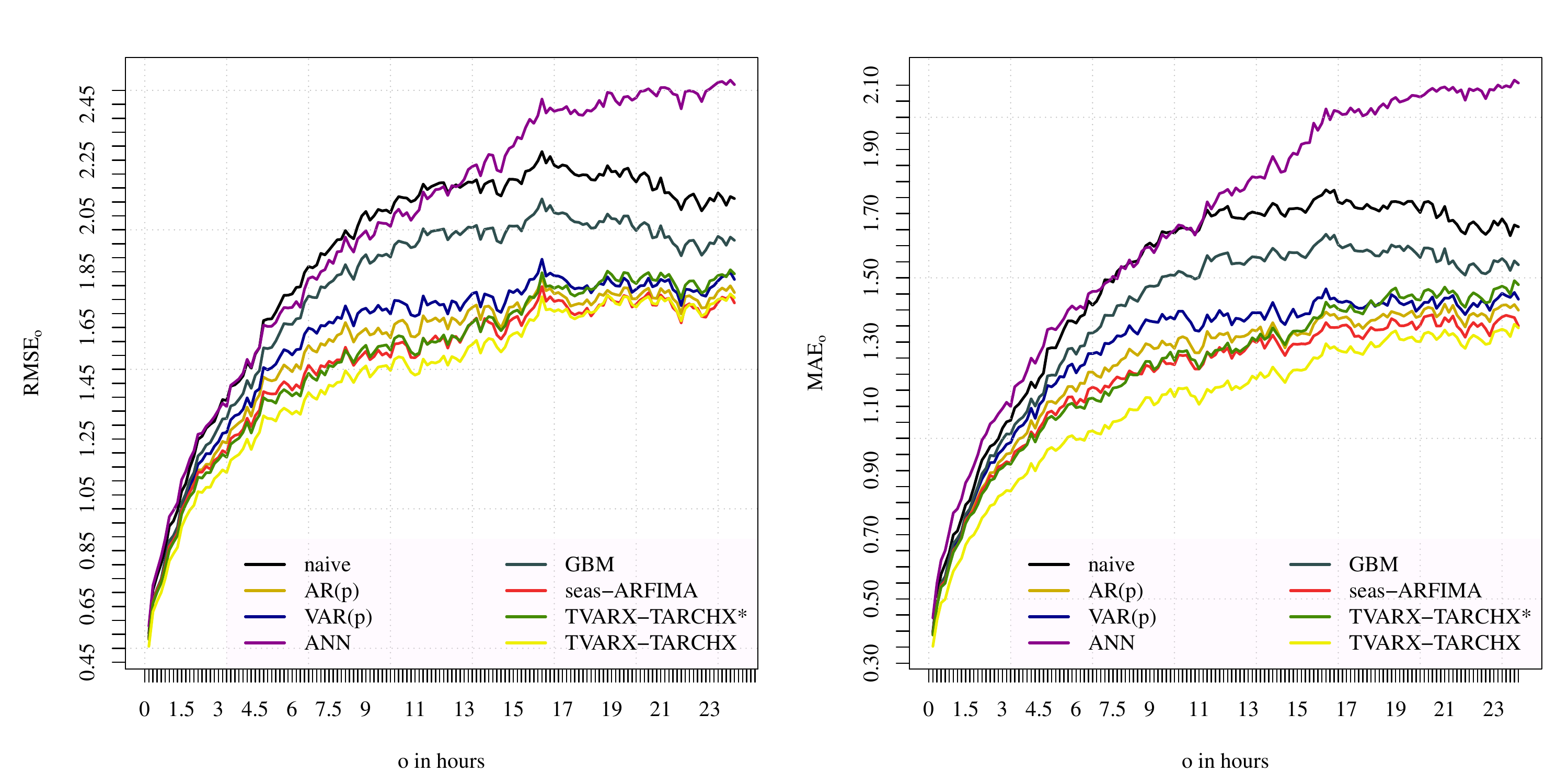}
  \caption{RMSE (first column) and MAE (second column) for the wind speed at station Berlin-Tempelhof and a forecasting horizon up to one day}\label{figure:meas4}
\end{figure}

Figures \ref{figure:meas3} and \ref{figure:meas4} depict the RMSE and MAE of the air pressure (first row) and the wind speed (second row) for the station Berlin-Tempelhof. For the RMSE of the air pressure a clear ranking of the applied forecasting method is not possible. However, the new forecasting technique always belongs to the best procedures. For the MAE criterion the periodic seasonal TVARX-TARCHX model turns out to be the best forecasting model. These results verify the previous findings for the station Berlin-Tegel. Note that the second forecast of the air pressure $\widehat{P}_{\tau^{(i)}+o}^\ast$ is clearly outperformed by the AR, the VAR and the seasonal ARFIMA-APARCH model. Regarding the MAE, we observe the best forecasts for the seasonal TVARX-TARCHX model. The results of the RMSE of the wind speed prediction are a bit different than those for Berlin-Tegel. Our novel forecasting model is not significantly better than the seasonal ARFIMA-APARCH model, if the forecasting horizon increases beyond eighteen hours. Nevertheless, the MAE provides a clear recommendation for the new wind speed prediction model. Certainly, the forecasting results $\widehat{W}_{\tau^{(i)}+o}^\ast$ of our periodic and seasonal TVARX-TARCHX outperform the considered benchmark models.
  
Finally, we are able to conclude that the forecasts $\widehat{P}_{\tau^{(i)}+o|\tau^{(i)}}$ and $\widehat{W}_{\tau^{(i)}+o|\tau^{(i)}}$ of the periodic seasonal TVARX-TARCHX model provide the best forecasting results. The Diebold-Mariano test approves this result. 

\begin{figure}[h]
  \centering
  \begin{subfigure}[b]{0.49\textwidth}
  \includegraphics[width=1\textwidth]{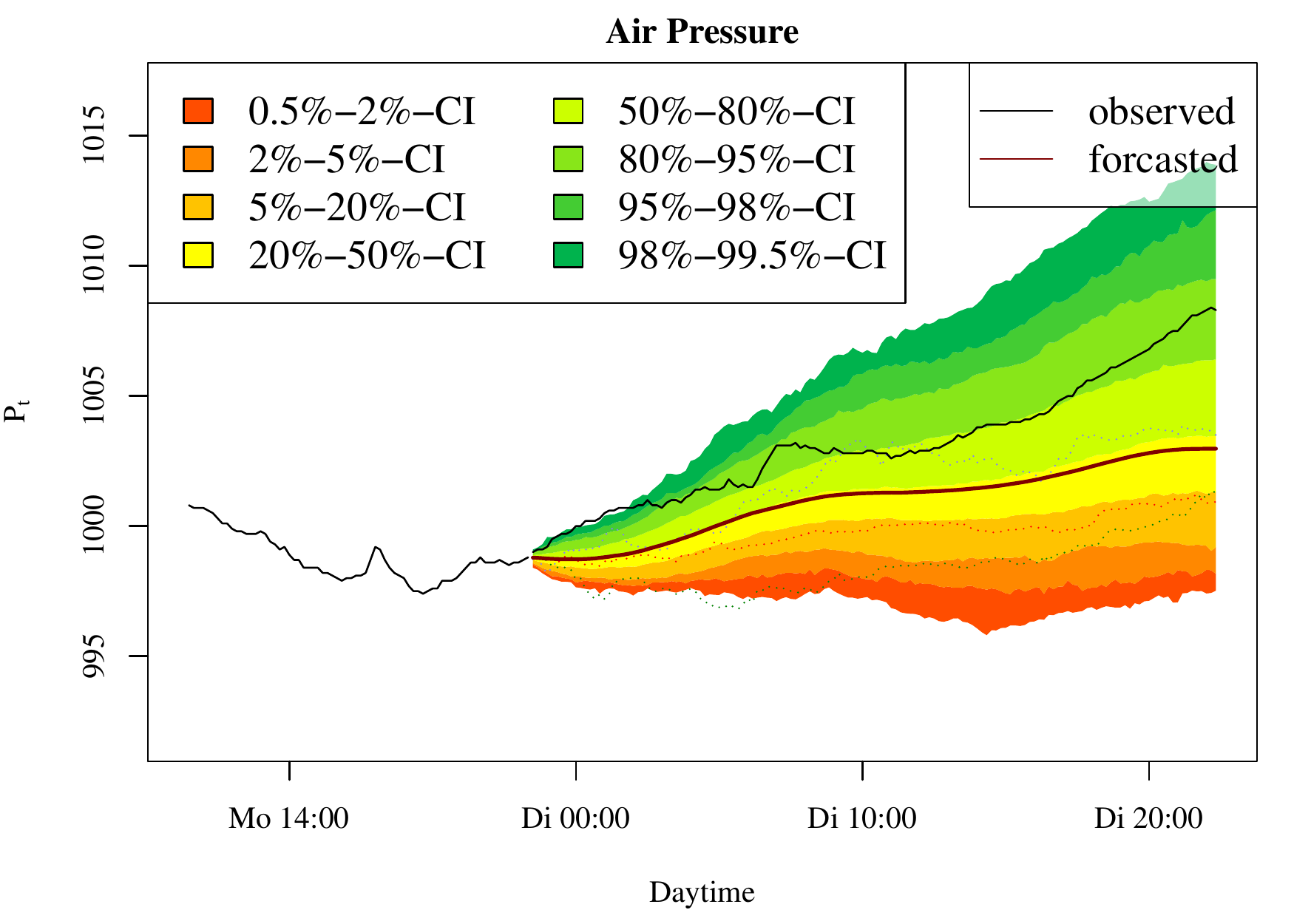} \\
  \includegraphics[width=1\textwidth]{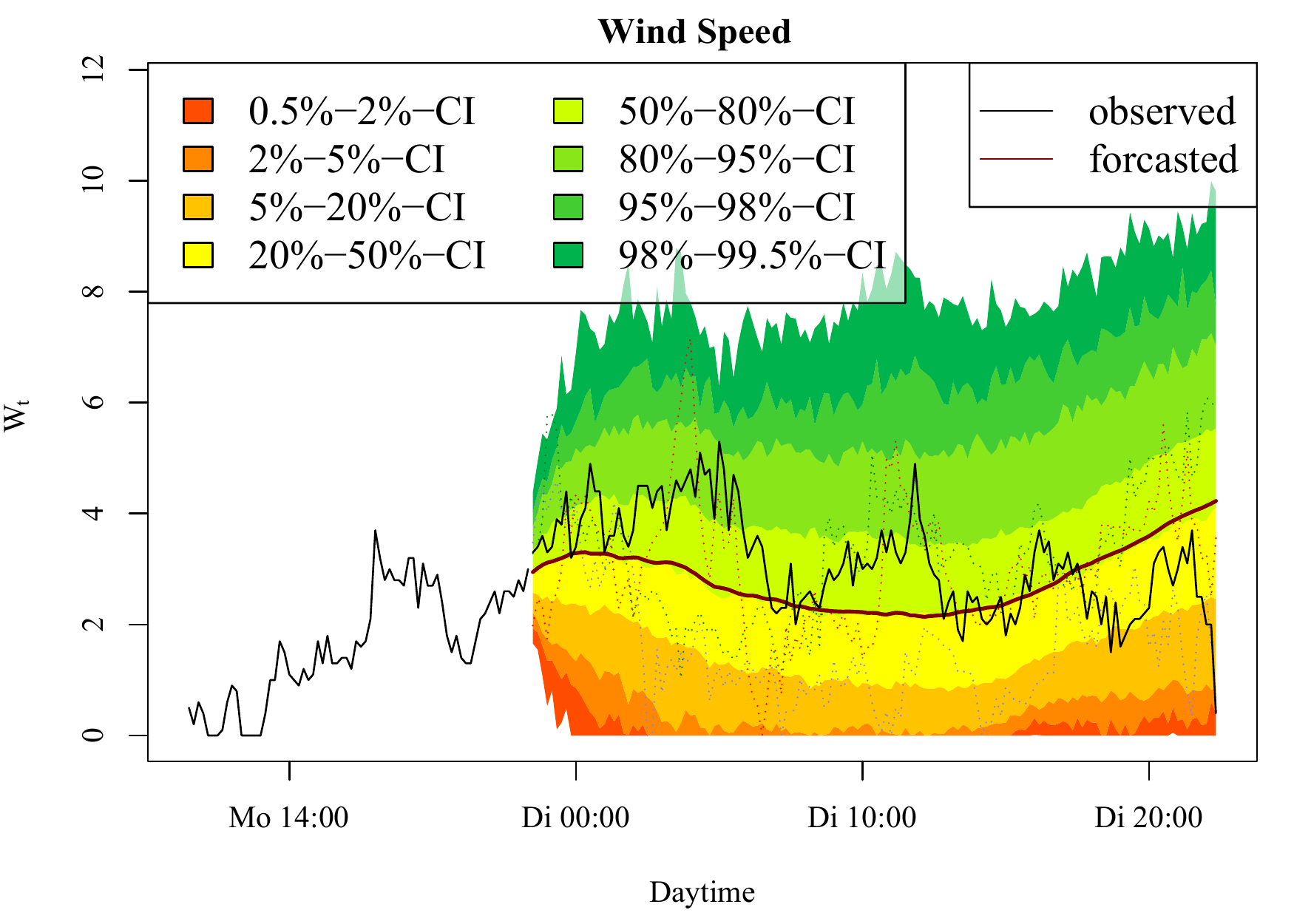}
    \caption{Bootstrap predictions for Berlin-Tegel.}
  \end{subfigure}
  \begin{subfigure}[b]{0.49\textwidth}
   \includegraphics[width=1\textwidth]{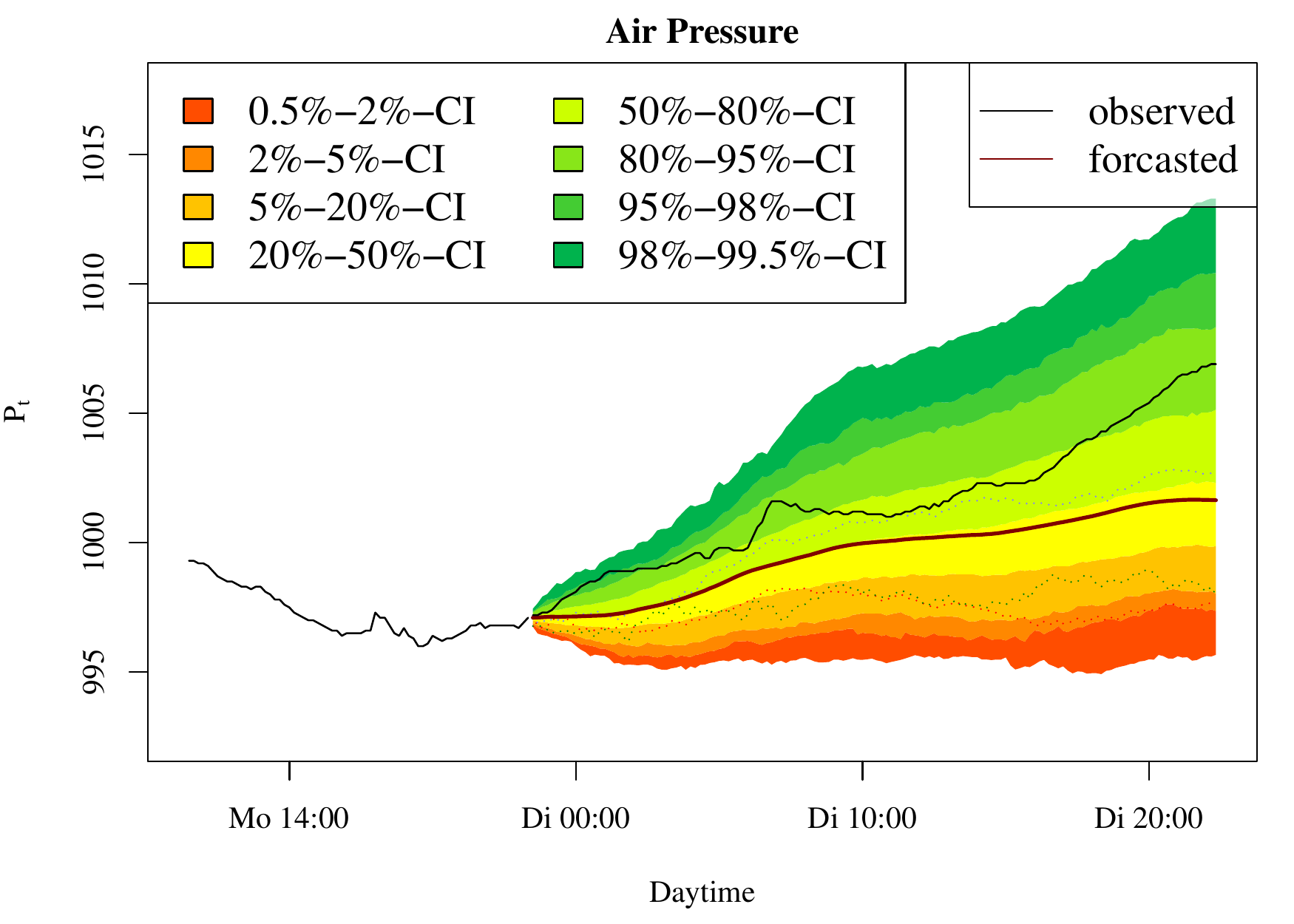} \\
   \includegraphics[width=1\textwidth]{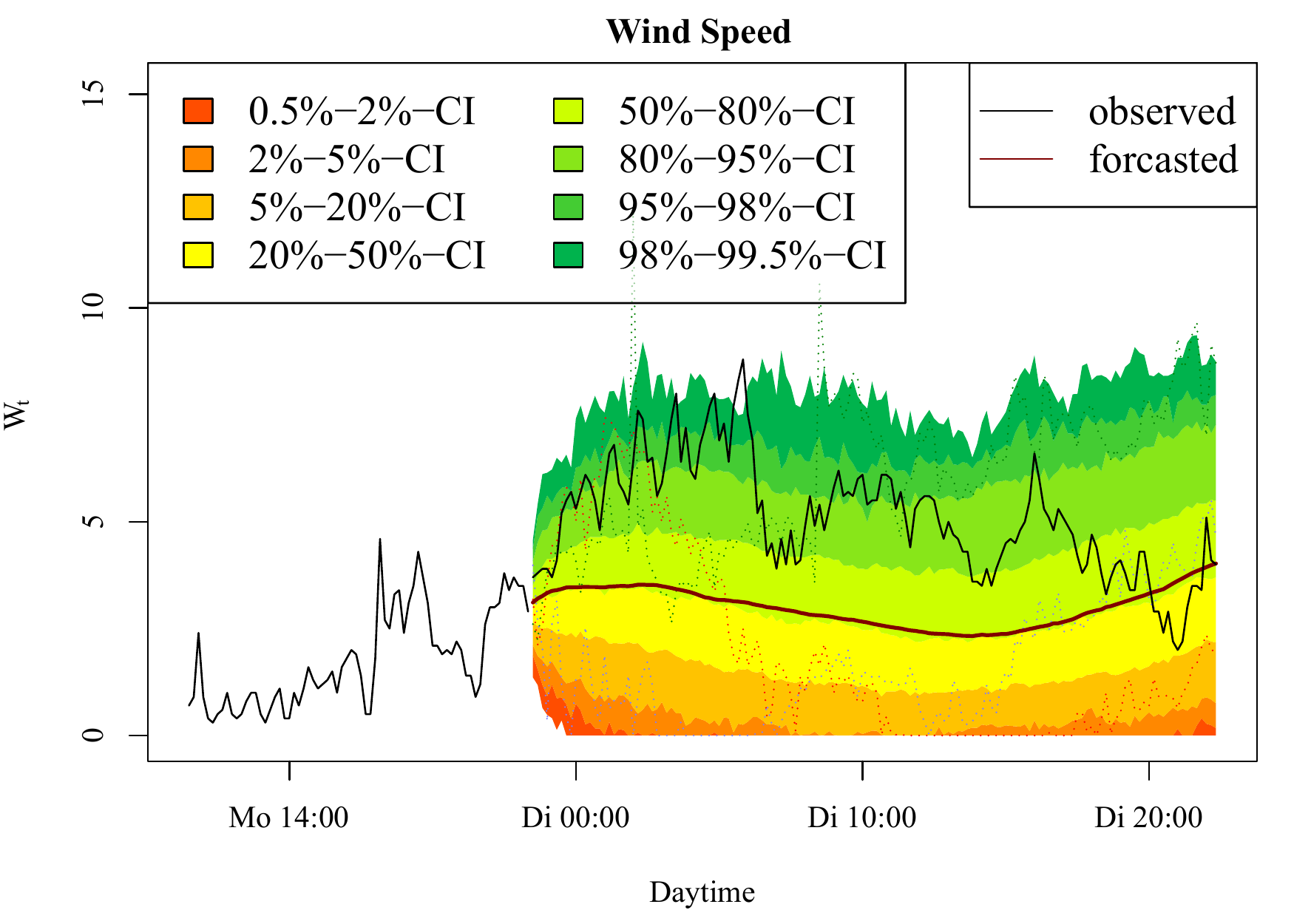}  
    \caption{Bootstrap predictions for Berlin-Tempelhof.}\label{graph:prob1}
 \end{subfigure}
  \caption{Bootstrap predictions for air pressure (first row) and wind speed (second row).}
  \label{graph:prob}
\end{figure}

Following \cite{ziel2016lasso} we derive residuals based bootstrap sample paths for the wind speed and air pressure. Primarily, we must evaluate the empirical quantiles of the bootstrap samples paths. Subsequently, we obtain estimates for the corresponding quantiles. Figure \ref{graph:prob} shows the probabilistic wind speed and air pressure forecast for the 99.5 percentiles of the stations Berlin-Tegel and Berlin-Tempelhof, starting at Monday 25th of November 2014, 22:30. The Figure reveals a daily periodicity and a heteroscedastic variance for both variables. Obviously, if the  wind speed increases we observe larger confidence intervals. The intervals for our air pressure forecasts get larger from hour to hour. Nevertheless, we observe that our prediction intervals are not so wide for the wind speed and the air pressure. Therefore, we observe situations where the actual value is larger than the 99.5\% confidence band. For the air pressure this situation is observed for short-term forecasting horizons. Figure \ref{graph:prob1} shows that the wind speed is highly volatile and might cross the upper confidence intervals between short- to medium-term. Therefore, the upper prediction intervals seem to be too conservative.

\begin{figure}[h]
  \includegraphics[width=1\textwidth,trim= .01cm .3cm .2cm .7cm,clip=true]{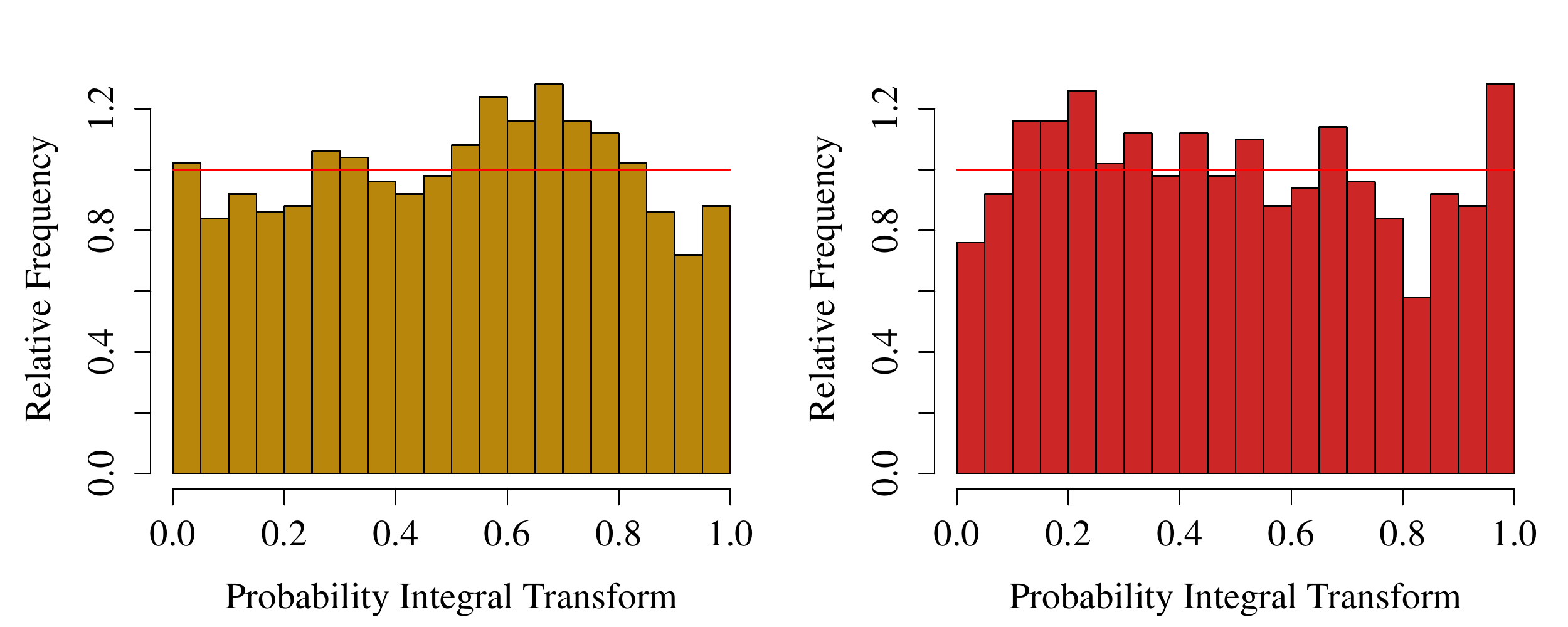} \\
  \includegraphics[width=1\textwidth,trim= .01cm .3cm .2cm .7cm,clip=true]{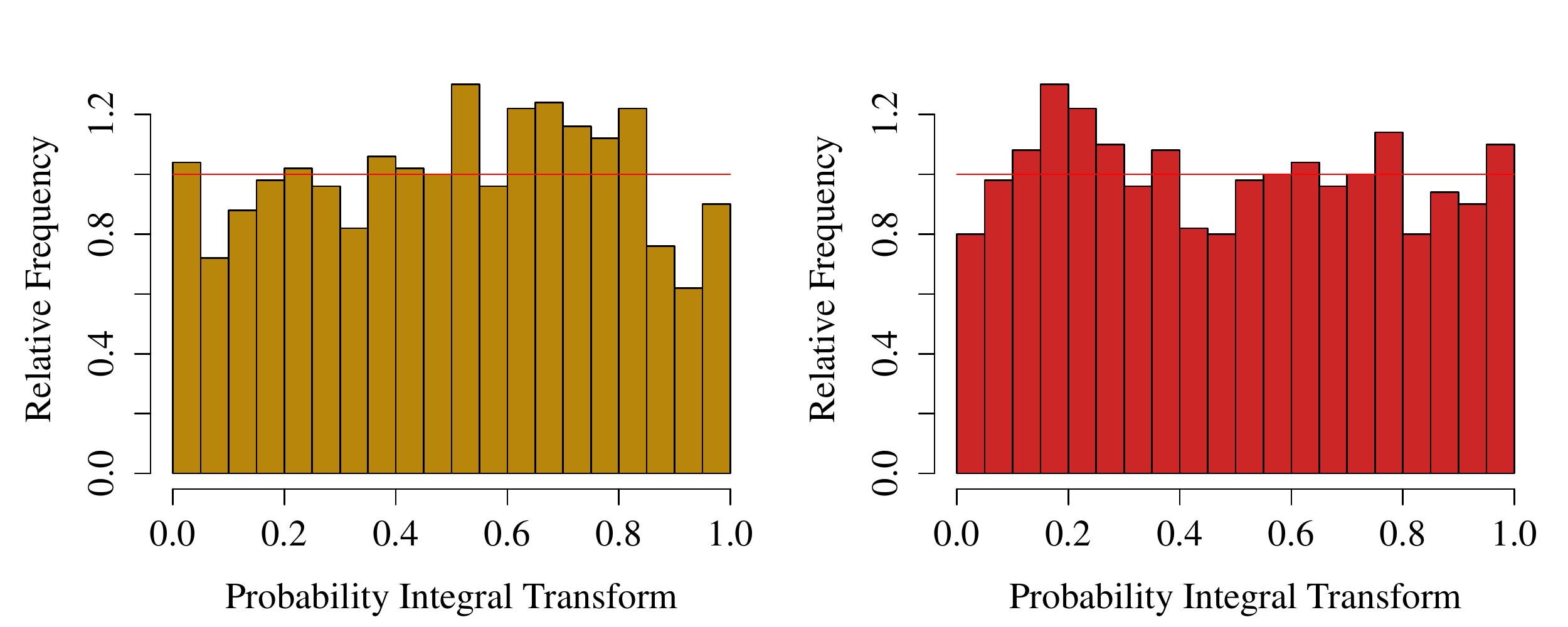} 
  \caption{Probability integral transform (PIT) histogram for wind speed (first column) and air pressure (second column) for the forecasting horizon of $o = 1$ up to $o =72$ for Berlin-Tegel (first row) and Berlin-Tempelhof (second row).}\label{figure:PIT}
\end{figure}

\noindent Figure \ref{figure:PIT} shows the PIT histogram for short- to medium-term forecasting horizons of ten minutes up to twelve hours and the iteratively re-weighted LASSO forecasting technique. The calibration and sharpness of our forecasts is good if forecasting horizons are between short- to medium-term. If we analyse very short- to short-term forecasts, the PIT histogram provides a certain degree of an under-confident model calibration (this result is omitted here to conserve space, but it is available upon request). These findings support our new approach for short-term predictions.

\begin{figure}[h]
  \includegraphics[width=.7\textwidth,clip=true]{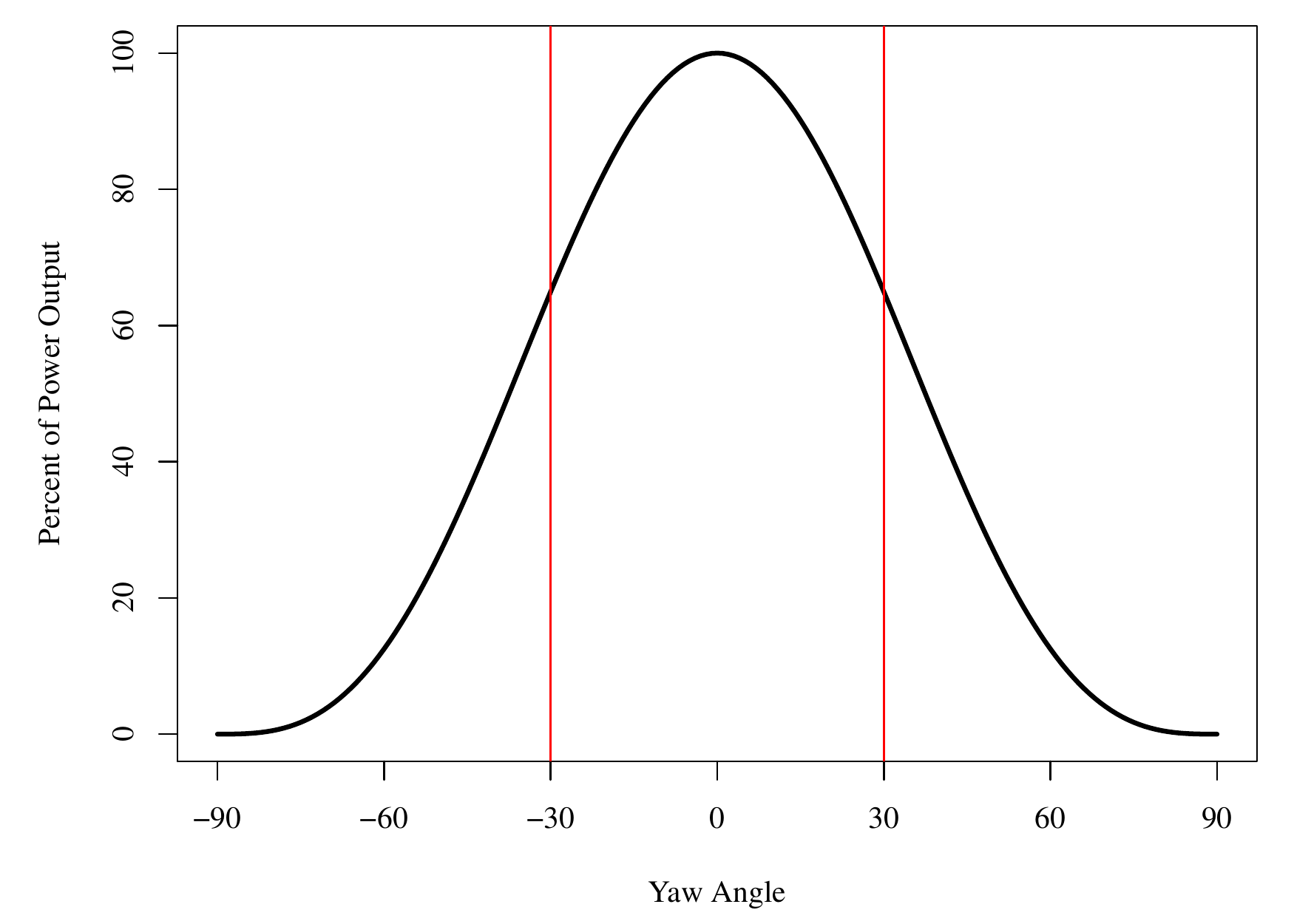}
  \caption{Loss of wind power related to a mismatch of yaw adjustment.}\label{graph:loss}
\end{figure}

\begin{figure}[h]
  \includegraphics[width=1\textwidth,clip=true]{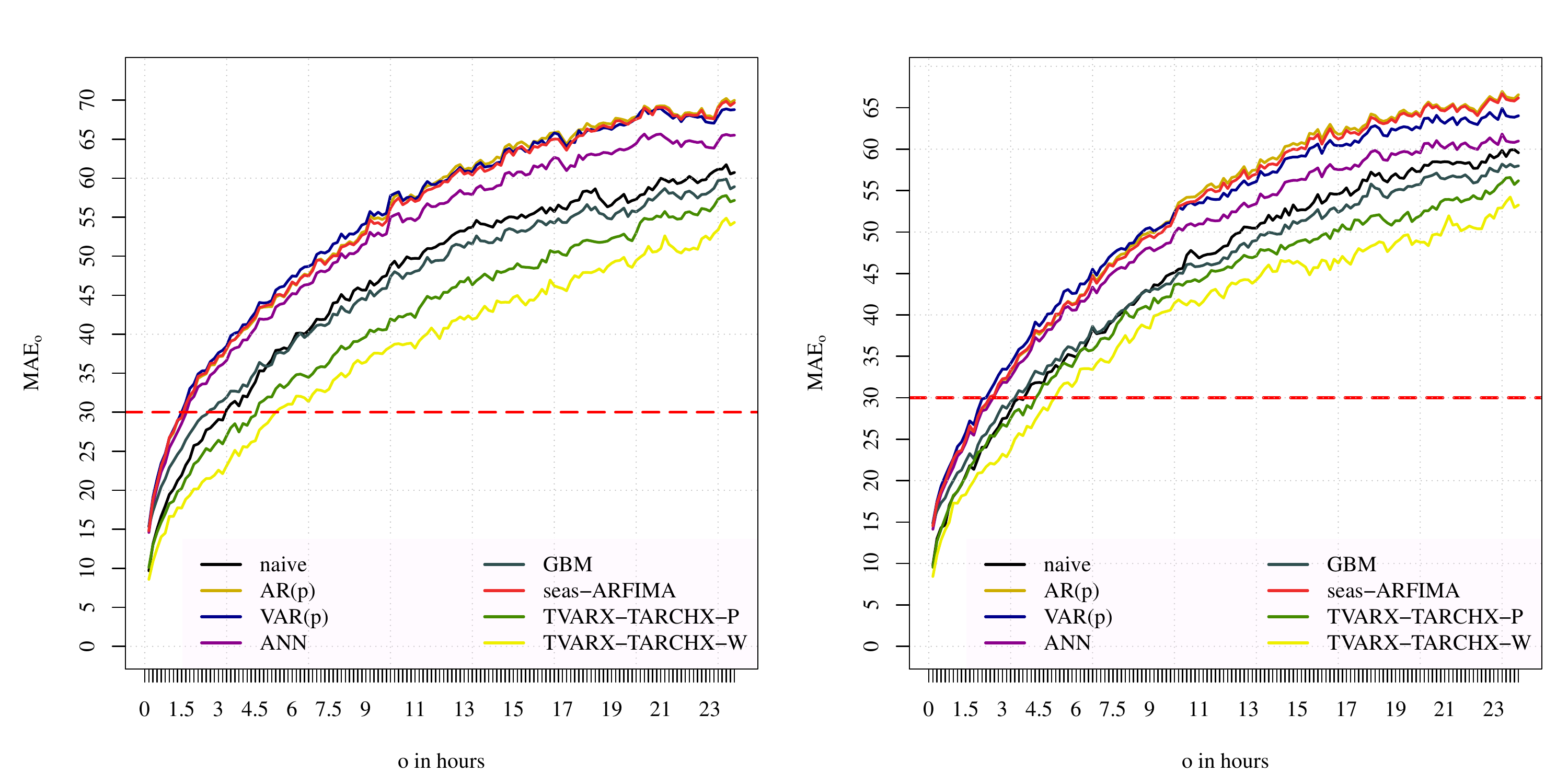}
  \caption{MAE for the wind direction (left side Berlin-Tegel, right side Berlin-Tempelhof).}\label{figure:meas5}
\end{figure}

The accuracy of the wind direction forecast has a direct influence on the efficiency of the wind turbine. Such turbines are actively controlled to turn directly into a predominant wind direction \citep[see][]{Burton2011}. Mostly, wind turbines use the information which is measured on the back of the nacelle. A mismatch between the observed wind direction and actual position of the nacelle, results in a loss of the produced wind power. However, the power and non-symmetrical loads could be maximized by minimizing the yaw angle. Hence, the wind direction has a huge variability and if the wind direction varies rapidly, the wind turbine will follow with a small delay. Therefore, each turbine has a slight yaw angle. \cite{johnson2006control} and \cite{wan2015effects} describe the problem of yaw angle misconfiguration. The approximated loss of wind power is given by \citep[see][]{johnson2006control}

\begin{equation}
 \Delta Pow_{t} = (cos( \widehat{\varepsilon}_{\tau^{(i)}+o)} ))^3,
\end{equation}

where $ Pow_{loss,t}$ is the loss of wind power related to a yaw error. The corresponding loss function which is related to a specfic yaw angle is shown in Figure \ref{graph:loss}. \cite{wan2015effects} describe the effect of yawing and argue that for specific wind speeds the effect of yawing can reduce the wind turbines output significantly. Moreover, they state that a yaw error which is larger than $30^\circ$ provides a very large loss of the wind power. \cite{hure2015optimal} focus on an optimal yaw control system related to very short-term wind speed predictions. However, we perform very short- to medium term predictions. Moreover, we obtain two different wind direction predictions $\widehat{D}_{\tau^{(i)}+o}^{\ast,W}$ and $\widehat{D}_{\tau^{(i)}+o}^{\ast,P}$ which are shown in Figure \ref{figure:meas5}.

Figure \ref{figure:meas5} provides the mean absolute error of the wind direction. Furthermore, we draw a red dotted horizontal line into our plot which corresponds to a yaw error of $30^\circ$. If we cross the dotted line, the wind power loss is too high. The square above the red dotted line represents a high wind power loss area. The wind direction forecast $\widehat{D}_{\tau^{(i)}+o}^{\ast,W}$ of our new model discloses the best results. The yaw error remains below $30^\circ$ up to a time span of five hours, which is a good result. The predictions of $\widehat{D}_{\tau^{(i)}+o}^{\ast,W}$ and $\widehat{D}_{\tau^{(i)}+o}^{\ast,P}$ are not similar, but they outperform all the other benchmark methods. The Diebold-Mariano test is able to confirm this result, but the wind direction predictions which are obtained by $\widehat{D}_{\tau^{(i)}+o|\tau^{(i)}}^{W}$ are significantly better than $\widehat{D}_{\tau^{(i)}+o|\tau^{(i)}}^{P}$. The na\"ive benchmark model provides the third best result, which is surprising. Nevertheless, our new approach is able to perform the best wind direction predictions. Finally, after investigating the impact of the accuracy of the wind direction forecast on the efficiency of a wind turbine we come up with the illustration of the border line to depict the area of wind power loss. 

Summing up, our forecasting approach leads to better predictions for the wind speed, the wind direction and the air pressure than the other considered forecasting methods. Note that in our study, we focus on the accuracy of short-term predictions. Our results are in line with recent research \citep[see][]{jung2014current}. The new model provides a computing resource-saving method which provides accurate forecasts. Our model is able to capture the joint stochastic process of the wind speed, the air pressure and the wind direction to calculate accurate short-term predictions. Moreover, we provide the medium-term results and additionally confidence intervals for our new approach. The results are consistently good, so that our model is appropriate for short- to medium-term forecasts, but there is still capacity for further enhancements. In future we want to include combined approaches and NWP models in our comparison study to discuss the medium-term accuracy.

\section{Conclusion}\label{sec:con}

\noindent The production and the capacities of wind power production have been increasing within recent years. The main driving force of wind power is, on the one hand, the energy turnaround and on the other hand, the availability and the appearance of wind speed. This article focuses on the problem of modelling and forecasting wind speed. Since \cite{ambach2015periodic} compare different univariate time series models for wind speed predictions, we observe a growing popularity of space time or multivariate wind speed predictions. \cite{zhu2014incorporating} include different meteorological variables to improve the wind speed predictions. Therefore, we conclude, that different meteorological dependent variables might increase the wind speed modelling performance and lead to better wind speed forecasts.

Many characteristics of wind speed have already been described in literature, but including different meteorological dependent variables, new insights are obtained. Our multivariate prediction model includes three different types of explanatory variables, the wind speed, the air pressure and the wind direction. We take the Cartesian coordinates of the wind direction to ensure that we model the wind direction in an appropriate way. The correlation structure of all dependent variables is analysed and we propose a model which performs simultaneous multivariate forecasts of each predicted variable. The novel prediction approach includes in the mean model possible non-linearities, persistence, a specific correlation structure of all dependent variables, periodic and seasonal effects. The conditional standard deviation of each response variable shows a presence of conditional heteroscedasticity which is described by our periodic threshold ARCH model. The proposed periodic seasonal TVARX-TARCHX model is estimated by means of an iteratively re-weighted LASSO algorithm which has the ability to incorporate heteroscedasticity. Finally, we derive a novel modelling approach which is able to capture almost all characteristics of our dependent variables. One exception is the correlation structure of the air pressure. This might be related to the specific structure of the process. Therefore, we obtain a possibility to extend the model performance if we find further explanatory variables which describe the air pressure. Moreover, there are further possible extensions if as input NWP models are considered. For this situation, we will develop a detailed comparison study.  

The novel periodic seasonal TVARX-TARCHX model includes air pressure, wind direction and wind speed to perform accurate short-term forecasts. It outperforms several benchmarks for almost each dependent variable. We observe accurate short-term prediction results which are related to our model enhancements, but the medium-term forecasts must be evaluated carefully. For our further research, we will extend our comparison study and include NWP models. The novel model approach achieves several advantages compared to other short-term prediction approaches. It is easy to apply, to implement and the calculation time is short. Our modelling approach is cost saving and needs only the explained variables. Certainly, we derive a plenty of modelling and forecasting improvements, but some enhancements still seem to be possible. We derive a couple of forecasts with our novel prediction model which might be combined with NWPs \citep[see][]{graff2014wind}. The obtained portfolio of predictions might provide even better forecasts than a single model. 


\end{document}